\documentclass[12pt]{article}

\usepackage{blindtext}
\usepackage[normalem]{ulem}
\usepackage{xcolor}
\usepackage{amsmath,amssymb}
\usepackage{subcaption}
\usepackage{multirow}
\usepackage{comment}
\usepackage{natbib}

\usepackage{algorithm}
\usepackage{algorithmicx}
\usepackage{enumitem}

\newtheorem{innercustomprop}{Proposition}
\newenvironment{customprop}[1]
  {\renewcommand\theinnercustomprop{#1}\innercustomprop}
  {\endinnercustomprop}

\newtheorem{innercustomlem}{Lemma}
\newenvironment{customlem}[1]
  {\renewcommand\theinnercustomlem{#1}\innercustomlem}
  {\endinnercustomlem}
\usepackage{xcolor}
\usepackage{hyperref}
\hypersetup{
    colorlinks=true,
    citecolor=blue,
    linkcolor=blue, 
    urlcolor=blue
}

\newtheorem{theorem}{Theorem}

\newtheorem{definition}{Definition}

\newtheorem{remark}{Remark}

\newcommand{\xx}{\boldsymbol{x}}
\newcommand{\XX}{\boldsymbol{X}}

\newcommand{\VV}{\mbox{\boldmath $V$}}
\newcommand{\UU}{\mbox{\boldmath $U$}}
\newcommand{\vv}{\mbox{\boldmath $v$}}

\newcommand{\bzero}{\mbox{\boldmath $0$}}

\newcommand{\BB}{\mbox{\boldmath $B$}}
\newcommand{\bA}{\mbox{\boldmath $A$}}
\newcommand{\MM}{\mbox{\boldmath $M$}}

\newcommand{\bbP}{\mbox{\boldmath $P$}}

\newcommand{\bbeta}{\boldsymbol{\beta}}
\newcommand{\bvarrho}{\boldsymbol{\varrho}}
\newcommand{\boleta}{\mbox{\boldmath $\eta$}}
\newcommand{\balpha}{\mbox{\boldmath $\alpha$}}

\newcommand{\bpi}{\mbox{\boldmath $\pi$}}

\newcommand{\bxi}{\mbox{\boldmath $\xi$}}

\newcommand{\bPhi}{\mbox{\boldmath $\Phi$}}

\newcommand{\bSigma}{\mbox{\boldmath $\Sigma$}}

\newcommand{\bzeta}{\mbox{\boldmath $\zeta$}}

\newcommand{\bgamma}{\mbox{\boldmath $\gamma$}}
\newcommand{\bTheta}{\mbox{\boldmath $\Theta$}}

\newcommand{\boeta}{\boldsymbol{\eta}}

\newcommand{\norm}[1]{\left\lVert#1\right\rVert}

\usepackage[figuresright]{rotating}

\renewcommand{\hat}{\widehat}

\newcommand{\argmax}{\operatornamewithlimits{argmax}}
\DeclareMathOperator*{\esssup}{ess \, sup}

\usepackage{lastpage}

\begin{document}

\title{Joint Estimation and Feature Selection in Overspecified Finite Mixture of Regressions}

\author{Abbas Khalili\thanks{The first three authors are listed alphabetically and contributed equally.} \footnote{Department of Mathematics and Statistics, McGill University, Montreal, Canada. \\ email: abbas.khalili@mcgill.ca} \and Pengqi Liu\footnotemark[1] \footnote{Department of Mathematics and Statistics, McGill University, Montreal, Canada. \\ email: pengqi.liu@mail.mcgill.ca} \and Huy Nguyen\footnotemark[1] \footnote{Department of Statistics and Data Science, The University of Texas at Austin, Austin, TX 78712, USA. email: huynm@utexas.edu} \and Nhat Ho  \footnote{Department of Statistics and Data Science, The University of Texas at Austin, Austin, TX 78712, USA. email: minhnhat@utexas.edu}} 

\date{}
\maketitle

\begin{abstract}
We investigate joint estimation and feature selection in sparse finite 
mixtures of regressions with an overspecified number of mixture components or mixture order.
More specifically, given a conservative fixed upper bound on the unknown 
mixture order, we 
deliberately overspecify the order and propose a double-penalized likelihood 
approach for simultaneous parameter estimation
and feature selection. 
We establish consistent recovery of the 
sparsity structure of 
the underlying mixing measure in terms of its regression coefficients, 
showing that the true sparsity 
of the mixture regression can be consistently identified even 
under order over-specification. 
The proposed methodology incorporates two complementary penalties:\ the first penalizes small (near-zero) mixing proportions, inducing an asymptotic Voronoi-type partition of the estimated regression vectors into clusters associated with the true sparse regression vectors; 
the second penalizes regression coefficients, 
resulting in sparsity through the selection of relevant features. 
Although the sparsity penalty is applied to individual regression coefficients, the combined effect of the two penalties yields an asymptotic group feature selection property 
where the same features tend to be selected or not selected 
within each Voronoi cluster. We develop a modified EM algorithm combined with 
proximal gradient for numerical implementations. 
Extensive simulations and a real data 
application demonstrate the effectiveness of the proposed approach.
\end{abstract}


\section{Introduction}
\label{sec:Introduction}

Finite mixture of regressions (FMR) 
is widely used in supervised learning problems that involve heterogeneous populations 
\citep{MCLACHLAN2000}. The objective is to model the relationship between a response 
variable \( Y \) and a set of features \( \XX = (X_1, \ldots, X_p)^\top \), when the 
data arise from a few unobserved (latent) subpopulations 
\citep{fruhwirth-schnatter2018}. The observed data are on  
\((\XX,Y) \), while the subpopulation memberships remain 
unobserved. Since the relationship between \( Y \) and \( \XX \) may vary across subpopulations,  
an FMR allows for different regression models within each subpopulation. 
 These models have found applications in various domains, including statistical genetics \citep{SchorkAllisonThiel1996, McLachlanLeeSuren2019}, marketing  
\citep{WedelKam2000}, economics, finance, social sciences 
\citep{fruhwirth2006finite, Campiani2016}, 
and machine learning \citep{goodfellow2016}.

Likelihood-based inference in FMRs, particularly maximum likelihood estimation (MLE),
has been extensively studied. When the number of latent subpopulations or mixture
order, $K_0$, is known, the MLE possesses standard asymptotic properties such as
$\sqrt{n}$-consistency and asymptotic normality \citep{GEER2000,Chen2023},
where $n$ denotes the sample size. These properties enable the construction of
approximate confidence intervals and the implementation of hypothesis tests for
the regression parameters. There has also been a surge of research on feature 
selection in sparse FMR models using continuous Lasso-type regularization techniques; 
see, for example, \cite{KhaliliChen2007,Stdler20101penalizationFM,Devijver2015,LioydNg2018},
among others. To establish estimation and sparsity consistency of the resulting
regularized estimators, these methods typically assume that the mixture order
$K_0$ is known. In contrast, statistical inference on the mixture order in FMR
models has been studied through hypothesis testing
\citep{ZhuZhang2004,KasaharaShim2015,ShenHe2015,Takeishi2023}
and model selection methods based on AIC/BIC and their variations
\citep{Bozdogan1987,BierCeleuxGov2000,SarSchw2008,fruhwirth-schnatter2018}.
Bayesian methods for order and feature selection in sparse FMRs include 
\cite{GuptaIbrahim2007} and \cite{Liuetal2015}, among others. In this manuscript, 
given the well-known difficulty of determining the true
mixture order $K_0$, we take a different perspective 
and rather than estimating $K_0$ we allow the order 
to be overspecified with a fixed upper bound $K>K_0$. 
We then study simultaneous parameter estimation and feature selection in sparse FMR models under 
this 
setting, which we refer to as an \emph{overspecified FMR model}.

The foundational works of \cite{CHEN1995}, \cite{CHEN2003}, \cite{HO2016weak}, and \cite{HO2016strong} 
established the consistency of the MLE in Wasserstein distance 
for the true mixing measure in overspecified finite mixture models.
More recently, \cite{HOJMLR2022} and \cite{DoDoLong2025} 
investigated the MLE in overspecified FMR models based on the Gaussian and 
exponential families, respectively. These works established consistency of the 
MLE in Wasserstein distance for the true mixing measure, with convergence rates of $n^{-1/4}$ 
or slower, depending on the underlying parametric family. 
The results, however, do not study 
feature selection in sparse FMRs under order overspecification.
Building on this line of work, we introduce a double-penalized likelihood framework that 
preserves the estimation consistency of the MLE under overspecification while additionally 
recovering the sparse structure of the underlying FMR, even when the mixture is fitted 
with a fixed overspecified order $K>K_0$.

In our proposed method, we apply
two penalty functions to the log-likelihood of an
overspecified FMR model. The first 
penalizes near-zero estimates of 
the mixing probabilities, 
preventing 
fitted mixture components from having vanishing mixing 
probabilities under overspecification.
The second penalty is applied to 
the regression coefficients to induce sparsity and facilitate feature selection. 
A key property of the proposed penalization method is that the first penalty asymptotically 
induces a Voronoi-type partition of the $K$ fitted regression vectors 
into $K_0$ clusters associated with the $K_0$ true sparse regression vectors.
Thus, the 
fitted mixture components of the overspecified model  
are naturally grouped around the underlying true
components; 
see Figure 
\ref{fig-sparse1}, and 
Theorem \ref{thm:paramConsistency}. 
In particular, the fitted mixing measure is shown to lie within an
$n^{-1/4}$-neighborhood, up to a logarithmic factor, of the true mixing 
measure in Wasserstein distance; see Proposition \ref{densityConsistency} 
in Appendix~\ref{sec:additional_results}.
This rate is also attained by the MLE in overspecified mixture and FMR models
studied in \cite{HOJMLR2022} and \cite{DoDoLong2025}. Within each Voronoi cluster (cell), 
the second penalty asymptotically induces sparsity in the fitted regression vectors, thereby
recovering the sparse structure of the true regression vectors; see Figure 
\ref{fig-sparse1} and Theorem \ref{thm:sparsity}.
The proposed estimators thus have similar estimation consistency properties as the MLE 
while additionally achieving consistent sparsity recovery, which is the main 
focus of our work. Interestingly, although the sparsity penalty is applied separately to individual
regression coefficients of the overspecified FMR, the resulting estimators exhibit an
asymptotic group feature selection property \citep{YUAN2006}, 
where the same features tend to be
selected or not selected across the fitted regression vectors within 
each Voronoi cluster; see Remark \ref{rmk3} toward the end of Section \ref{sec:asymp}.

Our main contributions are as follows. First, we propose a double-penalized
likelihood method for fitting sparse FMR models with a fixed overspecified order $K>K_0$,
where the two penalties, respectively, prevent mixture components
from having vanishing mixing probabilities and induce sparsity in the regression coefficients. 
Second, we establish theoretical consistency of the proposed method for both 
parameter estimation and feature selection. In particular, asymptotically, 
our results characterize a 
Voronoi-type clustering of the fitted mixture components of the overspecified model, 
establish recovery of the sparse structure of the true underlying FMR, and demonstrate 
a group feature selection property among the fitted components within each cluster. 
Third, we develop a computational procedure based on a modified EM algorithm combined with the
proximal gradient method \citep{fercoq2015accelerated} to obtain numerical
approximations to the proposed penalized estimators. Extensive simulation
studies and a real data analysis are conducted to empirically demonstrate the
performance of the proposed method and support our theoretical findings.

The manuscript is organized as follows. In Section \ref{sec:fmr-model}, we provide a formal definition of the FMR models, including well-known examples of this family of statistical models. Our proposed estimators are then introduced in Section \ref{sec:feature-selec} along with examples of different choices of penalty functions fulfilling our theoretical conditions. In Section \ref{sec:asymp}, we present the theoretical properties of the proposed estimators. Numerical implementation 
and simulation study on the performance of the methods are provided in Sections \ref{sec:numeric} and \ref{sec:simul}, respectively. To demonstrate our main result, a real data example is provided in Section \ref{sec:real_data}. A conclusion and discussion are given in Section \ref{sec:discussion}. 
Regularity conditions, proofs, additional simulations, and more results on 
the real data analysis are given in the Appendix. 

\noindent
{\bf Notation.} For any vector $\vv=(v_1,\ldots,v_p)\in\mathbb{R}^p$, with $p \in \mathbb{N}$, 
we denote $\|\vv\|_q:=(\sum_{i=1}^{p}|v_i|^q)^{1/q}$ as its $\ell_q$-norm value, for all $q\geq 1$, and $\|\vv\|$ indicates its Euclidean norm $\|\cdot\|_2$. For any $n\in\mathbb{N}$, 
let $[n]:=\{1,\ldots,n\}$. For a sequence $(A_n)_{n\geq 1}$ of positive random variables, the notation $A_{n} = \mathcal{O}_{P}(b_{n})$ indicates that $A_{n}/b_{n}$ is stochastically bounded, that is, for any $\epsilon>0$, 
there exists an $M>0$ such that $\mathbb{P}( A_{n}/b_{n} > M) < \epsilon $, for all sufficiently large $n$. 
For $x \in \mathbb{R}$, we define $x_{+} := \max \{0, x\}$.

\section{Sparse finite mixture of regressions ({\sc fmr})}
\label{sec:fmr-model}

Let $Y$ be a real-valued response and $\XX = (X_1, \ldots,X_p)^\top$ 
be the vector of features that may affect $Y$.
Let ${\cal F} =  \big \{f(y; \theta(\xx, \bbeta), \phi):  
(\bbeta, \phi) \in \Theta = \Omega_1 \times \Omega_2,~
\Omega_1 \subseteq \mathbb{R}^p, \Omega_2 \subseteq \mathbb{R}^+ \big \}$
denote a family of parametric densities with respect
to a $\sigma$-finite measure $\nu_Y$ over a set ${\cal Y}$. 
A well-known choice of $\mathcal{F}$ is the exponential family, including Gaussian, 
Poisson, Binomial, and negative Binomial. 
Furthermore, let ${\cal X} \subseteq \mathbb{R}^p$ be the feature space, 
and $P_{\XX}$ be an absolutely continuous probability distribution 
with respect to a $\sigma$-finite measure $\nu_{\XX}$ on ${\cal X}$. 

A pair of random variables $(\XX, Y)$ 
taking values in ${\cal X} \times {\cal Y}$ follows a 
finite mixture of regression (FMR) model with respect to 
${\cal F}$ if $\XX \sim P_{\XX}$ and the conditional 
density function of $Y$ given $\XX=\xx$,
for any $\xx \in {\cal X}$, is defined as
\begin{equation}
q_G(y| \xx) = 
\sum_{j=1}^{K} \pi_{j} f(y; \theta(\xx, \bbeta_j),  \phi_j)
=
\int f(y; \theta(\xx, \bbeta), \phi) ~dG(\bbeta, \phi),
\label{fmr_model}
\end{equation}  
where $\theta(\xx, \bbeta) := h(\xx^{\top} \bbeta)$ with a 
known link function $h$, and 
$G = \sum_{j=1}^K \pi_j \delta_{(\bbeta_j, \phi_j)}$ is called a mixing measure.
Furthermore, $\pi_1, \dots, \pi_K \geq 0$ are called mixing proportions, 
satisfying $\sum_{j=1}^K \pi_j = 1$, and $(\bbeta_j, \phi_j) \in \Theta$
are called atoms, where $\bbeta_j = (\beta_{j1},\ldots,\beta_{jp})^\top$, 
$j\in[K]$. 

We assume the true FMR model underlying the data is {\it sparse}
with the corresponding mixing measure $G_0 = \sum_{j=1}^{K_0} \pi^0_j 
\delta_{(\bbeta^0_j, \phi^0_j)}$ of order $K_0$. Without loss of generality, let 
\begin{equation}
\label{sparse-assum}
\bbeta^0_j = (\bbeta^0_{j,1}, \bbeta^0_{j,2}),~
\text{such that}~~\bbeta^0_{j,2}=\bzero,~
\text{for all} ~j\in[K_0],~\text{and}~\text{dim}(\bbeta^0_{j,1}) = p_j < p,
\end{equation}
and $\bbeta^0_{j,1}$ is the sub-vector of true nonzero regression coefficients. \\
We also let ${\cal G}_K = \{G: G = \sum_{j = 1}^K \pi_j \delta_{(\bbeta_j, \phi_j)},~
(\bbeta_j, \phi_j) \in \Theta, 
\pi_j \ge 0, \sum_{j=1}^K \pi_j = 1\}$
be the class of mixing measures with at most $K$ atoms. 

Our primary objective is to obtain a Wasserstein-consistent estimator 
of the true mixing measure $G_0$ that also consistently 
recovers the sparse structure of $G_0$
as specified in \eqref{sparse-assum}, without
requiring prior knowledge of the true mixture order $K_0$. Rather than estimating
$K_0$, we take a conservative approach by fitting an overspecified FMR model with
a fixed upper bound $K>K_0$. We show that, with an appropriate regularization
(penalization) of the log-likelihood of the overspecified model, both
Wasserstein-consistent estimation of $G_0$ and recovery of its sparse structure can be
achieved, which we will discuss in the sequel.

\section{Simultaneous estimation and feature selection in overspecified {\sc FMR}}
\label{sec:feature-selec}
Let $(\xx_i, y_i), i \in [n]$, be a random sample of observations from a sparse
FMR model with the true sparse mixing measure $G_0$ as specified in
\eqref{sparse-assum}. Since the true mixture order $K_0$ is unknown, we instead
fit an FMR model with a fixed upper bound $K>K_0$. The log-likelihood function
of the resulting overspecified model is given by
\begin{equation*}
\ell_n(G) 
= 
\sum_{i=1}^{n} \log q_G(y_i|\xx_i),
\end{equation*}
where $q_G$ is the conditional mixture density given in equation \eqref{fmr_model}. 
Our proposed method is based on the overspecified log-likelihood $\ell_n(G)$. 
Before formally presenting the method, we discuss an
observation that motivates our double-penalized likelihood
approach.

\subsection{When existing feature selection methods could fall short}
\label{one-penalty}

As in existing feature selection methods for FMRs, such as
\cite{KhaliliChen2007}, \cite{Stdler20101penalizationFM},
and \cite{LioydNg2018}, one may attempt
to perform simultaneous feature selection and parameter 
estimation in the overspecified FMR by applying a sparsity 
penalty only on the regression coefficients. 
This leads to the penalized estimator
\begin{equation}
    \label{smple}
 \widetilde G_n 
 = 
\sum_{j = 1}^K \tilde \pi_j \delta_{(\tilde \bbeta_j, 
\tilde \phi_j)}
 =
 \argmax_{G \in {\cal G}_K} \bigg \{ \ell_n(G) 
 -
 n\sum_{k=1}^{K} \sum_{j=1}^{p} r_{\lambda_n}(\beta_{kj})
 \bigg \}.
\end{equation}
Under Conditions (P1) and (P2) on the penalty $r_{\lambda_n}$ in
Section~\ref{sec:asymp}, and Conditions (R1)--(R4) on the family ${\cal F}$ in
Appendix~\ref{appendix:notation}, $\widetilde G_n$ is a consistent estimator
of $G_0$ under a generalized Wasserstein distance. This property is also
shared by the overspecified MLE, which maximizes the leading term
$\ell_n(G)$ in \eqref{smple}, as established in
\cite{HOJMLR2022,DoDoLong2025}. However, Wasserstein consistency of 
$\widetilde G_n$ does not necessarily imply
feature selection consistency. In particular, under overspecification, some
fitted atoms $(\tilde{\bbeta}_j,\tilde{\phi}_j)$ may have mixing probabilities
$\tilde{\pi}_j$ that vanish asymptotically and 
need not converge to any atom of
$G_0$ \citep{HO2016weak,HO2016strong}. Consequently, 
the sparsity patterns of
the regression vectors associated with such redundant components may not
correspond to the sparse structure of the true mixing measure. 
We also investigated this point via simulations. We generated 
$100$ random samples, each of size $n = 800$,
from a Gaussian FMR model with the true parameters: $K_0= 2$, 
$\pi_1^0 = 0.45$, dispersion parameters $\phi_1^0 = \phi_2^0 = 1$, and sparse regression vectors 
$\bbeta_1^0 = (2, 2.5, 3, 0, \ldots, 0)^\top$, 
$\bbeta_2^0 = (-2, 0, 2.5, 0, 1.5, 0, \ldots, 0)^\top$ 
of dimension $p=10$. We then fitted Gaussian FMRs with order $K = 4$ 
to each sample using $\widetilde G_n$ in \eqref{smple}. We observed 
that in $94$ out of $100$ samples at least one estimated mixing probability, $\tilde \pi_j$, were close to zero and the corresponding fitted vectors $\tilde \bbeta_j$ were distanced from the true vectors $\bbeta^0_1$ or $\bbeta^0_2$. 
This suggests that applying a sparsity penalty only to the regression coefficients 
as in \eqref{smple} may not be sufficient to guarantee feature selection consistency when $K>K_0$.

Alternatively, one may first obtain $\widetilde G_n$ 
and then remove fitted components whose estimated mixing probabilities $\tilde{\pi}_j$ are smaller than a pre-specified threshold $\eta_n>0$. 
Although such a post-hoc thresholding strategy may seem appealing, 
choosing an appropriate
$\eta_n$ can be nontrivial. If $\eta_n$ is too large, components carrying
non-negligible probability mass may be removed, whereas if $\eta_n$ is too small, redundant components may remain. Moreover, post-hoc thresholding changes the
fitted mixing measure $\widetilde G_n$: the retained mixing probabilities
generally no longer sum to one and hence do not define a valid mixing measure; with renormalization, the resulting estimator is no longer the maximizer of \eqref{smple} over the original $K$-component model, 
and furthermore, its theoretical properties are 
unknown to us. Nevertheless, 
we examined the performance of the thresholding estimator 
in the above simulations by varying the parameter $\eta_n$. 
As expected, its performance in terms of feature selection is 
very sensitive to $\eta_n$; a value $\eta_n = .15$, 
based on trail-error, resulted in satisfactory performance similar 
to our double-penalized estimator in \eqref{MPLE-Est}. 
Our proposed method, on the other hand, 
incorporates a penalty on the mixing probabilities directly into the penalized likelihood in \eqref{smple}, with the penalty diverging as any mixing probability approaches zero. This prevents near-zero estimates of the mixing probabilities and enables simultaneous parameter estimation and feature selection within a single estimation procedure, without requiring a second-stage thresholding of the estimated mixing probabilities.
Next, we introduce our method. 

\subsection{A new double-penalized likelihood method}
\label{two-penalty}

Motivated by the considerations above, we introduce a double-penalized
log-likelihood, 
\begin{equation}
L_n(G) = \ell_n(G) - \rho_K(\pi_1,\ldots,\pi_K) - n\sum_{k=1}^{K} \sum_{j=1}^{p} 
r_{\lambda_n}(\beta_{kj}),
 \label{pen-like}
\end{equation}
where $\rho_K$ is a nonnegative penalty 
such that $\rho_K(\pi_1,\ldots,\pi_K) \to \infty$ 
as $\min_{1 \leqslant k \leqslant K} \pi_k \to 0$. 
We define the overspecified maximum penalized likelihood 
estimator (OMPLE) of $G_0$ as 
\begin{equation}
\hat G_n = \sum_{j=1}^K \hat \pi_j \delta_{(\hat \bbeta_j, \hat \phi_j)}
=
\argmax_{G \in {\cal G}_K} L_n(G).
 \label{MPLE-Est}
\end{equation}
The penalty $\rho_K$ is designed to prevent the fitted mixing probabilities
$\hat \pi_j$'s from approaching zero, thereby resulting in a clustering structure of the 
fitted atoms $(\hat \bbeta_j, \hat \phi_j)$'s in the overspecified model. 
As in \eqref{smple}, the regression coefficient penalty $r_{\lambda_n}$, 
on the other hand, induces sparsity in the clustered fitted regression vectors 
$\hat \bbeta_j$'s. More discussion on these theoretical consequences of the two penalties 
is given below, and formally in Section \ref{sec:asymp}. General 
requirements on the penalties are given 
in Conditions (P0)--(P2) in Section~\ref{sec:asymp}.  
It is worth noting that the numerical implementation of \eqref{MPLE-Est}
does not require any additional computational burden compared to that of \eqref{smple}. 
The penalty $\rho_K$ will be incorporated into the M-step of the EM algorithm 
as outlined in Section \ref{sec:numeric}.

\vspace{.1in}
\noindent
{\bf Examples of penalty functions}: 
\vspace{.1in}

{\bf (i)} In our numerical experiments, we use the penalty: 
\(
\rho_K(\pi_1, \ldots, \pi_K) = -C_K \sum_{j=1}^K \log\pi_j
\),
for some tuning parameter $C_K>0$. This penalty 
was discussed in \cite{CHEN1996, CHEN2001modified} and \cite{LI2009}, 
for order selection and hypothesis testing in finite 
mixtures. The penalty satisfies Condition (P0), and it also 
results in convenient closed-form updates for $\pi_j$
in the M-step of the modified EM algorithm 
in Section \ref{sec:numeric}.

\vspace{.1in}
{\bf (ii)} 
Four well-known choices for the feature selection penalty 
are discussed below. 

\noindent
-- The Smoothly Clipped Absolute Deviation 
(SCAD; \citet{FAN2001}): 
\[ 
\label{scad}
r'_{\lambda_n}(\beta) = 
  \lambda_n ~I \{ |\beta| \leq \lambda_n \} +
  \frac{(a \lambda_n - |\beta|)_+}{a-1} ~I\{ |\beta| > \lambda_n \}, \qquad a > 2.
\] 
-- The Minimax Concave Penalty (MCP; \citet{ZHANG2010}): 
\[ 
\label{mcp}
 r'_{\lambda_n}(\beta) 
=  \big(\lambda_n - |\beta|/a \big)_+,
\qquad a > 1.
\] 
Here, $r'_{\lambda_n}$ is the first derivative of $r_{\lambda_n}$ 
with respect to $\beta$. 
 
\noindent
-- The other two penalties are the Lasso $r_{\lambda_n}(\beta) = \lambda_n |\beta|$ 
\citep{tibshirani1996regression}, and adaptive Lasso 
(AdaLasso) $r_{\lambda_n}(\beta) = \omega \lambda_n |\beta|$ with some 
pre-specified weight $\omega$ \citep{ZOU2006}. However, as further discussed 
following Theorem \ref{thm:sparsity}, they do not satisfy Conditions (P1) and (P2). Nevertheless, 
due to its ease of implementation, we included the Lasso in our simulations. 

Theorem~\ref{thm:paramConsistency} establishes that the estimated mixing
probabilities $\hat\pi_k$ are asymptotically bounded away from zero under the
penalty $\rho_K$.\ Consequently, the $K$ estimated atoms
$(\hat\bbeta_1,\hat\phi_1),\ldots,(\hat\bbeta_K,\hat\phi_K)$ exhibit an
asymptotic Voronoi-type clustering structure, and each cluster is associated
with one of the $K_0$ true atoms
$(\bbeta^0_1,\phi^0_1),\ldots,(\bbeta^0_{K_0},\phi^0_{K_0})$ of $G_0$.
Within each cluster, the estimated atoms converge to their
corresponding true atom 
at a quantifiable rate; see
Proposition~\ref{densityConsistency} in
Appendix~\ref{sec:additional_results}. The 
definition of the Voronoi-type partition is given in Section~\ref{sec:asymp}.
Theorem~\ref{thm:sparsity} establishes
that the OMPLE $\hat G_n$ asymptotically recovers the true sparse structure of
$G_0$ with probability tending to one. Thus, our key results imply that 
simultaneous parameter
estimation and feature selection can be achieved even when the unknown
mixture order is overspecified by a fixed finite upper bound $K>K_0$. 
In contrast, the singly penalized estimator $\widetilde G_n$ in
\eqref{smple} does not guarantee this feature selection property under
overspecification. The two penalties therefore play complementary roles:
$\rho_K$ establishes the clustering structure needed to handle
overspecification, while $r_{\lambda_n}$ recovers the sparse structure within
these clusters, asymptotically. 

\begin{figure}[t]
\centerline
{\includegraphics[width=8cm,height=8cm]{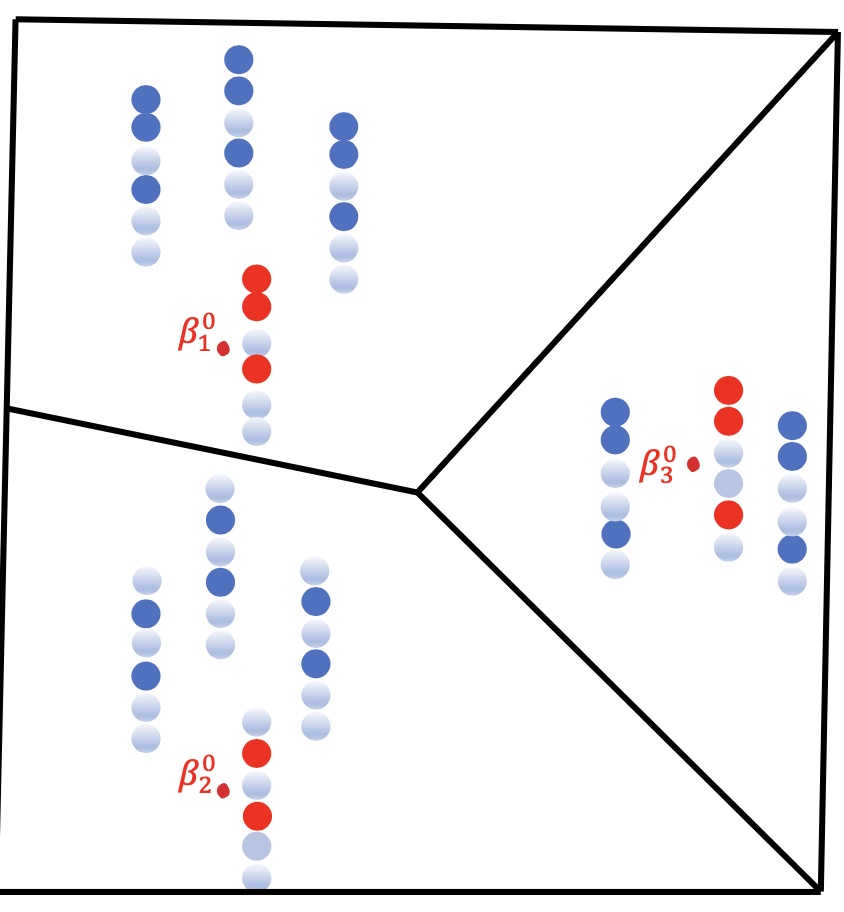}}
\caption{
An overspecified sparse {\sc fmr} with $K = 8$ components (blue, solid) is 
compared with the true sparse {\sc fmr} having $K_0 = 3$ components (red, solid), 
for a feature vector $\xx = (x_1, \ldots, x_6)^{\top}$. 
Light shading indicates zero entries in the true regression vectors 
$(\bbeta^0_1, \bbeta^0_2, \bbeta^0_3)$ and their estimates 
$(\hat \bbeta_1, \ldots, \hat \bbeta_8)$, while solid blue and red highlight 
their nonzero entries.
}
\label{fig-sparse1}
\end{figure}

Figure~\ref{fig-sparse1} provides a graphical illustration of the behavior of
the OMPLE in a setting where the true sparse {\sc FMR} has $K_0=3$
components, while the fitted model is overspecified with $K=8$ components.
The resulting cluster partition consists of three clusters, two of which
contain three fitted atoms each, located near the true atoms $\bbeta^0_1$ and
$\bbeta^0_2$, respectively, while the third cluster contains two fitted atoms
near $\bbeta^0_3$. The true atoms are represented in red, with light red
indicating zero regression coefficients $\beta^0_{kj}$ and solid red
indicating nonzero coefficients. Similarly, the estimated atoms are
represented in blue, with light blue indicating zero estimated 
coefficients $\hat\beta_{kj}$ and solid blue indicating nonzero 
coefficients.

\section{Asymptotic Study}
\label{sec:asymp}

In this section, we establish the key theoretical properties of the OMPLE under
a fixed upper bound $K>K_0$ and fixed covariate dimension $p$: consistency in 
estimation and feature selection.
We first introduce additional notation and state the regularity conditions.

\subsection{Preliminaries}
\label{subsec:note-reg}

We denote
$\hat{\bPhi}_k=(\hat\bbeta_k^{\top},\hat\phi_k)^{\top}$, $k\in[K]$, as
the atoms of the OMPLE $\hat G_n$, and let
$\bPhi_k^0=(\bbeta_k^{0\top},\phi_k^0)^\top$, $k\in[K_0]$, 
be the atoms of the true mixing measure $G_0$. 

Define the Voronoi-type partition 
\citep{GreenSibson1978, Burrough2015principles} 
of the estimated atoms
$\{\hat{\bPhi}_1,\ldots,\hat{\bPhi}_K\}$ induced by the 
true atoms $\{\bPhi_1^0,\ldots,\bPhi_{K_0}^0\}$ by
$\{(\widehat{\mathcal V}_k,\widehat{\mathcal I}_k):k\in[K_0]\}$, 
where, for $k\in[K_0]$,
\begin{eqnarray}
\widehat{\mathcal V}_k
&=&
\left\{
\hat{\bPhi}_j:
\|\hat{\bPhi}_j-\bPhi_k^0\|
<
\|\hat{\bPhi}_j-\bPhi_l^0\|,
\quad \forall l\neq k,\ j\in[K]
\right\},
\label{Voronoi}\\
\widehat{\mathcal I}_k
&=&
\left\{
j\in[K]:
\hat{\bPhi}_j\in\widehat{\mathcal V}_k
\right\}.
\label{Voronoi-index}
\end{eqnarray}
Here, $\widehat{\mathcal V}_k$ and $\widehat{\mathcal I}_k$ are called as 
the Voronoi cell and its corresponding index set, respectively.\ 
Theorem~\ref{thm:paramConsistency} establishes asymptotically 
such a Voronoi-type partition of the 
atoms $\hat\Phi_k$.

By the sparsity assumption in equation~\eqref{sparse-assum}, for each
$k\in[K_0]$, 
we partition
\(
\bbeta_k^0=(\bbeta_{k,1}^0,\bbeta_{k,2}^0)
\)
such that the subvector $\bbeta_{k,1}^0$, with $\dim(\bbeta_{k,1}^0)=p_k<p$, contains all 
nonzero regression coefficients in the $k$-th component, 
and $\bbeta_{k,2}^0=\bzero$. We further assume the minimum-signal strength 
condition, as $n \to \infty$, 
\begin{equation}
\label{min-signal}
\min_{k\in[K_0]}\min_{l \in [p_k]}
\left|(\bbeta_{k,1}^0)_l\right|/\lambda_n \to \infty
\end{equation}
where $(\bbeta_{k,1}^0)_l$ is the $l$-th 
entry of $\bbeta_{k,1}^0$. For each $k\in[K_0]$, 
$j\in\widehat{\mathcal I}_k$, we accordingly partition
\begin{equation}
\label{part-sparse}
\hat\bbeta_j=(\hat\bbeta_{j,1},\hat\bbeta_{j,2}),
\qquad
\dim(\hat\bbeta_{j,1})=p_k.
\end{equation}
It is important to note that this partitioning is defined from an oracle perspective, 
since it relies on knowledge of the true sparsity structure of $G_0$. 
We assume the parameter space 
$\Theta = \Omega_1 \times \Omega_2$, with 
$\Omega_1\subseteq\mathbb R^p,\,
\Omega_2\subseteq\mathbb R^+$, 
is compact. 

We now state the regularity conditions on the family ${\cal F}$ and the
penalties $(\rho_K,r_{\lambda_n})$.
Our theoretical analysis primarily focuses on strongly identifiable (SI)
models, as defined below, with the definition adapted from
\citet{CHEN1995,HO2016strong} to FMR models. 

\begin{definition}[Strong Identifiability (SI)]
\label{def:SI}
The parametric family
\(
{\cal F}
=
\big\{
f(y;\theta(\xx,\bbeta),\phi):(\bbeta,\phi)\in \Theta 
\big\}
\)
is said to be strongly identifiable of second order if
$f(y;\theta(\xx,\bbeta),\phi)$ is twice differentiable 
with respect to $(\bbeta,\phi)$ for all $y\in{\cal Y}$ 
and $\xx\in{\cal X}$, and the following condition holds 
for every integer $K\geq1$.
\begin{enumerate}
\item[(SI)]
Given distinct $(\bbeta_j,\phi_j)\in \Theta$,
if $\zeta_{j,1},\zeta_{j,2},\zeta_{j,3}\in\mathbb R$ and
$\boleta_j,\bgamma_j\in\mathbb R^p$ satisfy
\begin{eqnarray*}
\esssup_{y \in {\cal Y}, \xx \in {\cal X}} 
 \bigg | \sum_{j=1}^K  \bigg \{ 
	\zeta_{j,1} f(y; \theta(\xx, \bbeta_j), \phi_j) + 
	\zeta_{j,2} \frac{\partial f(y; \theta(\xx, \bbeta_j), \phi_j)}{\partial \phi} 
 +   
	\boleta_j^\top \frac{\partial f(y; \theta(\xx, \bbeta_j), \phi_j)}{\partial \bbeta}
 &  &  \\
 \zeta_{j,3} \frac{\partial^2 f(y; \theta(\xx, \bbeta_j), \phi_j)}{\partial \phi^2} 
+	\bgamma_j^\top \frac{\partial^2 
 f(y; \theta(\xx, \bbeta_j),  \phi_j)}{\partial \bbeta \partial \bbeta^\top} \bgamma_j
+ 2 \bgamma_j^\top \frac{\partial^2 
 f(y; \theta(\xx, \bbeta_j),  \phi_j)}{\partial \bbeta \partial \phi} \zeta_{j,3}
\bigg \}
\bigg |   =  0 &&
\end{eqnarray*}
then
$\zeta_{j,1}=\zeta_{j,2}=\zeta_{j,3}=0$ and
$\boleta_j=\bgamma_j=\bzero\in\mathbb R^p$, for all $j\in[K]$.
\end{enumerate}
\end{definition}

Several well-known parametric families satisfy the SI condition. 
For example, the FMR based on the following distributions: 
Gaussian distribution with mean  
$\theta(\xx, \bbeta_j) = \mu_j(\xx) = \xx^\top \bbeta_j$ 
and known variance $\phi_j$, or 
$\mu_j(\xx) = \xx^\top \bbeta_j$ with zero 
intercept and unknown variance  \citep{HOJMLR2022}; 
Poisson distribution with 
$\theta(\xx, \bbeta_j) = \log\{\mu_j(\xx)\} =  \xx^{\top} \bbeta_j$; 
and Binomial with 
$\theta(\xx, \bbeta_j) 
=
\text{logit} \{ P(Y=1|\xx) \}
=
\text{logit} \{ p(\xx) \}
= 
\xx^{\top} \bbeta_j$ 
and the number of 
trials satisfying 
$m_{\text{trial}} \ge 3K - 1$ 
\citep{DoDoLong2025}. 
Note that the Gaussian model with an unknown intercept
in the mean and an unknown variance
does not satisfy the SI condition. 
Nevertheless, as demonstrated in our simulations, 
the OMPLE $\hat G_n$ still performs well for such models. 
We also require the regularity conditions (R1)–(R4)
in Appendix~\ref{appendix:notation}
on the family ${\cal F}$.
These conditions
are commonly used to establish consistency of 
the overspecified MLE in finite mixture and FMR models.

The following conditions are on the penalties 
$\rho_K$, $r_{\lambda_n}$, and the tuning parameter $\lambda_n$.

\begin{enumerate}
\item[(P0)] 
For every integer $m \in \mathbb{N}$, 
the function $\rho_m(t_1, \ldots, t_m)$, 
defined as $\rho_m:(0,1]^m\rightarrow \mathbb{R}^{+}$,
is symmetric in its arguments and Lipschitz
continuous on any compact subset of $(0,1]^m$. 
For every fixed $m$ and every fixed
$(t_1,\ldots,t_m)\in(0,1]^m$,
\(
\rho_m(t_1,\ldots,t_m)/n \to 0,
\)
as $n \to \infty$. 
Moreover, 
$\rho_m(t_1,\ldots,t_m)\rightarrow \infty$
as
$\min_{1\leq j\leq m}t_j\rightarrow 0$. Also, 
for any integers $1\leq K_0\leq K$, any
$(t_1,\ldots,t_K)\in(0,1]^K$, and any
$(t_1^*,\ldots,t_{K_0}^*)\in(0,1]^{K_0}$ satisfying
\(
t_k\leq t_k^*, k=1,\ldots,K_0,
\)
we have
\(
\rho_K(t_1,\ldots,t_K)
\geq
\rho_{K_0}(t_1^*,\ldots,t_{K_0}^*).
\)

\vspace{.1in}
The key assumption on the penalty $\rho_m$ in
Condition~(P0) is that it diverges when any of its arguments
approaches zero. This property discourages near-zero estimated mixing
probabilities $\hat\pi_k$ in the OMPLE in equation~\eqref{MPLE-Est},
thereby facilitating the asymptotic Voronoi-type partition of the fitted
atoms established in Theorem~\ref{thm:paramConsistency}. The other
requirements in (P0) are used in establishing the estimation consistency
and feature selection properties of the OMPLE.

\item[]
The following conditions are on the feature selection penalty 
$r_{\lambda_n}$ and $\lambda_n$:

\item [(P1)]  
For any $\lambda_n > 0$ and $n \ge 1$, 
$r_{\lambda_n}(\beta)$ is 
symmetric in $\beta \in \mathbb{R}$, 
nondecreasing in $|\beta|$, and 
$r_{\lambda_n}(0) = 0$. Moreover, 
for all $\beta \in \mathbb{R}$, 
$\lim_{n \to \infty} r_{\lambda_n}(\beta) = 0$. 
Additionally, there exists a constant $\delta>0$ such that, for all
sufficiently large $n$,
\(
r_{\lambda_n}(\beta)=c_n,
\)
for all  $|\beta|>\delta\lambda_n$,
where $\lim_{n\to\infty}c_n=0$ and $\lim_{n\to\infty} \lambda_n=0$.

\item[(P2)]
Denote the shrinking neighborhood 
${\cal N}_n 
=
\big \{\beta: 0 < \beta \leq \gamma_n^{\frac 1 2} \log n 
\big \}$.\ 
We assume $r_{\lambda_n}$ is differentiable on ${\cal N}_n $, and  
\[
\text{lim}_{n \to \infty} \text{inf}_{{\cal N}_n}~
\bigg \{\gamma_n^{-1/2} \ 
\frac{\partial r_{\lambda_n}(\beta)}{\partial \beta} 
\bigg \} 
= \infty~~,~~\text{with}~\gamma_n \asymp (\log n/n)^{\frac 1 2}.
\]
\end{enumerate}

Condition~(P1) is used in 
Theorem~\ref{thm:paramConsistency} and Proposition
\ref{densityConsistency} to establish consistency in parameter and density
estimation. 
In addition to standard properties of continuous penalty
functions, such as being zero at zero and nondecreasing in $|\beta|$,
Condition~(P1) ensures that the penalty becomes asymptotically negligible
for coefficients outside a shrinking neighborhood of zero. In particular,
for sufficiently large $n$, the penalty is equal to $c_n$ for coefficients
larger than $\delta\lambda_n$ in absolute value, with $c_n\to0$.
Together with the minimum-signal condition~\eqref{min-signal}, this ensures
that the true nonzero regression coefficients eventually lie in this region,
so that the penalty for these coefficients is asymptotically negligible 
compared to the log-likelihood.
Condition~(P2), on the other hand, requires the penalty to increase
sufficiently rapidly in a shrinking neighborhood of $\beta=0$, resulting in 
sparsity of the OMPLE as established in
Theorem~\ref{thm:sparsity}. The implications of Conditions~(P1)--(P2) 
for the four feature selection penalties considered in
Section~\ref{sec:feature-selec} are discussed in
Remarks~\ref{rmk1} and~\ref{rmk2} below.
We are now ready to state our main results.

\subsection{Main Results}
\label{subsce:main-resul}

Recall the family 
${\cal F} = \big \{f(y; \theta(\xx, \bbeta), \phi):  
(\bbeta, \phi) \in \Omega_1 \times \Omega_2,~
\Omega_1 \subseteq \mathbb{R}^p, \Omega_2 \subseteq \mathbb{R}^+ \big \}
$, where both parameter spaces $\Omega_1$ and $\Omega_2$ are assumed to be compact. 
In Theorem \ref{thm:paramConsistency}, 
we guarantee that the OMPLE $\hat G_n$ 
is a Wasserstein-consistent estimator of the true mixing measure $G_0$, and also 
that its atoms asymptotically form the Voronoi-type partition introduced in 
\eqref{Voronoi}-\eqref{Voronoi-index}.

\setcounter{theorem}{0}
\begin{theorem}
\label{thm:paramConsistency}
Suppose that the penalty functions $\rho_K$ and $r_{\lambda_n}$ satisfy
Conditions~(P0) and (P1), respectively. Under Condition~(SI) and
(R1)--(R2) on the family ${\cal F}$, the {\sc OMPLE}
$\hat G_n$ has the following properties. As $n\to\infty$, 
\begin{enumerate}
   \item[(i)] $W_r(\hat{G}_n, G_0) \to 0$, almost surely, 
    for all $r \geqslant 1$.
\item[]
If, in addition, Condition~(R3) and the minimum-signal condition
    \eqref{min-signal} hold, then 
    \item[(ii)]
    The fitted mixing proportions $\hat\pi_j$, $j\in[K]$, 
    are bounded away from zero in probability.
    Moreover, for each $k\in[K_0]$,
    \(
    \sum_{j\in\widehat{\mathcal I}_k}\hat\pi_j
    =\pi_k^0+o_p(1)
    \).
\footnote{
In the proof of Theorem 1-(ii) we show that: 
for any $\epsilon>0$, there exists $b_{\epsilon}>0$ and $N_\epsilon$
such that for all $n \ge N_\epsilon$, we have 
\(
\Pr\left(
\hat\pi_k\geq b_\epsilon,\ k\in [K]
\right) >1-\epsilon.
\)}
 
    \item[(iii)]
    For every $l\in[K]$, there exists a
    $k\in[K_0]$ such that
    \(
    \|\hat{\bPhi}_l-\bPhi_k^0\|
    \overset{p}{\to}0.
    \)
  Thus, with probability tending to one, 
    $\big \{(\widehat{\mathcal V}_k, \widehat{\cal I}_k):k\in[K_0] \big \}$
    is a Voronoi partition of $\hat{\bPhi}_1,\ldots,\hat{\bPhi}_K$. 
    
\end{enumerate}
\end{theorem}

The Wasserstein consistency established in Theorem~\ref{thm:paramConsistency} is a crucial 
property that is also shared by the overspecified MLE studied in 
previous work \citep{HO2016strong, HOJMLR2022, DoDoLong2025}. A notable 
distinction, however, is about the behavior of the estimated atoms. 
The overspecified MLE may contain redundant atoms that do not converge to 
any of the atoms of $G_0$. In contrast, due to the penalty 
$\rho_K$, every atom of the OMPLE is asymptotically associated with 
an atom of $G_0$. Consequently, the estimated atoms of the OMPLE 
asymptotically form the Voronoi-type partition described in \eqref{Voronoi}-\eqref{Voronoi-index}.

In addition to the results of Theorem~\ref{thm:paramConsistency}, 
we 
have also characterized the convergence rate of the OMPLE 
in the Wasserstein distance, which is similar to the MLE rate. 
By Proposition \ref{thm:ho_long_results} in 
Appendix \ref{sec:additional_results} 
on the relation between the Hellinger and 
Wasserstein distances, i.e., 
$h(p_{\hat{G}_n}, p_0) \gtrsim W_2^2(\hat{G}_n, G_0)$,
and by Proposition \ref{densityConsistency} on the parametric convergence 
rate of the estimated density to the true density under the Hellinger distance, 
i.e., $h(p_{\hat{G}_n},p_0)=\mathcal{O}_p((\log(n)/n)^{1/2})$, 
we obtain 
$W_2(\hat{G}_n,G_0) =\mathcal{O}_p((\log(n)/n)^{1/4})$.
Thus, the OMPLE $\hat G_n$ falls in an $n^{-1/4}$-neighborhood 
of $G_0$, up to a logarithmic factor. 

The aforementioned properties, however, do not 
imply consistency of the OMPLE  
in feature selection, which we will study in Theorem~\ref{thm:sparsity}. 
Recall the partitioning $\hat\bbeta_j=(\hat\bbeta_{j,1},\hat\bbeta_{j,2})$
in \eqref{part-sparse}. 
Given the above results, we are ready to state our sparsity theorem.

\begin{theorem}
\label{thm:sparsity}
Suppose that the penalty functions $\rho_K$ and 
$r_{\lambda_n}$ satisfy Conditions (P0) and (P1)-(P2), respectively.\ Also, assume that the 
family ${\cal F}$ satisfies Conditions (SI) and (R1)–(R4), 
and that the minimum-signal condition \eqref{min-signal} holds. 
Let $\mathbb{P}_0$ be the true joint probability distribution of $(\XX, Y)$. 
Then, as $n \to \infty$, 
\[
\mathbb{P}_0\bigg(
\bigcap_{k=1}^{K_0} 
\bigcap_{j \in \hat {\cal I}_k}
\{ \hat{\bbeta}_{j,2} = \bzero\}
\bigg ) \rightarrow 1.
\]
\end{theorem}

Taken together, Theorems~\ref{thm:paramConsistency} and~\ref{thm:sparsity} establish the feature selection consistency of the OMPLE $\hat G_n$. Specifically, Theorem~\ref{thm:paramConsistency}-(iii) guarantees consistent estimation of the component parameters, and hence of the corresponding true nonzero regression coefficients, 
as $n \to \infty$,
\[
\|\hat \bbeta_{j,1} - \bbeta^0_{k,1}\|= o_p(1)~~,~~\forall k \in[K_0], j \in \hat {\cal I}_k. 
\]
Meanwhile, Theorem~\ref{thm:sparsity} guarantees that the true zero elements of the regression vectors, $\bbeta^0_{k,2} = \bzero$, are correctly identified with probability tending to one. Consequently, although the fitted mixture model is overspecified, with a fixed upper bound $K>K_0$, the OMPLE 
consistently recovers the sparsity structure of $G_0$ according to the partition in \eqref{part-sparse}. More specifically, for each $k\in[K_0]$, all fitted atoms belonging to the $k$-th Voronoi cell, that is, all $\hat{\bbeta}_j\in\hat{\cal V}_k$, asymptotically have the same sparsity structure as the corresponding true atom $\bbeta_k^0$. Thus, the OMPLE consistently estimates the true nonzero regression coefficients while simultaneously identifying the true zero regression coefficients. 

\setcounter{theorem}{0}
\begin{remark}
\label{rmk1}
For SCAD and MCP penalties, Condition (P1) is satisfied as long as 
$\lambda_n = o(1)$. This condition is used in Theorem \ref{thm:paramConsistency}  
to establish estimation consistency of the OMPLE \( \hat G_n \) for \( G_0 \). 
It is also used in Propositions \ref{thm:ho_long_results} and \ref{densityConsistency} 
to quantify the convergence rate of \( \hat G_n \) to \( G_0 \), 
and the estimated density \( q_{\hat G_n} \) for \( q_{G_0} \), 
under the Wasserstein distance.\ 
On the other hand, Condition (P2) is required for the sparsity of the OMPLE. 
In particular, setting $\lambda_n \asymp \gamma_n^{1/2} \log n$ with 
$\gamma_n \asymp (\log n/n)^{\frac 1 2}$, guarantees that both Conditions 
(P1) and (P2) are met, resulting in sparsity of the OMPLE. 
\end{remark}

\begin{remark}
\label{rmk2}
From the practical point of view, our proposed methodology works for
the Lasso and AdaLasso penalties. In fact, due to its simplicity and 
ease of implementation, we include an evaluation of the OMPLE using the 
Lasso in our simulation study. 
However, it appears that no choice of the 
tuning parameter $\lambda_n$ (and the weight $\omega$ in AdaLasso)
can simultaneously satisfy both conditions (P1) and (P2) 
for these two penalties.
\end{remark}

\begin{remark}
\label{rmk3}
 An interesting phenomenon is that, although the penalty in 
 \eqref{pen-like} is applied separately to individual regression 
 coefficients $\beta_{kj}$, the double penalization yields an 
 asymptotic groupwise sparsity property within each Voronoi cell, 
 analogous to group feature selection \citep{YUAN2006}, despite 
 the absence of an explicit group penalty.
 More specifically, under the sparsity assumption 
 \eqref{sparse-assum} for the true mixing 
measure $G_0$, we have $\beta^0_{kl} = 0$, for all $k\in[K_0]$ and $l=p_k+1,\ldots, p$. 
Theorem \ref{thm:sparsity} implies that all the estimated regression coefficients 
in the overspecified FMR corresponding to 
each zero coefficient $\beta^0_{kl} = 0$ will be estimated as zero, that is, 
$\{\hat \beta_{jl} = 0: j \in \hat {\cal I}_k \}$, 
with probability tending to one. To see this, recall that the event in Theorem \ref{thm:sparsity},
$\forall k \in [K_0]~,~\forall j \in \hat {\cal I}_k: 
\{\hat \bbeta_{j, 2} = \bzero\} = \{\hat \beta_{jl} = 0: l=p_{k}+1, \ldots, p\}$, can also be rewritten as $\forall k \in [K_0]~,~\forall~ l= p_{k}+1, \ldots, p: 
\{\hat \beta_{jl} = 0: j \in \hat {\cal I}_k \}$, that is, asymptotic group variable selection. 
\end{remark}

\vspace{-0.2in}

\section{Numerical implementation}
\label{sec:numeric}
In this section, we describe a modified EM algorithm together 
with the proximal gradient method \citep{fercoq2015accelerated}
to obtain a numerical approximation to 
$\hat G_n$ in equation~\eqref{MPLE-Est}.
Given an upper bound $K$ for the mixture order, 
the EM algorithm \citep{DEMPSTER1977} works with the penalized 
complete log-likelihood function 
\begin{equation}
    L^{c}_n(G) = \sum_{i=1}^{n} \sum_{k=1}^{K} Z_{ik} \left[\text{log } \pi_k + \text{log }f(y_i; \theta(\xx_i, \bbeta_k), \phi_k)\right] - \rho_K(\pi_1,\ldots, \pi_K) - n\sum_{k=1}^{K}\sum_{j=1}^{p}
    r_{\lambda_n}(\beta_{kj}), 
   \label{PenComLogLike} 
\end{equation}
where $Z_{ik}$'s are latent variables showing the 
component membership of 
the $i$th observation $y_i$, such that $Z_{ik} = 1$ if the $i$th observation belongs 
to the $k$th component, and $0$ otherwise. Also, $\sum_{k=1}^K Z_{ik} = 1$, for each 
$i\in[n]$, indicating that each observation belongs to exactly one component.\ 
For convenience, we replace the mixing measure $G$ in equation \eqref{PenComLogLike} 
by a corresponding vector $\bPhi$ of all the parameters 
$(\bbeta_k, \phi_k, \pi_k)$, $k \in[K]$, 
and write $L^{c}_n(\bPhi)$. 

Since $Z_{ik}$'s are missing, 
the modified EM algorithm maximizes the conditional expected value (with respect to 
$Z_{ik}$) of $L^{c}_n(G)$ 
by iterating between two steps as described below. Let $\bPhi^{(m)}$ be the updated 
parameter vector at the $m$-th iteration. At the $(m+1)$-th iteration, the modified 
EM algorithm proceeds as follows.

\textbf{E-Step.} 
Given the observed data $(\xx_i, y_i),i \in [n]$, and the current estimate 
$\bPhi^{(m)}$, we compute the conditional expectation of $L_n^{c}(G)$ with respect to $Z_{ik}$ which results in 
\begin{equation}
Q(\bPhi; \bPhi^{(m)}) = 
    \sum_{i=1}^{n} \sum_{k=1}^{K} \tau_{ik}^{(m)} \left[\log \pi_k + \log f(y_i; \theta(\xx_i, \bbeta_k), \phi_k)\right] 
    - \rho_K(\bpi) - n\sum_{k=1}^{K}\sum_{j=1}^{p}r_{\lambda_n}(\beta_{kj})
    \label{Q-whole}
\end{equation}
where $\bpi = (\pi_1, \ldots, \pi_K)^\top$, and 
for $i \in [n]$ and $k \in[K]$, 
\begin{equation}
\label{weight}
    \tau_{ik}^{(m)} = 
    E\{Z_{ik} | (\xx_{1:n}, y_{1:n}), \bPhi^{(m)}\} = 
    \frac{\pi_{k}^{(m)} \, f(y_i; \theta(\xx_i, \bbeta_k^{(m)}), \phi_k^{(m)})}{\sum_{l=1}^{K} \pi_{l}^{(m)}\,f(y_i; \theta(\xx_i, \bbeta_l^{(m)}), \phi_l^{(m)})}
\end{equation}

\textbf{M-Step.} Let 
$\bPhi^{(m+1)} = \argmax_{\bPhi} Q(\bPhi; \bPhi^{(m)})$. 
The updated mixing probabilities are 
\begin{equation}
  \bpi^{(m+1)} = (\pi_1^{(m+1)}, \ldots, \pi_K^{(m+1)})^{\top} = \underset{\pi_k > 0, \sum_k \pi_k = 1}{\text{argmax}} \bigg\{\sum_{i=1}^{n}\sum_{k=1}^{K} \tau_{ik}^{(m)} ~\log \pi_k - \rho_K(\bpi) \bigg\}
    \label{mix_prop_update}
\end{equation}
For example, if $\rho_K(\bpi) = -C_K \sum_{k=1}^{K} \log \pi_k$, 
for some constant $C>0$, then
\begin{equation*}
    \pi_k^{(m+1)} = \frac{\sum_{i=1}^{n} \tau_{ik}^{(m)} + C_K}{n + KC_K}, ~~~ k \in [K].
\end{equation*}

We obtain the updates for the 
vectors $\bbeta_1, \ldots, \bbeta_K$ 
as follows. Define $\bPhi_k = (\bbeta_k, \phi_k)$. 
The maximization of $Q$ in equation~\eqref{Q-whole} with respect to each $\bbeta_k$ 
is equivalent to maximizing (scaled by $1/n$),
\begin{eqnarray}
    Q_k(\bPhi_k;\bPhi^{(m)})  =  \mathcal{L}_k(\bPhi_k) - \sum_{j=1}^{p} r_{\lambda_n}(\beta_{kj})
    =
    \frac{1}{n}\sum_{i=1}^{n} \tau_{ik}^{(m)} \text{log }f(y_i; \theta(\xx_i, \bbeta_k), \phi_k) - \sum_{j=1}^{p} r_{\lambda_n}(\beta_{kj})~
    \label{Comp-Q}
\end{eqnarray}
resulting in the updates $\bbeta^{(m+1)}_k$, for all $k \in[K]$.  
Clearly, these updates do not have closed forms, and therefore require numerical approximation methods. 

The maximization can be done using either proximal gradient or 
Newton-Raphson-type algorithms in which the leading term in each $Q_k$ is locally 
approximated by a quadratic function of $\bbeta_k$ \citep{Nesrerov2004}, 
except the Gaussian {\sc fmr} which is already quadratic in $\bbeta_k$. Note that the regularization 
function $r_{\lambda_n}$ is singular at $\beta=0$, and thus the standard Newton-Raphson 
method for approximating $Q_k$ by quadratic functions of $\bbeta_k$ cannot be used. 
To remedy this issue, there are multiple options discussed in the literature including the local quadratic approximation (LQA) or local linear approximation (LLA) to $r_{\lambda_n}$ \citep{FAN2001,ZOULI2008}, 
or coordinate-wise maximization 
in which no local approximation to $r_{\lambda_n}$ is needed. We use 
the latter method which performs well in our simulations. 
Hence, applying a proximal-coordinate method to \eqref{Comp-Q}, each $\beta_{kj}$ is updated by minimizing 
\begin{equation}
\label{CoordObj}
    -\nabla_j \mathcal{L}_k(\tilde{\bPhi}_k)(\beta_{kj} - \tilde{\beta}_{kj}) - \frac{1}{2} \nabla^2_{jj} \mathcal{L}_k(\tilde{\bPhi}_k)(\beta_{kj} - \tilde{\beta}_{kj})^2 +\frac{\xi}{2}(\beta_{kj} - \tilde{\beta}_{kj})^2 +  r_{\lambda_n}(\beta_{kj}),
\end{equation}
where $\nabla_j \mathcal{L}_k$ is the first order derivative of $\mathcal{L}_k$ in equation \eqref{Comp-Q}
with respect to $\beta_{kj}$, $\nabla^2_{jj} \mathcal{L}_k$ is the $j$-th diagonal element of the Hessian matrix $\nabla^2 \mathcal{L}_k$, and $\tilde{\bPhi}_k$ is the latest update of the estimate. The quadratic term $\frac{\xi}{2}(\beta_{kj} - \tilde{\beta}_{kj})^2$ penalizes the distance between $\beta_{kj}$ and $\tilde{\beta}_{kj}$. The idea is that we only trust the quadratic approximation of $\mathcal{L}_k(\bPhi_k; \bPhi^{(m)})$ in a small neighborhood of the latest 
iterate $\tilde{\bPhi}_k$, so we penalize the $l_2$-distance between $\bbeta_k$ and $\tilde{\bbeta}_k$ 
when updating $\bbeta_k$. Since we do not apply the penalty to the intercept, its update is  
\begin{equation}
    \beta_{k0}^{(m, s+1)} = \tilde{\beta}_{k0} + [\nabla_0 
    \mathcal{L}_k(\tilde{\bPhi}_k)]/[\xi - \nabla^2_{00} \mathcal{L}_k(\tilde{\bPhi}_k)],
 ~~ k \in [K].
    \label{UpIntercept}
\end{equation}

There are also analytical updates 
of $\beta_{kj}$ as follows. For some 
$\eta>0$, consider the soft-threshold function 
$S(x; \eta) =  \text{sign}(x) ~(|x| - \eta)_+$. For the Lasso, 
we have the updates 
\begin{equation}
    \beta_{kj}^{(m, s+1)} = S \left(\tilde \beta_{kj} + 
    \nabla_j \mathcal{L}_k(\tilde{\bPhi}_k)/[\xi -\nabla^2_{jj} \mathcal{L}_k(\tilde{\bPhi}_k)]  
    ; \lambda_n/[\xi -\nabla^2_{jj} \mathcal{L}_k(\tilde{\bPhi}_k)]  
    \right)~,~
    k \in [K], j \in [p].
    \label{Lasso}
\end{equation} 
The updates for SCAD and MCP 
are more involved and are given in Appendix~\ref{sec:additional_computation}. 

Finally, the dispersion parameters $\phi_k$ are updated by solving the system of equations
\begin{equation}
    \frac{\partial}{\partial \phi_k} \left(\sum_{i=1}^{n} \tau_{ik}^{(m)} \text{log }f(y_i; \theta(\xx_i, \bbeta_k^{(m+1)}), \phi_k) \right) = 0~,~k \in [K].
    \label{Dispersion1}
\end{equation}

Starting from an initial value $\bPhi^{(0)}$, 
the algorithm alternates 
between the E- and M-step until some convergence 
criterion is met. 
One may stop the algorithm when 
$\big \Vert \bPhi^{(m+1)} - \bPhi^{(m)} \big \Vert < \epsilon$, 
for some small value $\epsilon > 0$. Algorithm \ref{overfitting_alg} 
summarizes the modified EM algorithm.
Additional numerical details including tuning parameter selection 
$\lambda_n$ in $r_{\lambda_n}$, and the choice of initial value $\bPhi^{(0)}$
are given in Appendix \ref{sec:add-numeric}.


\begin{algorithm}[H]
\normalsize
\caption{The modified EM algorithm for obtaining the OMPLE $\hat{G}_n$ numerically}
\label{overfitting_alg}
\begin{itemize}[leftmargin=0pt]
    \item[]\textbf{Input:} $\bPhi^{(0)}$, the data $\{(\xx_i, y_i), i \in[n]\}$, a fixed fitted order $K > K_0$, and fixed $\lambda_n$.
    \item[]\textbf{Set} $m \gets 0$.
    \item[]\textbf{Repeat:}
    \item[]\begin{itemize}[label=•]
        \item \textbf{E-Step:} Compute the weights $\tau_{ik}^{(m)}, i \in[n]; \, k \in[K]$, using 
        \eqref{weight}. 
        \item \textbf{M-Step:} for all $k\in[K]$, 
        \begin{enumerate}
            \item[] \textbf{(I) Mixing probabilities:} 
            update to $\pi_k^{(m+1)}$, using 
            \eqref{mix_prop_update}.
            \item[] \textbf{(II) Regression parameters:} 
            \begin{enumerate}
                \item[] 
                \begin{enumerate} 
                    \item  $s \leftarrow 0$; $\bbeta_k^{(m,s)} \leftarrow \bbeta_k^{(m)}$; $\mathcal{B} = [p]$.
                    \item Update the intercept $\beta_{k0}^{(m,s)}$ using 
                    \eqref{UpIntercept}. Then, successively for $j \in \mathcal{B}$, update $\beta_{kj}^{(m,s)}$ by minimizing 
                    \eqref{CoordObj} if the updated $\bbeta_k^{(m,s)}$ strictly increases the objective function \eqref{Comp-Q} to obtain $\bbeta_k^{(m,s+1)}$. Set $\mathcal{B} \leftarrow \text{Supp}(\bbeta_k^{(m,s+1)})$ and $s \leftarrow s+1$.
                    \item Repeat step ii until $\big\Vert \bbeta_k^{(m,s+1)} - \bbeta_k^{(m,s)} \big\Vert < \delta$, for some $\delta > 0$ or $s = s_{\max}$ for some pre-specified maximum number of iterations $s_{\max}$.
                    \item Output $\bbeta_k^{(m+1)} = \bbeta_k^{(m,s+1)}$. 
                \end{enumerate} 
            \end{enumerate}
            \item[] \textbf{(III) Dispersion parameters:} update $\phi_k$'s according to \eqref{Dispersion1}.
        \end{enumerate}
        \item \textbf{Set} $m \gets m+1$.
    \end{itemize}
    \item[]\textbf{Until} $\Vert \bPhi^{(m+1)} - \bPhi^{(m)} \Vert < \epsilon$ or $m$ reaches a pre-specified $M_{\max}$. 
    \item[]\textbf{Output} $\bPhi^{(m+1)}$.
\end{itemize}
\end{algorithm}

\section{Simulation Study}
\label{sec:simul}

In this section, we assess the performance of OMPLE through simulations. 
The $p$-dimensional design points 
$\xx_i, i \in [n]$, are generated from a 
multivariate Gaussian with mean $\bzero$, and covariance matrix 
$\bSigma = \{\sigma_{ij}= 0.5^{|i-j|}: i, j \in[p]\}$.
Given $\xx_i$'s, 
the response $y_i$ is then generated from 
the FMR models considered below. 
For each simulated sample $(\xx_i, y_i), i\in[n]$,  
we fit FMR models using the OMPLE $\hat G_n$ in equation 
\eqref{MPLE-Est}, considering both the true order $K_0$ and 
also overspecified orders $K>K_0$. 

For the penalty on the mixing probabilities, we use 
\(
\rho_K(\pi_1,\ldots,\pi_K)
    =-C_K\sum_{j=1}^{K}\log \pi_j.
\)
Following the recommendation of \citet{CHEN2001modified}, we set
$C_K=|2\log M|$, where \(M=\max_{k\in[K_0]}\|\bbeta_k^0\|.\)
Since the true coefficients $\bbeta_k^0$ are unknown in practice, we estimate $M$ by
\(
\tilde M=\max_{k\in[K]}\|\tilde{\bbeta}_k\|,
\)
where $\tilde{\bbeta}_k$ is the single-penalized estimator in \eqref{smple}. 

The penalty $r_{\lambda_n}$ is chosen as one of the Lasso, SCAD, or MCP penalties, 
with its tuning parameter $\lambda_n$ selected using a BIC-type criterion following 
\cite{Stdler20101penalizationFM}. Further details on the tuning procedure are 
provided in Appendix~\ref{sec:tuning}.

To assess the performance of the estimators, 
we use the following metrics based on the Wasserstein distance.

\begin{enumerate}
    \item[(i)] {\bf Wasserstein distance (mean squared error)}: We consider
\begin{equation}
W_2(\hat{G}_n, G_0) \equiv W_2= 
\left(\text{inf}_{\mathbf{q} \in \mathcal{Q}(\hat{\bpi}, \bpi^0)} \sum_{j=1}^{K} \sum_{k=1}^{K_0} q_{jk} \Vert (\hat{\bbeta}_{j}, \hat{\phi}_j) - (\bbeta_{k}^{0}, \phi_{k}^{0}) \Vert^2\right)^{1/2},
\label{SIMW2}
\end{equation} 
where $\mathcal{Q}(\hat{\bpi}, \bpi^0)$ denotes the set of all 
joint discrete probability distributions 
$\mathbf{q} = \{q_{jk} : j \in [K],   k \in [K_0] \}$ 
such that $\sum_{j=1}^{K}q_{jk} = \pi_{k}^{0}$ 
and $\sum_{k=1}^{K_0} q_{jk} = \hat{\pi}_j$. 
After computing 
$\Vert (\hat{\bbeta}_{j}, \hat{\phi}_j) - (\bbeta_{k}^{0}, \phi_{k}^{0}) \Vert^2$, for $j \in[K]$ 
and $k \in[K_0]$, 
we use the CVX package in MATLAB to compute the 
weights $q_{jk}$'s. 
This optimization task is formulated as a linear programming problem \citep{dantzig2003linear} and is efficiently solved using the CVX package.

\item[(ii)] {\bf Sensitivity/Specificity for feature selection}:  
    Let CIZ = \# Correctly Identified Zero,  CIN = \# Correctly Identified Nonzero,  IIZ = \# Incorrectly Identified Zero and IIN = \# Incorrectly Identified Nonzero regression coefficients. The sensitivity (SE) and specificity (SP) are respectively defined as $\text{SE} = \frac{\text{CIN}}{\text{CIN}+\text{IIZ}}$ and $\text{SP} = \frac{\text{CIZ}}{\text{CIZ}+\text{IIN}}$. 
    We assess correctly identified zeros and nonzeros by adapting the idea of Wasserstein distance. Precisely, we compute 
\begin{eqnarray}
W_{(1-\text{SE})}(\hat{G}_n, G_0) 
\equiv W_{(1-\text{SE})} 
& = & \sum_{j=1}^{K} \sum_{k=1}^{K_0} q_{jk} (1 - \text{SE})_{jk}
\label{SIMNSE}\\
W_{(1-\text{SP})}(\hat{G}_n, G_0) 
\equiv 
W_{(1-\text{SP})}
& = & \sum_{j=1}^{K} \sum_{k=1}^{K_0} q_{jk} (1 - \text{SP})_{jk}
\label{SIMNSP}
\end{eqnarray}
where $q_{jk}$'s are obtained from (\ref{SIMW2}), $(1-\text{SE})_{jk}$ is equal to $\frac{\text{IIZ}}{\text{CIN}+\text{IIZ}}$, and $(1-\text{SP})_{jk}$ is equal to 
$\frac{\text{IIN}}{\text{CIZ}+\text{IIN}}$, both 
computed by comparing $\hat{\bbeta}_{j}$ with $\bbeta_{k}^{0}$. 
\end{enumerate}

Our simulation results are based on $R=100$ replications, 
with the outcomes presented in the form of boxplots below, 
and also summarized as empirical means 
(standard deviations) of the performance 
measures 
in the tables given in Appendix~\ref{sec:additional-simul}. 

\newpage

\noindent
{\bf Example 1 (Gaussian FMR)}. 
In this model, we have the conditional distribution
\[
Y_i|\xx_i \sim \sum_{j=1}^{K_0} \pi_j~ {\cal N}(\mu_j(\xx_i), \phi_j)~,~i\in[n]
\]
such that  
\(
\mu_j(\xx_i) = \xx^{\top}_i \bbeta_j,~ j\in[K_0].
\)
We first consider the following true parameter settings: 
$K_0 = 2$, mixing probability $\pi^0_1 = 0.45$, 
dispersion parameters $\phi^0_{1} = \phi^0_{2} = 1$, 
and sparse regression vectors: 
\(
    \bbeta_{1}^{0} = (2, 2.5, 3, 0, \ldots, 0)^\top,~
    \bbeta_{2}^{0} = (-2, 0, 2.5, 0, 1.5, 0, \ldots, 0)^\top 
\),
with three dimensions $p = 10, 50, 100$. Here, as the dimension
$p$ increases, the number of nonzero regression coefficients remains fixed while the number of zero coefficients grows.

\begin{figure}[htbp]
\centering
\subfloat{%
\includegraphics[width=0.8\textwidth, height=6.5cm, keepaspectratio=false]{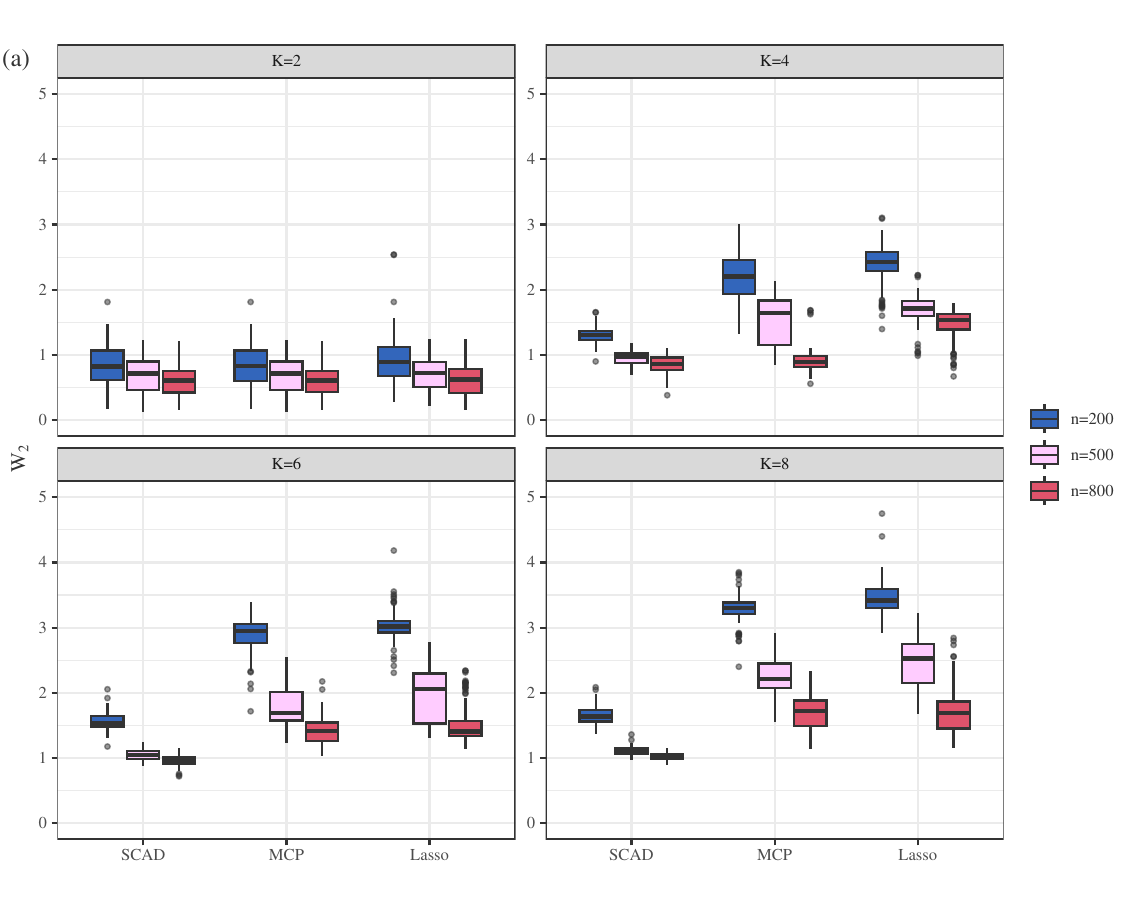}
}    
\vspace*{0.0001cm} 
\subfloat{%
\includegraphics[width=0.8\textwidth, height=6.5cm, keepaspectratio=false]{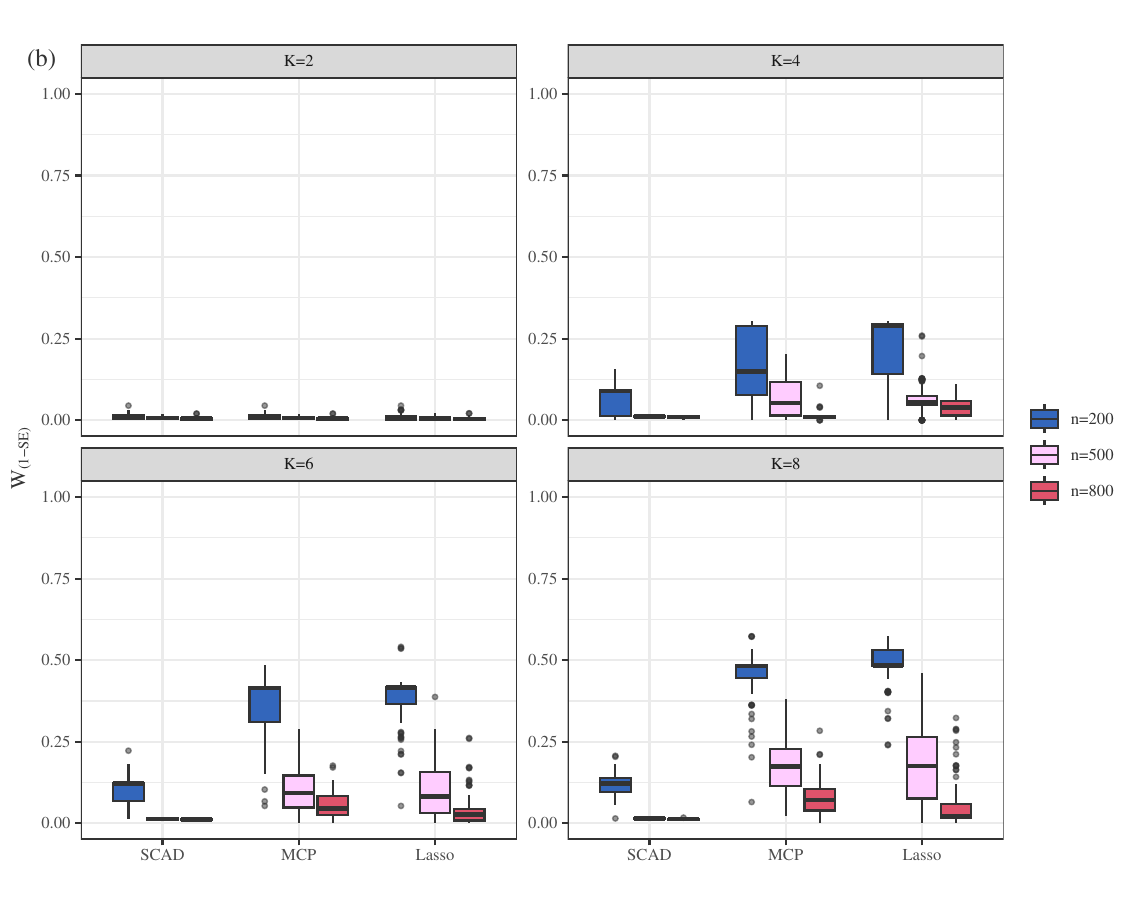}
}
\vspace*{0.0001cm} 
\subfloat{%
\includegraphics[width=0.8\textwidth, height=6.5cm, keepaspectratio=false]{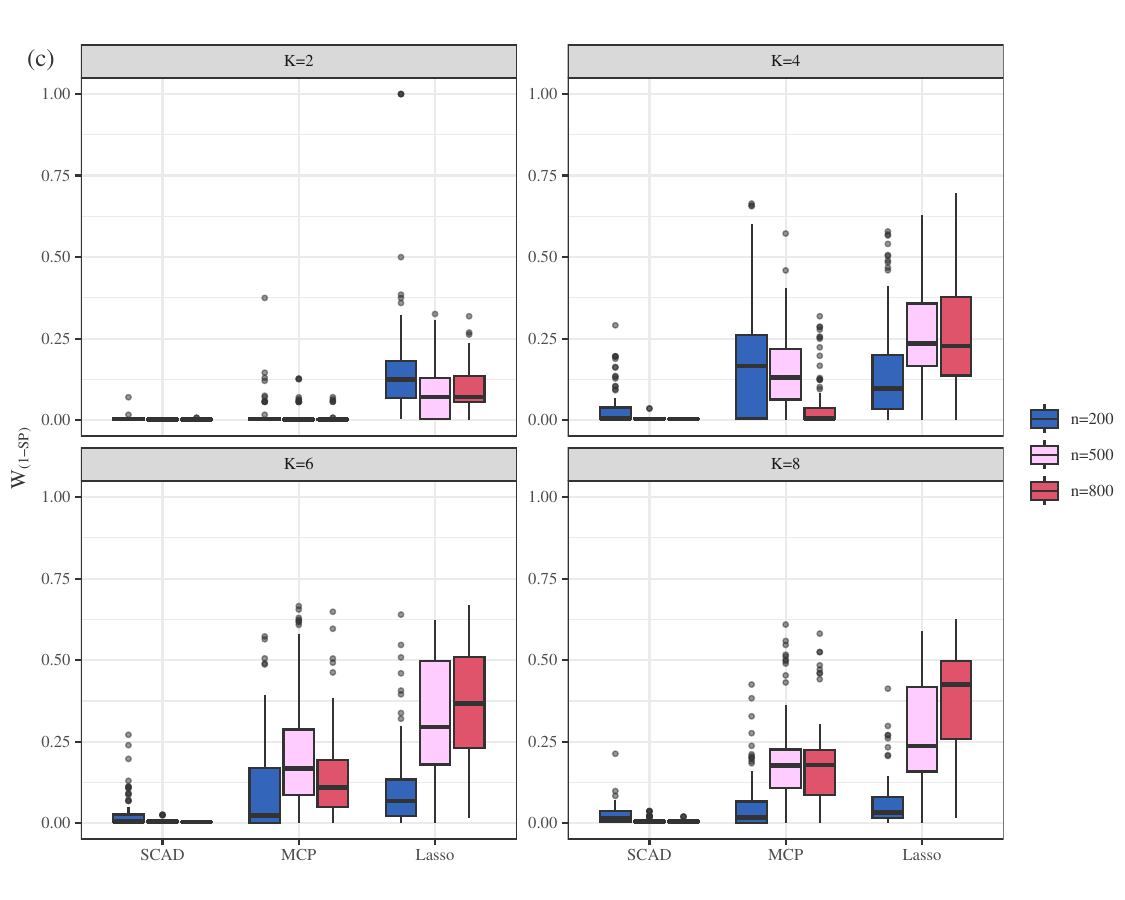}
}
 \caption{Simulation results for 
Gaussian FMR with $K_0 = 2$ and $p = 10$. Boxplots of:\\ (a) $W_2$; (b) $W_{\text{(1-SE)}}$; (c) $W_{\text{(1-SP)}}$.
The mixture order upper bounds are: $K = 2, 4, 6, 8$.
}
 \label{Gaussian_overfit_plot_p10}
\end{figure}

We generate $(\xx_i, y_i), i\in [n]$, from the above model with sample sizes $n = 200, 500, 800$ for $p = 10$; 
$n = 500, 800, 1000$ for $p = 50$, and 
$n = 1000, 1500$ for $p = 100$. 
We fit Gaussian FMR using the OMPLE 
$\hat G_n$ in \eqref{MPLE-Est},
with the true order $K_0= 2$ and 
the overspecified orders $K=4, 6, 8$. 
The boxplots for dimension $p=10$ are presented in Figure~\ref{Gaussian_overfit_plot_p10}, and those for 
$p=50$ and $p=100$ are shown in 
Figures~\ref{Gaussian_overfit_plot_p50} and~\ref{Gaussian_overfit_plot_p100}, respectively, in Appendix~\ref{sec:additional-simul}. The empirical means and standard deviations of the performance metrics 
are reported in Tables~\ref{K02_p10}--\ref{K02_p100}
in Appendix~\ref{sec:additional-simul}. A summary of our key observations is provided below.

Figure \ref{Gaussian_overfit_plot_p10} 
shows that, for a fixed sample size, increasing the upper bound $K$, and thus the number of parameters, 
makes both parameter estimation ($W_2$; Figure \ref{Gaussian_overfit_plot_p10}-(a)) and feature selection 
(sensitivity, $W_{(1-\text{SE})}$; Figure \ref{Gaussian_overfit_plot_p10}-(b)) more challenging. 
This effect is particularly pronounced for MCP and Lasso penalties. 
As the sample size increases to $n=800$, the performance 
improves substantially, especially for these two penalties. 
While the Lasso shows gains in feature selection 
indicated by lower values of $W_{(1-\text{SE})}$, 
it does so at the cost of producing less sparse models, 
as indicated by higher values of $W_{(1-\text{SP})}$
(Figure \ref{Gaussian_overfit_plot_p10}-(c)). 
Overall, SCAD has the best performance, 
followed by MCP and then Lasso. 
In 
Figures \ref{Gaussian_overfit_plot_p50} and \ref{Gaussian_overfit_plot_p100} 
in Appendix~\ref{sec:additional-simul}
which correspond to 
covariate dimensions ($p=50, 100$), 
we consider larger sample sizes ($500 \le n \le 1500$). 
This enables the OMPLE to maintain a performance pattern 
similar to that observed in Figure \ref{Gaussian_overfit_plot_p10}.
The summary statistics of the performance measures 
in Tables~\ref{K02_p10}, \ref{K02_p50}, and \ref{K02_p100} in Appendix~\ref{sec:additional-simul} also support the above 
observations. 

We also conducted simulations for a Gaussian FMR model 
with $K_0 = 3$, dimensions $p=5, 10, 50, 100$,
and the parameter settings are given 
in Appendix~\ref{sec:additional-simul}. 
We fit FMRs using the OMPLE $\hat G_n$ 
in equation~\eqref{MPLE-Est},
considering the true order $K_0= 3$ and 
also overspecified orders $K=6, 9, 12$. 
We considered sample sizes in the range \( 500 \le n \le 3000 \), 
depending on the dimension \( p \) and the upper bound \( K \).
We report results based only on the SCAD, as its
overall performance is generally better than the 
other two penalties in the scenarios considered here.
The results, presented in Appendix~\ref{sec:additional-simul}, 
corresponding to $p=5, 10$ are given in Table \ref{SCADK03_p5_10}, and those for $p=50, 100$ 
are given in Table \ref{SCADK03_p50_100}. 
The results indicate that for 
sufficiently large $n$ relative to 
the total number of parameters in the overspecified FMR, 
OMPLE performs reasonably well in both parameter 
estimation and feature selection.\\

\begin{figure}[htbp]
\centering
\subfloat{%
\includegraphics[width=0.8\textwidth, height=6.5cm, keepaspectratio=false]{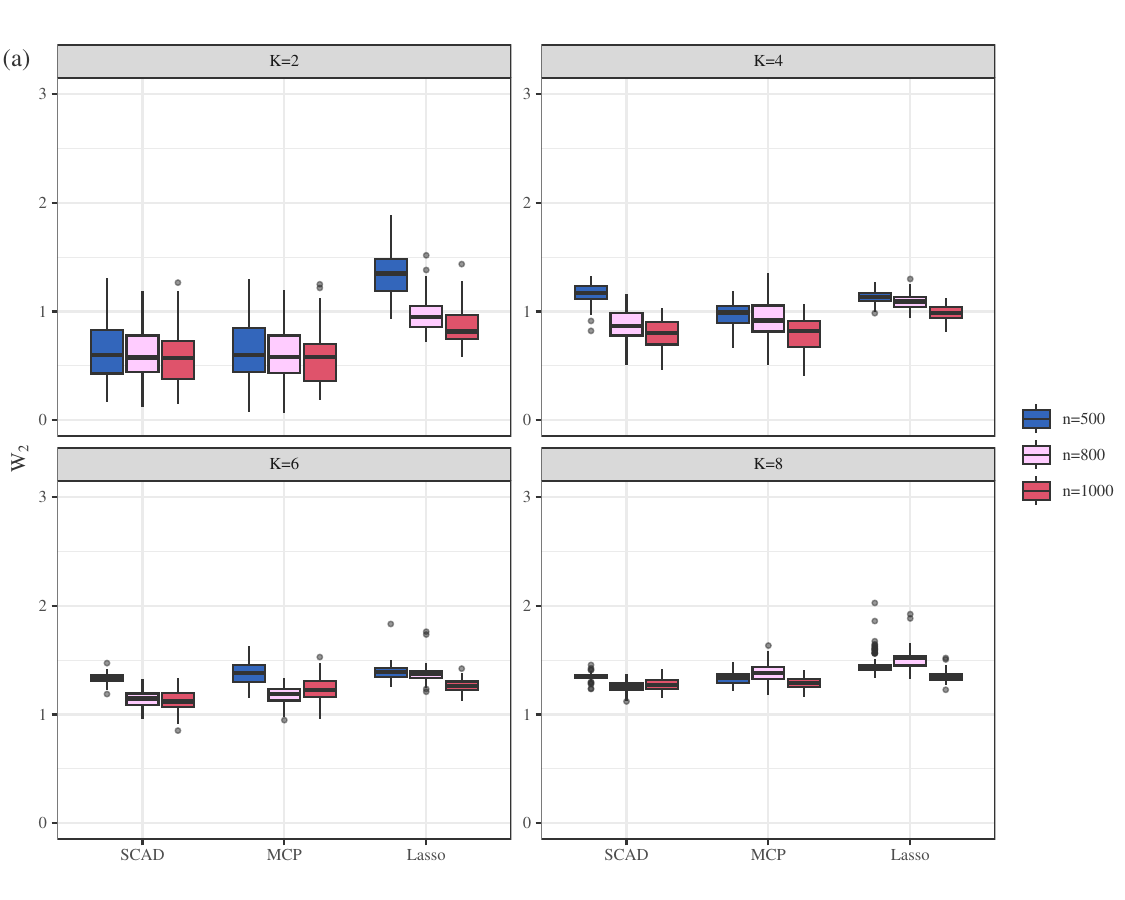}
}    
\vspace*{0.0001cm} 
\subfloat{%
\includegraphics[width=0.8\textwidth, height=6.5cm, keepaspectratio=false]{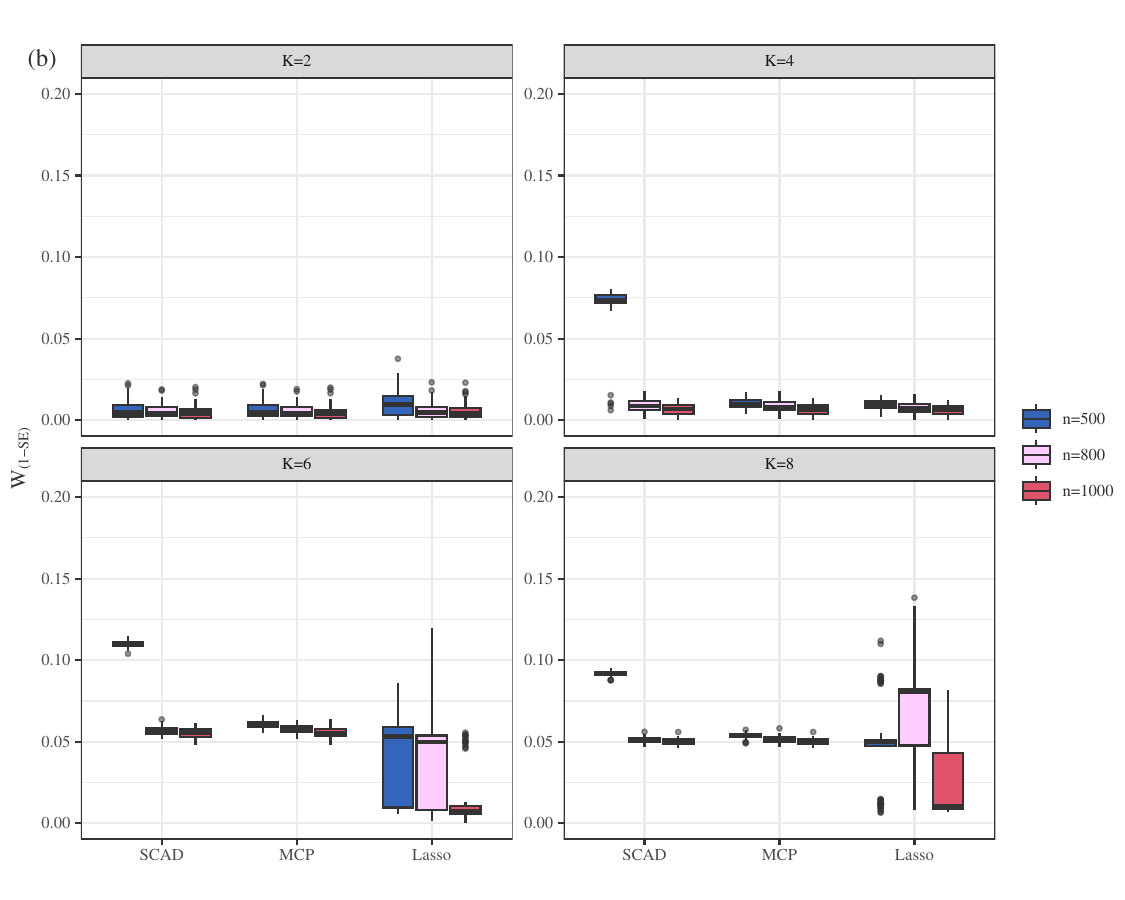}
}
\vspace*{0.0001cm} 
\subfloat{%
\includegraphics[width=0.8\textwidth, height=6.5cm, keepaspectratio=false]{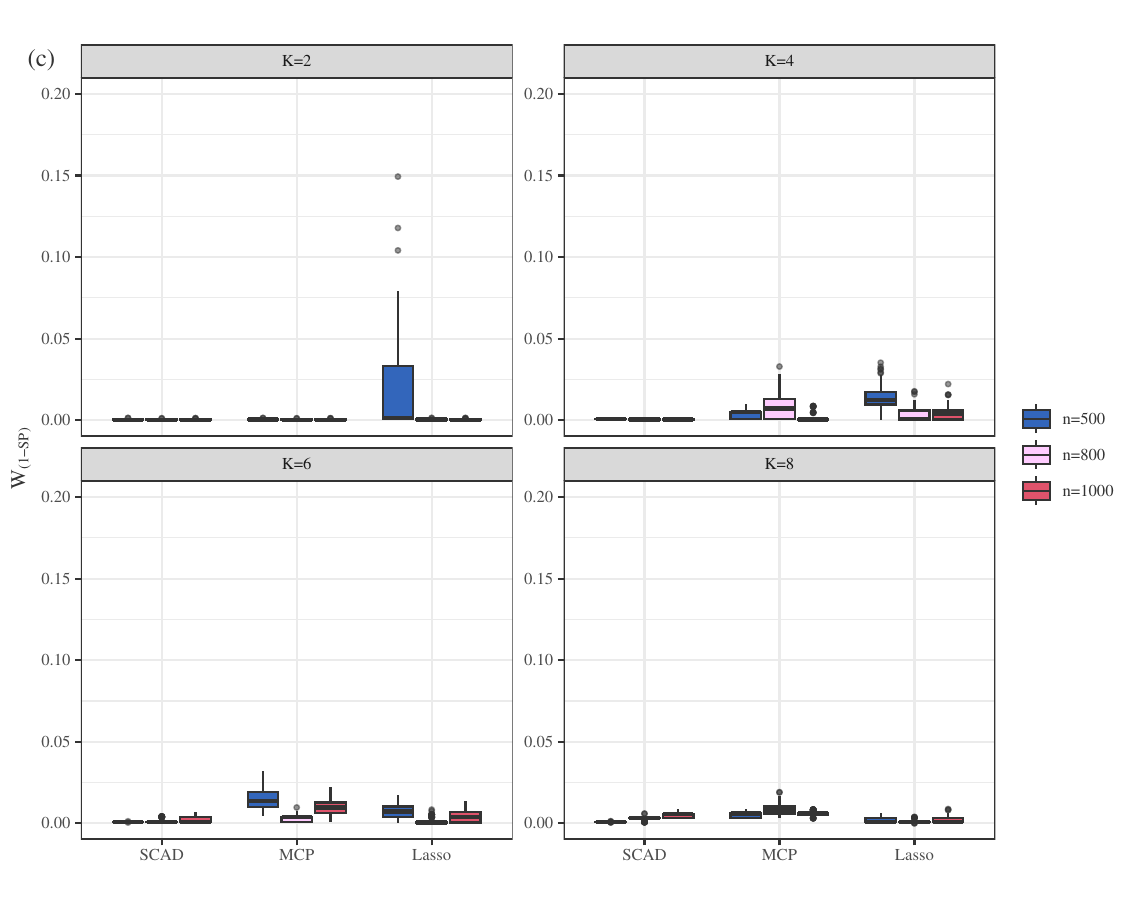}
}
 \caption{Simulation results for 
Binomial FMR with $K_0 = 2$ and $p = 50$. 
Boxplots of:\\ (a) $W_2$; (b) $W_{\text{(1-SE)}}$; (c) 
$W_{\text{(1-SP)}}$.
The mixture order upper bounds are: $K = 2, 4, 6, 8$.}
 \label{Binomial_overfit_plot}
\end{figure}

\newpage
\noindent
{\bf Example 2. (Binomial FMR)}
In this model, we have
\[
Y_i|\xx_i \sim \sum_{j=1}^{K_0} 
\pi_j
~ 
\text{Bin}(m_{\rm trial}, \, p_j(\xx_i))~,~i\in[n]
\]
such that  
$
\text{logit} \{p_j(\xx_i)\}
= \xx^{\top}_i \bbeta_j~ j\in[K_0]$.
We considered the values of the true parameters: 
$K_0 = 2$, $\pi^0_1 = 0.45$, 
number of Bernoulli trials $m_{\rm trial}= 30$, 
and sparse regression vectors:
\(
\bbeta_1^0 = (2, 2.5, 3, 0, \ldots, 0)^\top, 
    \bbeta_2^0 = (-2, 0, 2.5, 0, 1.5, 0, \ldots, 0)^\top
\), with dimension $p=50$, and sample sizes $n = 500, 800, 1000$.
We fit Binomial FMRs using the OMPLE 
$\hat G_n$ in equation~\eqref{MPLE-Est},
considering the true order $K_0= 2$ and 
also overspecified orders $K=4, 6, 8$. 
The results based on the three penalties SCAD, MCP, and Lasso
are given in the form of Boxplots in 
Figure \ref{Binomial_overfit_plot}. 
Below is a summary of our observations. 

The OMPLEs based on SCAD and MCP show nearly similar performance 
in terms of parameter estimation error ($W_2$) and feature selection accuracy 
($W_{(1\text{-SE})}$ and $W_{(1\text{-SP})}$). Interestingly, compared to the 
Gaussian model, the Lasso also 
performs quite well for this model, particularly in feature 
selection. 
By fixing the sample 
size $n$ and increasing the mixture order upper bound $K$ 
(equivalently the number of parameters), as expected, 
both estimation and feature selection become more difficult. 
In summary, the Lasso shows comparable performance with SCAD and MCP in 
feature selection, as $n$ increases.

From our simulations in Examples 1 and 2, 
we observe that the OMPLE performs well overall, 
both in parameter estimation and feature selection 
in FMR models with an overspecified mixture order.

\section{Real Data Analysis}
\label{sec:real_data}
In this section, we illustrate the application of the OMPLE by analyzing the Aquatic Toxicity dataset, available at \url{https://archive.ics.uci.edu/dataset/505/qsar+aquatic+toxicity}. This dataset is used to model the relationship between the acute aquatic toxicity of organic compounds 
and their molecular descriptors. The dataset consists of 
$n = 546$ organic compounds, each characterized by 
$p = 8$ molecular descriptors (covariates, $\xx$), 
and a continuous 
response ($y$) as concentration of toxicity that 
causes 50\% mortality in test Daphnia Magna over a test 
period of 48 hours \citep{cassotti2014prediction}.
The covariates are all numeric: 
the total topological polar surface area (TPSA(tot)), 
the Van der Waals surface area of atoms that are acceptors of hydrogen bonds (SAacc), the count of hydrogen atoms bonded to heteroatoms (H-050), the octanol–water partition coefficient calculated from the Moriguchi model (MLOGP), expressing the lipophilicity of a molecule \citep{moriguchi1992simple}, a topological index encoding molecular size and branching (RDCHI), an index for molecular polarisability (GATS1p), where lower 
values indicate uniform polarizability across bonds, 
the number of nitrogen atoms in the molecule (nN), 
and the number of carbon atoms that belong to strongly 
electrophilic functional groups (C-040).

We fitted Gaussian FMRs with orders $K \in[5]$, to the data
$(\log y_i, \xx_i), i \in [546]$. Table~\ref{real_data_BIC} 
shows the BIC and log-likelihood values for each model,
and the optimal mixture order according to BIC can be selected as 
$\hat K_0 = 3$. Note that these values are computed based on the OMPLE 
and not the ordinary MLE. The log‑likelihood rises sharply as $K$ increases from $1$ to $3$ and then decreases. This non‑monotone pattern reflects the influence of the penalty in equation~\eqref{pen-like}: although increasing $K$ potentially also increases model complexity, the fitted parameters for $K > 3$ are heavily regularized, so the improvement in fit no longer offsets the shrinkage induced by the penalty $r_{\lambda_n}$. Consequently, $K =3$ achieves the largest log‑likelihood and the lowest BIC at the same time, indicating that the fitted model  with $K=3$ captures the heterogeneity without overfitting. For the fitted Gaussian FMR with $K = 3$, the selected covariates in the first component are TPSA(tot), MLOGP, and GATS1p; in the second component, MLOGP; while there is no selected covariate in the third component. Since the primary aim of this example is to illustrate the sparsity result in Theorem~\ref{thm:sparsity}, an interpretation of the fitted model using the selected covariates is deferred to Appendix~\ref{appendix:plots_real_data}.

\begin{table}[h]
    \centering
    \begin{tabular}{|c|c|c|c|c|c|}
    \hline
     $K$ & 1 & 2 & 3 & 4 & 5  \\
    \hline
     BIC & 435.69 & 328.65 & {\bf 308.01} & 339.76 & 376.11\\
     \hline
     log-likelihood & -198.94 & -132.81 & {\bf -113.04} & -116.31 & -121.88\\
     \hline
    \end{tabular}
    \caption{BIC and log-likelihood for the Gaussian FMRs fitted to the Aquatic Toxicity data.}
    \label{real_data_BIC}
\end{table}

Table \ref{real_data_coef} displays the estimates 
$(\hat \bbeta_j, \hat \pi_j, \hat \phi_j), j \in [K]$, 
corresponding to the five fitted models $K \in [5]$, and the first element of each $\hat \bbeta_j$ is the estimated intercept.\ To avoid introducing additional notation, 
we use the same symbol 
$\hat{\bbeta}_j$ across all models, although these estimates differ 
for each value of $K$. 
We can see that, for the first three models, increasing $K$ leads to the emergence of entirely new components 
in the fitted model. On the other hand, for \( K = 4 \), we observe Voronoi cells \( \{\hat{\bbeta}_1\} \), \( \{\hat{\bbeta}_2, \hat{\bbeta}_3\} \), and \( \{\hat{\bbeta}_4\} \); and for \( K = 5 \), the cells are \( \{\hat{\bbeta}_1\} \), \( \{\hat{\bbeta}_2, \hat{\bbeta}_3, \hat{\bbeta}_4\} \), and \( \{\hat{\bbeta}_5\} \). 
These structures closely resemble those in the fitted model 
with \( K = 3 \). 
The fitted patterns are consistent with the groupwise sparsity phenomenon described in 
Remark \ref{rmk3} after Theorem~\ref{thm:sparsity}.

\begin{table}[ht]
\footnotesize
    \centering
    \begin{tabular}{|c|c|}
    \hline
    \multirow{1}{4em}{$K = 1$} & $\hat{\bbeta}_1 = (\underline{1.46}, \underline{0.16}, 0, 0, \underline{0.29}, 0, \underline{-0.08}, 0, 0)^T$, $\hat{\pi}_1 = 1$, $\hat{\phi}_1 = 0.12$\\
    \hline
    \multirow{2}{4em}{$K = 2$} & $\hat{\bbeta}_1 = (\underline{1.42}, \underline{0.37}, 0, 0, \underline{0.50}, 0, \underline{-0.11}, 0, 0)^T$, $\hat{\pi}_1 = 0.41$, $\hat{\phi}_1 = 0.14$\\ 
    & $\hat{\bbeta}_2 = (\underline{1.52}, 0, 0, 0, \underline{0.12}, 0, 0, 0, 0)^T$, $\hat{\pi}_2 = 0.59$, $\hat{\phi}_2 = 0.05$\\
    \hline
    \multirow{3}{4em}{$K = 3$} & $\hat{\bbeta}_1 = (\underline{1.39}, \underline{0.40}, 0, 0, \underline{0.59}, 0, \underline{-0.10}, 0, 0)^T$, $\hat{\pi}_1 = 0.31$, $\hat{\phi}_1 = 0.13$\\ 
    & $\hat{\bbeta}_2 = (\underline{1.44}, 0, 0, 0, \underline{0.22}, 0, 0, 0, 0)^T$, $\hat{\pi}_2 = 0.38$, $\hat{\phi}_2 = 0.03$\\
    & $\hat{\bbeta}_3 = (\underline{1.62}, 0, 0, 0, 0, 0, 0, 0, 0)^T$, $\hat{\pi}_3 = 0.31$, $\hat{\phi}_3 = 0.06$\\
    \hline
    \multirow{4}{4em}{$K = 4$} & $\hat{\bbeta}_1 = (\underline{1.38}, \underline{0.42}, 0, 0, \underline{0.64}, 0, \underline{-0.07}, 0, 0)^T$, $\hat{\pi}_1 = 0.25$, $\hat{\phi}_1 = 0.14$\\ 
    & $\hat{\bbeta}_2 = (\underline{1.53}, 0, 0, 0, \underline{0.08}, 0, 0, 0, 0)^T$, $\hat{\pi}_2 = 0.25$, $\hat{\phi}_2 = 0.05$\\
    & $\hat{\bbeta}_3 = (\underline{1.43}, 0, 0, 0, \underline{0.22}, 0, 0, 0, 0)^T$, $\hat{\pi}_3 = 0.27$, $\hat{\phi}_3 = 0.03$\\
    & $\hat{\bbeta}_4 = (\underline{1.64}, 0, 0, 0, 0, 0, 0, 0, 0)^T$, $\hat{\pi}_4 = 0.23$, $\hat{\phi}_4 = 0.07$\\
    \hline
    \multirow{5}{4em}{$K = 5$} & $\hat{\bbeta}_1 = (\underline{1.36}, \underline{0.42}, 0, 0, \underline{0.67}, 0, \underline{-0.06}, 0, 0)^T$, $\hat{\pi}_1 = 0.20$, $\hat{\phi}_1 = 0.14$\\ 
    & $\hat{\bbeta}_2 = (\underline{1.54}, 0, 0, 0, \underline{0.07}, 0, 0, 0, 0)^T$, $\hat{\pi}_2 = 0.20$, $\hat{\phi}_2 = 0.06$\\
    & $\hat{\bbeta}_3 = (\underline{1.41}, 0, 0, 0, \underline{0.21}, 0, 0, 0, 0)^T$, $\hat{\pi}_3 = 0.21$, $\hat{\phi}_3 = 0.03$\\
    & $\hat{\bbeta}_4 = (\underline{1.50}, 0, 0, 0, \underline{0.10}, 0, 0, 0, 0)^T$, $\hat{\pi}_4 = 0.20$, $\hat{\phi}_4 = 0.05$\\
    & $\hat{\bbeta}_5 = (\underline{1.66}, 0, 0, 0, 0, 0, 0, 0, 0)^T$, $\hat{\pi}_5 = 0.19$, $\hat{\phi}_5 = 0.07$\\
    \hline
    \end{tabular}
    \caption{Gaussian FMR models fitted to the Aquatic Toxicity data.}
    \label{real_data_coef}
\end{table}

To assess the stability of the selected 
$K=3$ and the associated features, we performed a bootstrap analysis with $B=50$ resamples. 
The results, with details presented in Appendix~\ref{appendix:plots_real_data}, 
confirm the stability of both the BIC order selection and the chosen features.

\section{Summary and Discussion}
\label{sec:discussion}

A double-penalized likelihood method 
is proposed for joint estimation and feature selection in finite 
mixture regression (FMR) models with an unknown mixture order 
$K_0$, which is deliberately overspecified by a fixed upper bound 
$K> K_0$. The proposed estimators, called OMPLE, are 
shown to have a sparsity property, effectively 
mimicking the assumed sparsity of the true mixing 
measure $G_0$ underlying an FMR model. 
Note that the ordinary MLE in overspecified FMR models 
is known to be estimation consistent for $G_0$ in Wasserstein distance, 
a property also shared by the OMPLE, but 
the MLE does not have the sparsity property. 
Due to double penalization, the OMPLE possesses an asymptotic 
group feature selection property, even though the penalization is applied to 
individual regression coefficients rather than groups.\ 
Our study provides valuable insights both asymptotically and empirically 
into the performance of the double regularization methods for 
fitting sparse FMR models when the mixture order is unknown and deliberately 
overspeficed by a fixed upper bound.

There are several promising directions for future research building 
on the current work. One 
extension is to weakly identifiable models, such as 
Gaussian FMR with unknown intercept and variance \citep{HOJMLR2022}.
As demonstrated in our simulations, the proposed methods also perform well in these settings; however, a rigorous theoretical framework remains to be developed as a future direction. 
Another interesting extension is to allow the mixing probabilities \( \pi_j \) to depend on the 
feature vector \( \xx \), giving rise to the so-called mixture of experts models \citep{Jacob_Jordan-1991,nguyen2023demystifying,nguyen2024general}. Also, while our current study includes simulation results for relatively large dimensions (\( p = 50, 100 \)), a valuable direction for future work is to develop a theoretical framework that accommodates high-dimensional feature spaces.

\appendix
\section{Notation and Regularity Conditions}
\label{appendix:notation}
\subsection{Notation}

Given a fixed value $K \in \mathbb{N}$, recall that 
\[
{\cal G}_K = \{G: G = \sum_{j = 1}^K \pi_j \delta_{(\bbeta_j, \phi_j)},~
(\bbeta_j, \phi_j) \in \Theta = \Omega_1 \times \Omega_2, 
\pi_j \ge 0, \sum_{j=1}^K \pi_j = 1\}
\]
is the class of all mixing measures with at most $K$ components, where $\Omega_1 \subseteq \mathbb{R}^p, \Omega_2 \subseteq \mathbb{R}^+$.

From the conditional density function $q_{G}(y|\xx)$ in equation~\eqref{fmr_model}, define the probability density function of the joint distribution of $\VV = (\XX, Y)$ with 
respect to a joint $\sigma$-finite measure $\nu=\nu_{\XX}\otimes\nu_{Y}$ as  
\begin{equation}
\label{joint-fmr}
p_G(\vv) \equiv p_G(\xx, y) = 
q_G(y|\xx) f_{\XX}(\xx),
\end{equation}
where $G$ is a mixing measure in ${\cal G}_K$ and
$f_{\XX}(\xx)$ is the joint pdf of $\XX$. Furthermore, let ${\cal X} \subseteq \mathbb{R}^p$ be compact, for an integer $p \geq 1$, 
and let $P_{\XX}$ be a probability distribution over ${\cal X}$, which is absolutely
continuous with respect to a $\sigma$-finite measure
$\nu_{\XX}$ on ${\cal X}$. Let $f_{\XX} = dP_{\XX}/d\nu_{\XX}$.
Here, we assume that the density $f_{\XX}$ does not depend on $G$,
and is therefore irrelevant in our estimation procedure. 
Let $p_0 = p_{G_0} = 
q_{G_0}(\cdot|\xx) f_{\XX}$ 
be the true joint density of $\VV = (\XX, Y)$
with the corresponding 
density of the true finite mixture model 
$q_{G_0}(\cdot|\xx)$, and with its corresponding probability distribution $P_0$. 

Let $\hat p_n = p_{\hat G_n} = q_{\hat G_n}(\cdot|\xx) f_{\XX}$ be the estimated 
joint density based on the OMPLE $\hat G_n$, 
and
define the empirical measure 
$P_n = \frac 1 n \sum_{i=1}^n \delta_{(\XX_i, Y_i)}$.

As we work with conveniently independent and identically distributed (iid) 
random variables $\VV_i, i\in[n]$ in the subsequent appendices, we also consider the joint log-likelihood 
\begin{equation}
\label{joint-like}
\ell_n(G) = \sum_{i=1}^n \log p_G(\vv_i) = \sum_{i=1}^n
\log\{q_G(y_i|\xx_i) f_{\XX}(\xx_i) \}.
\end{equation}
Next, we use two distances as described below for analyzing theoretical properties of 
the overspecified density $p_{\hat G_n}$ and the mixing measure $\hat G_n$ with respect to the true density 
$p_0 = p_{G_0}$ and the true mixing measure $G_0$, respectively.  

For any two mixing measures 
$G = \sum_{j=1}^{K} \pi_j \delta_{(\bbeta_j, \phi_j)}$ 
and $G' = \sum_{k=1}^{K'} \pi'_k \delta_{(\bbeta'_k, \phi'_k)}$,
 their Wasserstein distance of order 
$r \geqslant 1$ 
is given by
\begin{equation}
\label{Wasser-dist}
W_r(G,G') = \left(\text{inf}_{\mathbf{q} \in \mathcal{Q}(\bpi, \bpi')} \sum_{j=1}^{K} \sum_{k=1}^{K'} q_{jk} ||(\bbeta_j, \phi_j)-(\bbeta'_k, \phi'_k)||^r \right)^{\frac{1}{r}},
\end{equation} 
where $\mathcal{Q}(\bpi, \bpi')$ denotes the set of joint probability distributions $\mathbf{q} = \{q_{jk} : 1 \leqslant j \leqslant K,  1 \leqslant k \leqslant K' \}$ such that $\sum_{j=1}^{K}q_{jk} = \pi'_k$ and $\sum_{k=1}^{K'}q_{jk} = \pi_j$, and $\|\cdot\|$ is the $l_2$-norm. 

Also, for any two mixing measures $G, G' \in {\cal G}_K$, 
the Hellinger distance between their corresponding mixture densities 
$(p_G, p_{G'})$ is given by
\begin{equation}
\label{Hellin-dist}
    h(p_G, p_{G'}) 
    = 
    \bigg\{ \int \bigg (\sqrt{p_G} - \sqrt{p_{G'}} \bigg )^2  d\nu \bigg\}^{1/2}.
\end{equation}

We denote the set of all joint mixture densities with respect to mixing measures $G \in {\cal G}_K$ as
\begin{equation}
\label{famden}
{\cal P}_K = \left\{
p_G(\xx, y) = q_G(y|\xx) f_{\XX}(\xx) = 
\bigg [\int f(y; \theta(\xx, \bbeta), \phi) ~dG(\bbeta, \phi) \bigg ] f_{\XX}(\xx) : 
G \in {\cal G}_K
\right\}.
\end{equation}

Furthermore, we also denote $
p_{\bar G} = p_{\frac{G+G_0}{2}} = 
\frac{p_G + p_0}{2} = \bar p_G $, for any $p_G \in {\cal P}_K$. 

Next, for each mixing measure $G \in {\cal G}_K$, we introduce 
the corresponding empirical process
\begin{equation}
\label{nuDef}
\nu_n(G) = 
\sqrt n\int_{\{p_0 > 0\}} \frac 1 2 \log\bigg\{\frac{\bar p_G}{p_0}\bigg \} d(P_n - P_0)
=
\sqrt n\int_{\{p_0 > 0\}} \frac 1 2 \log\bigg\{\frac{p_G + p_0}{2p_0}\bigg \} d(P_n - P_0).
\end{equation}

For some $0<b_0<1$, we further consider two 
subsets of mixing measures defined as
\begin{align}
\label{tildeGSet1}
{\cal G}_K(b_0) &= \left\{G \in {\cal G}_K \setminus {\cal G}_{K_0-1}: 
G = \sum_{j=1}^K \pi_j \delta_{(\bbeta_j,\phi_j)}, \ \pi_j \geq b_0\right\}, \\
{\cal G}_K(b_0; \gamma) &= \left\{G \in {\cal G}_K(b_0): h(p_G, p_{G_0}) 
\leq \gamma\right\}, 
\quad \forall \gamma > 0.
\label{tildeGSet2}
\end{align}

Given an arbitrary mixing measure 
$G = \sum_{j=1}^K \pi_j \delta_{\bPhi_j} \in {\cal G}_K$, with 
$\bPhi_j = (\bbeta_j, \phi_j)$, and
the true mixing measure 
$G_0 = \sum_{k=1}^{K_0} \pi^0_k \delta_{\bPhi^0_k}$, 
with $\bPhi^0_k = (\bbeta^0_k, \phi^0_k)$,
we define 
\begin{align*}
    \bPhi = (\bPhi_1, \dots, \bPhi_K),& \quad  \bpi = (\pi_1, \dots, \pi_K)^\top,\\
    \bPhi^0 = (\bPhi^0_{1}, \dots, \bPhi^0_{K_0}),& \quad \bpi^0 = (\pi^0_{1}, \dots, \pi^0_{K_0})^\top,
\end{align*}
and denote $ \rho_K(\bpi)=\rho_K(\pi_1, \ldots, \pi_K)$ for ease of presentation.
For any $G \in \mathcal{G}_{K}$, we define 
\begin{eqnarray}
\label{delta1}
\Delta_{1n}(G) & = & \frac{1}{n} \left\{\rho_{K_0}(\bpi^0) - \rho_K(\bpi) \right\},\\
\Delta_{2n}(G) &= & 
\bigg \{
 \sum_{k=1}^{K_0} \sum_{j=1}^{p} r_{\lambda_n}(\beta^0_{kj})
 - 
 \sum_{k=1}^{K} \sum_{j=1}^{p} 
 r_{\lambda_n}(\beta_{kj})
 \bigg \} \nonumber \\
 & = & 
 \bigg \{
 \sum_{k=1}^{K_0} \sum_{j=1}^{p_k} r_{\lambda_n}(\beta^0_{kj})
 - 
 \sum_{k=1}^{K} \sum_{j=1}^{p} 
 r_{\lambda_n}(\beta_{kj})
 \bigg \}
 \label{delta2}
\end{eqnarray}
using $r_{\lambda_n}(0) = 0$, and $p_k$ is the number 
of true nonzero regression coefficients in 
each component $k\in[K_0]$ of the true mixing measure 
$G_0$. 

For any 
$\bPhi = (\bbeta, \phi) = (\beta_1, \dots, \beta_p, \phi)^\top = (\eta_1, \ldots, \eta_{p+1})^\top$,
where $\bbeta \in \Omega_1 \subseteq \mathbb{R}^p$
and $\phi \in \Omega_2 \subseteq \mathbb{R}^+$,
and any mixing measure $G \in {\cal G}_K$, we denote
\begin{eqnarray}
\label{U0}
U(\bPhi; G) \equiv
U(y;\theta(\xx,\bbeta),\phi, G) &=& \frac {1} {q_G(y|\xx)}f(y;\theta(\xx,\bbeta),\phi)\\
\label{U1}
U_{\kappa_1\ldots\kappa_M}(\bPhi; G) \equiv 
U_{\kappa_1\ldots\kappa_M}(y;\theta(\xx,\bbeta),\phi, G) &=& \frac {1} {q_G(y|\xx)}\frac{\partial^{M} f(y; \theta(\xx,\bbeta),\phi)}{\partial \eta_{\kappa_1}\dots\partial \eta_{\kappa_M}}
\end{eqnarray}
for all $\kappa_1, \ldots, \kappa_M=1, \dots, p+1$, 
and any integer $M\ge 1$. 

Lastly, for any matrix $\bA = (a_{ij})_{i\in[d_1],  j\in[d_2]}$, we let the notation
$
\norm{\bA}_{\text F} 
= 
\left( \sum_{i=1}^{d_1} \sum_{j=1}^{d_2} a_{ij}^2 \right)^{\frac 1 2}
$ stand for its Frobenius norm. For any real symmetric matrix $\BB$, its respective minimum and maximum eigenvalues are denoted as $\varrho_{\text{min}}(\BB)$ and
$\varrho_{\text{max}}(\BB)$. In addition, for any set $S$, we denote by $|S|$ its cardinality, whereas $\text{diam}(S):=\sup\{\|x-y\|:x,y\in S\}$ stands for its diameter. For any $a, b \in \mathbb{R}$, we denote $a\vee b:=\max\{a,b\}$ and $a\wedge b:=\min\{a,b\}$.

\subsection{Regularity Conditions}
\label{appendix:reg_conditions}
Firstly, to ensure the consistency of maximum likelihood estimators, we assume a uniform law of large numbers on the empirical process $\nu_n(G)$ in equation \eqref{nuDef}, 
which is given below.

\noindent
\emph{(R1) Uniform Law of Large Numbers:}
The empirical process $\nu_n(G)$ satisfies in 
\begin{align*}
    \sup_{G\in\mathcal{G}_K}n^{-1/2}|\nu_n(G)|\xrightarrow{a.s.} 0, \quad \text{as } n\to\infty.
\end{align*}
It can be validated that the condition will be satisfied if the kernel density $f(y;\theta(\xx,\bbeta),\phi)$ is continuous w.r.t $(\bbeta,\phi)$ for $\nu$-almost every $(\xx,y)$ \citep{GEER2000}.

Secondly, we require a condition called the second-order uniform Lipschitz assumption, stated below, to control the remainder term in the Taylor expansion of the density \( f(y;\theta(\xx,\bbeta),\phi) \). This condition is essential for establishing the Hellinger lower bound in Proposition~\ref{thm:ho_long_results}.

\noindent
\emph{(R2) Second-order Uniform Lipschitz Condition:} There exist positive constants $C$ and $\delta$ such that for any $\bgamma\in\mathbb{R}^d$, $\bPhi_1 = (\bbeta_1, \phi_1),\bPhi_2 = (\bbeta_2, \phi_2)$, $\xx\in\mathcal{X}$, and $y\in\mathcal{Y}$, we have
\begin{align*}
    \Big|\bgamma^{\top}\Big(\frac{\partial^2f(y;\theta(\xx,\bbeta_1), \phi_1)}{\partial\bPhi\partial\bPhi^{\top}}-\frac{\partial^2f(y;\theta(\xx,\bbeta_2),\phi_2)}{\partial\bPhi\partial\bPhi^{\top}}\Big)\bgamma\Big|\leq C\|\bPhi_1-\bPhi_2\|^{\delta}_1\|\bgamma\|^2_2.
\end{align*}

Next, given the strong identifiability (SI) condition, it is necessary to employ the following condition (R3) to leverage a result on the likelihood ratio statistics from \cite{DACUNHA1999} for the proof of Theorem~\ref{thm:sparsity} in Appendix~\ref{sec:theorem_2_proof}.

\noindent
\emph{(R3) Smoothness:} For almost every $\xx \in {\cal X}$, there exists a function $h_1(\cdot;\xx)\in L^1(\nu_{\xx})$ such that $|\log f(y;\theta(\xx,\bbeta),\phi)|\leq h_1(y;\xx)$ $\nu$-almost everywhere. Furthermore, the density $f(y;\theta(\xx,\bbeta),\phi)$ admits partial derivatives up to order five with respect to $\bPhi=(\bbeta,\phi)$. In addition, for all $M\leq 5$, all $\kappa_1,\kappa_2,\ldots,\kappa_M$, and any atom $\bPhi_0$ of $G_0$, we have
\begin{align*}
    U_{\kappa_1,\kappa_2,\ldots,\kappa_M}(\bPhi_0;G_0)\equiv U_{\kappa_1,\kappa_2,\ldots,\kappa_M}(\cdot;\theta(\cdot,\bbeta_0),\phi_0,G_0)\in L^3(P_0(\xx)),
\end{align*}
where $P_0(\xx)$ is the probability distribution corresponding to the density $q_{G_0}(\cdot|\xx)$. We can also find $h_2(\cdot;\xx)\in L^3(P_0(\xx))$ and $\epsilon>0$ such that for all $y\in\mathcal{Y}$, we have
\begin{align*}
    \sup_{\|\bPhi-\bPhi_0\|\leq\epsilon}|U_{\kappa_1,\kappa_2,\ldots,\kappa_5}(y;\theta(\xx,\bbeta),\phi,G_0)|\leq h_2(y;\xx).
\end{align*}
Finally, to control the order of terms related to the likelihood ratio statistics, we introduce the following uniform boundedness condition.

\noindent
\emph{(R4) Uniform Boundedness:} For almost every $\xx$, there exist constants $\epsilon_1,\epsilon_2>0$ and functions $h_3(\cdot;\xx),h_4(\cdot;\xx)\in L^2(P_0(\xx))$ such that for all $y\in\mathcal{Y}$, $|U(y;\theta(\xx,\bbeta),\phi,G)|\leq h_3(y;\xx)$, and for every $\kappa\in[d]$, $|U_{\kappa}(y;\theta(\xx,\bbeta),\phi,G)|\leq h_4(y;\xx)$, uniformly for all $G$ such that $W_2(G,G_0)<\epsilon_1$, and for all $\bPhi\in\Theta$ such that $\|\bPhi-\bPhi_{0k}\|<\epsilon_2$, for some $k\in[K_0]$.

In summary, the {\sc FMR} models with respect to the probability density/mass functions of many popular distributions, namely Gaussian, Binomial, and Poisson distributions, all satisfy the above regularity conditions \citep{HOJMLR2022,DoDoLong2025}. 

\section{Proof of the Main Results}
\label{sec:theorem_1}
\subsection{Proof of Theorem~\ref{thm:paramConsistency}} 
\label{sec:theorem_1_proof}
{\bf Part (i)}. 
First of all, we show that
the {\sc OMPLE} $\hat{G}_n$ satisfies  
\begin{align}
    \label{lemma_bound}
    h^2 \left(\frac{\hat{p}_n + p_0}{2}, p_0 \right) 
\leqslant \frac{1}{\sqrt{n}} \nu_n(\hat{G}_n) 
+ \frac{1}{4} \left[\Delta_{1n}(\hat{G}_n) + 
\Delta_{2n}(\hat G_n)
\right],
\end{align}
for all $n\in\mathbb{N}$, where $\Delta_{1n}$ and $\Delta_{2n}$
are given in equation~\eqref{delta1} and \eqref{delta2}, respectively.  
Indeed, by the concavity of the log function, we have
\begin{equation}
\log \left\{\frac{\hat{p}_n + p_0}{2 p_0} \right\} 1 \{p_0 > 0 \} \geqslant \frac{1}{2} \log \left\{\frac{\hat{p}_n}{p_0} \right\} 1\{p_0 > 0\},
\label{logconcave}
\end{equation}
where $1\{\cdot\}$ is the indicator function. 
Since $\hat{G}_n$ is a maximizer of $L_n(G)$, we have 
\begin{eqnarray*}
0 & \leqslant & \frac{1}{n} \left\{L_n(\hat{G}_n) - L_n(G_0) \right\} = 
\frac{1}{n}\{\ell_n(\hat G_n)  - \ell_n(G_0)\}
+
\Delta_{1n}(\hat{G}_n) + \Delta_{2n}(\hat G_n)
\\
& = & 
\frac{1}{n} \sum_{i=1}^n \log 
\bigg \{\frac{q_{\hat G_n}(y_i|\xx_i) f_{\XX}(\xx_i)}{q_{G_0}(y_i|\xx_i)f_{\XX}(\xx_i) } \bigg \}
+\Delta_{1n}(\hat{G}_n) + 
\Delta_{2n}(\hat G_n)
\\
& = &
\frac{1}{n} \sum_{i=1}^n \log 
\bigg \{\frac{p_{\hat G_n}(\xx_i, y_i)}{
p_{G_0}(\xx_i, y_i)} \bigg \}
+\Delta_{1n}(\hat{G}_n) + 
\Delta_{2n}(\hat G_n)
\\
 & = &  
\int \log \left\{\frac{\hat{p}_n}{p_0} \right\} dP_n + \Delta_{1n}(\hat{G}_n) + 
\Delta_{2n}(\hat G_n). 
\end{eqnarray*}

Thus, by the result in equation \eqref{logconcave} and the above inequality, 
we arrive at  
\begin{align*}
-\frac{1}{4} &\left[\Delta_{1n}(\hat{G}_n) + 
\Delta_{2n}(\hat G_n)
\right] 
 \leqslant \int_{\{p_0 > 0\}} \frac{1}{4} \log \left\{\frac{\hat{p}_n}{p_0} \right\} dP_n \\
& \leqslant \int_{\{p_0 > 0\}} \frac{1}{2} \log \left\{\frac{\hat{p}_n + p_0}{2p_0} \right\} d(P_n - P_0) + \int_{\{p_0 > 0\}} \frac{1}{2} \log \left\{\frac{\hat{p}_n + p_0}{2p_0} \right\} dP_0 \\
&= \int_{\{p_0 > 0\}} \frac{1}{2} \log \left\{\frac{\hat{p}_n + p_0}{2p_0} \right\} d(P_n - P_0) - \frac{1}{2} \text{KL} \left(\frac{\hat{p}_n + p_0}{2}, p_0 \right) \\
&\leqslant \frac{1}{\sqrt{n}} \nu_n(\hat{G}_n) - h^2 \left(\frac{\hat{p}_n + p_0}{2}, p_0 \right).
\end{align*}
The last inequality occurs since the inequality $h^2(p_1, p_2) \leqslant \frac{1}{2} \text{KL}(p_1, p_2)$ holds for any two densities $p_1$ and $p_2$ 
with respect to the measure $\nu$. Then, we obtain the result in equation~\eqref{lemma_bound}. As a result, it follows that
\begin{align*}
h^2(\hat{p}_n, p_0) & \lesssim
h^2 \left(\frac{\hat{p}_n + p_0}{2}, p_0 \right) 
\leqslant \frac{1}{\sqrt{n}} \nu_n(\hat{G}_n) 
+ \frac{1}{4} \left[\Delta_{1n}(\hat{G}_n) + 
\Delta_{2n}(\hat G_n) 
\right]\\
&\leqslant \text{sup}_{G \in \mathcal{G}_{K}} \frac{1}{\sqrt{n}} |\nu_n(G)| + \frac{\rho_{K_0}(\bpi^0)}{4n} + \frac{1}{4} \sum_{k=1}^{K_0} \sum_{j=1}^{p_k} 
r_{\lambda_n}(\beta^0_{kj}).
\end{align*}

Since the elements of $\bpi^0$ are bounded away from 0, 
$\frac{\rho_{K_0}(\bpi^0)}{4n} = o(1)$, by Condition (P0). 
Also, by condition (P1), 
$\lim_{n \to \infty} r_{\lambda_n}(\beta) = 0$, as $n \to \infty$. 
Thus, $\frac{1}{4} \sum_{k=1}^{K_0} \sum_{j=1}^{p_k} r_{\lambda_n}(\beta^0_{kj}) = o(1)$. 
By the condition (R1) on uniform law of large numbers, 
we have 
$\text{sup}_{G \in \mathcal{G}_{K}} \frac{1}{\sqrt{n}} |\nu_n(G)| \xrightarrow{a.s.} 0$, 
as $n \to \infty$. Therefore, we obtain 
$h(\hat{p}_n, p_0) \xrightarrow{a.s.} 0$.

On the other hand, for any $r \geqslant 1$, 
using the interpolation equations in \cite{VILLANI2003} and part (ii)
of Proposition \ref{thm:ho_long_results}, under conditions (SI) and (R2)
and the compactness 
of the parameter space $\Theta$, we have
\begin{equation}
\label{w-h-dist}
W^r_r(\hat{G}_n, G_0) \leqslant \left(\text{diam}^{r-2}(\Theta) \lor 1 \right) W^2_2(\hat{G}_n, G_0) \lesssim h(\hat{p}_n, p_0) \xrightarrow{a.s.} 0.
\end{equation} 
Hence, $W_r(\hat{G}_n, G_0) \to 0$, almost surely, for all $r \geqslant 1$. This completes the proof of 
{\bf Part (i)}. \\

\noindent
{\bf Part (ii)}. The above result  
implies that, with probability tending to one for large $n$, 
the estimate $\hat{G}_n$ has at least as many atoms as $G_0$. 
Thus, for every $k = 1,\ldots,K_0$, there exists an index $1 \leqslant s \leqslant K$ such that $\big \Vert \hat{\bPhi}_s - \bPhi^0_{k} \big \Vert \xrightarrow{p} 0$, as $n \to \infty$. For each $k = 1,\ldots,K_0$, denote such $s$ as $s_{k}$.
Therefore, there exists a subset $S \subset \{1, \ldots, K\}$
such that $|S| \ge K_0$ and $s_k \in S$ with the above property. 
Then, for $k = 1,\ldots,K_0$, $j = 1,\ldots,p_k$, we have 
$|\hat{\beta}_{s_{k}j}| \geqslant \delta  = \frac{1}{2}\min_{1 \le j \le p_k} |\beta^0_{kj}| > 0$, and by compactness of $\Theta$, there must exist $D > 0$ such that $|\hat{\beta}_{s_{k}j}| \leqslant D$ in probability, 
for large $n$. Let $\bar{G}_n$ be the MLE over $\mathcal{G}_K$ so that 
$\ell_n(\hat{G}_n) \le \ell_n(\bar{G}_n)$. Thus, we have that 
\begin{align}
0 &\leqslant L_n(\hat{G}_n) - L_n(G_0) \nonumber\\
& = \left\{\ell_n(\hat{G}_n) - \ell_n(G_0) \right\} + \{\rho_{K_0}(\bpi^0) - \rho_K(\hat{\bpi})\} 
+ n \left\{\sum_{k=1}^{K_0}\sum_{j=1}^{p_k} r_{\lambda_n}(\beta^0_{kj}) 
- \sum_{k=1}^{K} \sum_{j=1}^{p} r_{\lambda_n}(\hat \beta_{kj}) \right\} \nonumber\\
& = \left\{\ell_n(\hat{G}_n) - \ell_n(G_0) \right\} 
+ \{\rho_{K_0}(\bpi^0) - \rho_K(\hat{\bpi})\}  
+ n~ \Delta_{2n}(\hat G_n) \nonumber \\
& \le 
\left\{\ell_n(\bar{G}_n) - \ell_n(G_0) \right\} 
+ \{\rho_{K_0}(\bpi^0) - \rho_K(\hat{\bpi})\}  
+ n~ \Delta_{2n}(\hat G_n)
\label{ineq1}
\end{align}
where 
$\Delta_{2n}(\cdot)$, evaluated at $\hat G_n$, is given in 
\eqref{delta2}. 
Since $r_{\lambda_n} \ge 0$, by Condition (P1) and 
the minimum-signal condition \eqref{min-signal}, 
with probability tending to one, for large $n$,
\begin{align}
n \Delta_{2n}(\hat G_n)
& \le 
n \left\{\sum_{k=1}^{K_0}\sum_{j=1}^{p_k} 
r_{\lambda_n}(\beta^0_{kj})  -  
\sum_{s \in S} \sum_{j=1}^{p_s}  
r_{\lambda_n}(\hat \beta_{sj}) 
\right\} = 
n \left\{\sum_{k=1}^{K_0}\sum_{j=1}^{p_k} 
c_n  -  
\sum_{s \in S} \sum_{j=1}^{p_s}  
c_n 
\right\} \le 0.
\label{ineq2}
\end{align}
Thus, combining \eqref{ineq1} and \eqref{ineq2}, for large $n$, we have that 
\begin{align*}
    0 & \le L_n(\hat{G}_n) - L_n(G_0)
     \le \ell_n(\bar{G}_n) - \ell_n(G_0) + \rho_{K_0}(\bpi^0) - \rho_K(\hat{\bpi})
\end{align*}
and therefore, 
\begin{equation*}
   0 \le \rho_K(\hat{{\bpi}}) \le \ell_n(\bar{G}_n) - \ell_n(G_0) + \rho_{K_0}({\bpi}_0).
\end{equation*}
with probability tending to 1. Note that $\rho_{K_0}(\bpi^0) = O(1)$. Also,
under Conditions (SI) and (R3), we use the result of Lemma \ref{thm:lrs} 
on the boundedness of the likelihood ratio statistic \citep{DACUNHA1999}, that is, 
$\ell_n(\bar{G}_n) - \ell_n(G_0) = O_p(1)$. Thus, for large $n$, 
\begin{equation}
\label{ineq4}
    0 \le \rho_K(\hat{\bpi}) \le \ell_n(\bar{G}_n) - \ell_n(G_0) + \rho_{K_0}(\bpi^0) = O_p(1).
\end{equation}
which implies that $\rho_K(\hat \bpi) = O_p(1)$,  
as $n \to \infty$. Thus, for any $\epsilon>0$, there exists
$M_\epsilon<\infty$ such that, for all sufficiently large $n$,
\(
\Pr\{\rho_K(\hat{\bpi})\le M_\epsilon\}>1-\epsilon.
\)
By Condition (P0), which requires $\rho_K(\bpi)\to\infty$ as
$\min_{1\leq k\leq K}\pi_k\to0$, there exists $b_{\epsilon}$
such that for sufficiently large $n$, 
\(
\Pr\left(
\hat\pi_k\geq b_\epsilon,\ k=1,\ldots,K
\right) \ge \Pr\{\rho_K(\hat{\bpi})\le M_\epsilon\}>1-\epsilon.
\)
This implies that, by {\bf Part (i)} and Lemma \ref{lem:ho}, 
for all $k = 1,\ldots, K_0$, up to relabeling,
\[
\sum_{j \in \hat {\cal I}_{k}} \hat{\pi}_j = \pi^0_{k} + o_p(1),
\]
which completes the proof of {\bf Part (ii)}. \\

\noindent 
{\bf Part (iii)}. By {\bf Part (i)} and Lemma \ref{lem:ho}, we also have that 
for every $l = 1,\ldots,K$, 
there exists $k = 1,\ldots, K_0$ such that 
$\big \Vert \hat{\bPhi}_l - \bPhi^0_{k} \big \Vert \xrightarrow{p} 0$, 
or equivalently for all $k\in[K_0]$, 
$\text{max}_{j \in \hat {\cal I}_{k}} \big \Vert \hat{\bPhi}_j 
- 
\bPhi^0_{k} \big \Vert \xrightarrow{p} 0$, 
as $n \to \infty$. 
$\hfill \blacksquare$

\subsection{Proof of Theorem \ref{thm:sparsity}}
\label{sec:theorem_2_proof}
First, it follows from Theorem~\ref{thm:paramConsistency}-(i) that the OMPLE $\hat G_n$ 
is a consistent estimator of $G_0$ with respect to $W_r$, and therefore has at least $K_0$ components, 
with probability tending to one, 
as $n \to \infty$. Thus, given $\epsilon > 0$, Theorem~\ref{thm:paramConsistency}-(ii) and Proposition~\ref{densityConsistency} reveal that we can find positive constants $b_\epsilon$ and $N_\epsilon$ that satisfy
\[
\bbP(\hat G_n \in {\cal G}_K(b_\epsilon;\gamma_n)) \geq 1 - \epsilon,
\]
for all $n\geq N_\epsilon$, where ${\cal G}_K(b_\epsilon;\gamma_n)$ is defined in equation \eqref{tildeGSet2}, and $\gamma_n \to 0$ is a sequence satisfying 
$\gamma_n \gtrsim (\log n / n)^{1/2}$ as stated in 
Proposition~\ref{densityConsistency}. Then, we have
\begin{eqnarray}
& &\mathbb{P} \big(\hat{G}_n \text{ does not recover sparsity of } G_0 \big) \nonumber\\
&=& \mathbb{P} \big(\hat{G}_n \text{ does not recover sparsity of } G_0~
\text{and}~
\hat{G}_n \notin \mathcal{G}_K(b_\epsilon,\gamma_n)\big) \nonumber\\
& + &  
\mathbb{P} (\hat{G}_n \text{ does not recover sparsity of } G_0~
\text{and}~
\hat{G}_n \in \mathcal{G}_K(b_\epsilon,\gamma_n)) \nonumber\\
& \leqslant & \epsilon + \mathbb{P} \big(\hat{G}_n \text{ does not recover sparsity of } G_0 ~\text{and}~ \hat{G}_n \in \mathcal{G}_K(b_\epsilon,\gamma_n)\big)\nonumber\\
&\leqslant& \epsilon + \mathbb{P}\bigg(\sup_{G \in \mathcal{G}_{K, \mathcal{S}^{c}}(b_\epsilon,\gamma_n)} L_n(G) > \sup_{G \in \mathcal{G}_{K, \mathcal{S}}(b_\epsilon,\gamma_n)} L_n(G)\bigg)\nonumber\\
&\leqslant& \epsilon + \mathbb{P}\bigg(\sup_{G \in \mathcal{G}_{K, \mathcal{S}^{c}}(b_\epsilon,\gamma_n)} 
\big \{ L_n(G) - L_n(\check{G}_{K_0,S}) \big\} > 0 \bigg),
\label{Ineq1}
\end{eqnarray}
where $\check{G}_{K_0,S}$ is introduced below, and  
\begin{equation*}
\mathcal{G}_{K, \mathcal{S}^{c}}(b_\epsilon,\gamma_n) = \{G \in \mathcal{G}_{K}(b_\epsilon,\gamma_n): G \text{ does not recover the sparsity of } G_0\}
\end{equation*}
and
\begin{equation*}
\mathcal{G}_{K, \mathcal{S}}(b_\epsilon,\gamma_n) = \{G \in \mathcal{G}_{K}(b_\epsilon,\gamma_n): G \text{ recovers the sparsity of } G_0\}
\end{equation*}
such that 
\[
\mathcal{G}_K(b_\epsilon,\gamma_n) = 
\mathcal{G}_{K, \mathcal{S}}(b_\epsilon,\gamma_n) \bigcup 
\mathcal{G}_{K, \mathcal{S}^{c}}(b_\epsilon,\gamma_n).
\]
Next, we describe $\check{G}_{K_0, S}$. Let $G = \sum_{j=1}^{K} \pi_j \delta_{(\bbeta_j, \phi_j)} \in \mathcal{G}_{K, \mathcal{S}^{c}}(b_\epsilon,\gamma_n)$ be an arbitrary mixing measure of the fixed order at most $K$ and at least $K_0$ such that $\pi_j \geqslant b_\epsilon$ for all $j \in[K]$. Moreover, $G$ does not recover the sparsity of $G_0$. From the result of Proposition~\ref{thm:ho_long_results}, 
we have $W_2(G, G_0) = O(\gamma_n^{1/2})$.
Let $\varrho_k = \sum_{j \in \mathcal{I}_k} \pi_j, k\in[K_0]$, and $\check{G}_{K_0,S} 
= \sum_{k=1}^{K_0} 
\varrho_k~\delta_{\big((\check \bbeta^{(1)}_k, \bzero),\,\check{\phi}_k\big)}$ 
be the maximizer of $L_n(G)$ over the set of mixing measures in $\mathcal{G}_{K_0}$ with 
mixing proportions fixed at $\varrho_1,\ldots,\varrho_{K_0}$ and the zero coefficients of 
$G_0$ are exactly recovered.
Since $\mathcal{G}_{K, \mathcal{S}}(b_\epsilon,\gamma_n)$ is the space of sparse 
mixing measures with at least $K_0$ and at most $K$ components, by construction 
for any arbitrary $G \in \mathcal{G}_{K, \mathcal{S}^{c}}(b_\epsilon,\gamma_n)$, we have 
$\check{G}_{K_0,S} \in \mathcal{G}_{K, \mathcal{S}}(b_\epsilon,\gamma_n)$. 
The sparse mixing measure $\check{G}_{K_0,S}$ also depends on $G$, i.e. 
$\check{G}_{K_0,S} := \check{G}_{K_0,S}(G)$. The reason for the last inequality 
\eqref{Ineq1} is as follows. We have that, 
\[
  \sup_{G \in \mathcal{G}_{K, \mathcal{S}}(b_\epsilon,\gamma_n)} L_n(G)\geq L_n(\check{G}_{K_0,S}(G))~~,~~ 
    \forall G\in \mathcal{G}_{K, \mathcal{S}^{c}}(b_\epsilon,\gamma_n)
\]
which implies 
\[
 \sup_{G \in \mathcal{G}_{K, \mathcal{S}}(b_\epsilon,\gamma_n)} L_n(G)\geq \sup_{G\in \mathcal{G}_{K, \mathcal{S}^{c}}(b_\epsilon,\gamma_n)}L_n(\check{G}_{K_0,S}(G)).
\]
Thus, based on probability statement above \eqref{Ineq1},
\[
\mathbb{P} \big(\hat{G}_n \text{does not recover sparsity of} G_0 \big) \le 
\]
\[
\epsilon + \mathbb{P}\bigg (\sup_{G \in \mathcal{G}_{K, \mathcal{S}^{c}}(b_\epsilon,\gamma_n)} L_n(G) > \sup_{G \in \mathcal{G}_{K, \mathcal{S}}(b_\epsilon,\gamma_n)} L_n(G)\bigg) \le 
\]
\[
\epsilon + \mathbb{P}\bigg(
\sup_{G \in \mathcal{G}_{K, \mathcal{S}^{c}}(b_\epsilon,\gamma_n)} L_n(G) > 
    \sup_{G \in \mathcal{G}_{K, \mathcal{S}^{c}}(b_\epsilon,\gamma_n)} L_n(\check{G}_{K_0,S}(G))
\bigg) \le
\]
\[
\epsilon + \mathbb{P}\bigg(\sup_{G \in \mathcal{G}_{K, \mathcal{S}^{c}}(b_\epsilon,\gamma_n)} 
\big \{ L_n(G) - L_n(\check{G}_{K_0,S}(G)) \big\} > 0 \bigg)
\]
and hence \eqref{Ineq1} is verified. 

Now, it suffices to show that the probability in equation \eqref{Ineq1} tends 
to zero as $n$ goes to infinity. For notational simplicity, 
throughout the proof we write $\check{G}_{K_0,S}(G) = \check{G}(G)$, for any 
arbitrary $G \in \mathcal{G}_{K, \mathcal{S}^{c}}(b_\epsilon,\gamma_n)$. 
Using the same proof techniques of Propositions~\ref{thm:ho_long_results} and \ref{densityConsistency}, under Condition (R3), we have 
$W_2(\check{G}(G),G_0) = O_p(\gamma^{1/2}_n)$.  
Also, since $\check{G}(G)$ has $K_0$ components with the positive mixing probabilities, 
every atom of this mixing distribution all falls in $\gamma^{1/2}_n$-neighborhood 
of a corresponding atom of $G_0$. We now proceed as follows. 

Denote $\bpi = (\pi_1, \dots, \pi_K)^\top$ and 
$\bvarrho = (\varrho_1, \dots, \varrho_{K_0})^\top$. We consider the difference, 
\begin{equation*}
    L_n(G) - L_n(\check{G}(G))  =  \left\{\ell_n(G)-\ell_n(\check{G}(G))\right\} - \left\{\rho_K(\bpi)-\rho_{K_0}(\bvarrho)\right\} - 
    \left\{\mathcal{P}(G) - \mathcal{P}(\check{G}(G))\right\}
\end{equation*}
for any $G \in \mathcal{G}_{K, \mathcal{S}^{c}}(b_\epsilon,\gamma_n)$,
where 
\begin{eqnarray*}
    \mathcal{P}(G) - \mathcal{P}(\check{G}(G))
     & =  & 
n \sum_{k=1}^{K} \sum_{j=1}^{p} r_{\lambda_n}(\beta_{kj})
-
n\sum_{k=1}^{K_0} \sum_{j=1}^{p} r_{\lambda_n}(\check \beta_{kj}).
\end{eqnarray*}
By Condition (P0) on $\rho_K$ in Section \ref{sec:asymp} of the paper, we have 
\(
\left\{\rho_K(\bpi)-\rho_{K_0}(\bvarrho)\right\} \ge 0.
\)
Thus, 
\begin{equation}
    \label{dif-plike}
    L_n(G) - L_n(\check{G}(G))  \le  \left\{\ell_n(G)-\ell_n(\check{G}(G))\right\}  
    - \left\{\mathcal{P}(G) - \mathcal{P}(\check{G}(G))\right\}
\end{equation}
for any $G \in \mathcal{G}_{K, \mathcal{S}^{c}}$. Subsequently, 
we show that the penalized log-likelihood 
difference in \eqref{dif-plike} is negative, for 
sufficiently large $n$, by bounding the two terms on the right hand side.\\ 

\noindent
\textbf{Step 1: Bounding below the second term on the right hand side of \eqref{dif-plike}}:
Due to the sparsity of $\check G(G)$ and that $r_{\lambda_n}(0) = 0$, 
by Condition (P1) on $r_{\lambda_n}$ in Section \ref{sec:asymp}, 
\begin{eqnarray}
    \mathcal{P}(G) - \mathcal{P}(\check{G}(G))
& = &
n\sum_{k=1}^{K} \sum_{j=1}^{p} r_{\lambda_n}(\beta_{kj})
-
n\sum_{k=1}^{K_0} \sum_{j=1}^{p_k} r_{\lambda_n}(\check \beta_{kj})
\nonumber\\
& = & 
n \sum_{k=1}^{K_0} \bigg(\sum_{s \in \mathcal{I}_k} \sum_{j=1}^{p} r_{\lambda_n}(\beta_{sj}) - \sum_{j=1}^{p_k} r_{\lambda_n}(\check \beta_{kj}) \bigg)
\nonumber \\
& = &
n \sum_{k=1}^{K_0} \bigg(\sum_{s \in \mathcal{I}_k} \bigg [ \sum_{j=1}^{p_k} 
r_{\lambda_n}(\beta_{sj})
+
\sum_{j>p_k}^{p} 
r_{\lambda_n}(\beta_{sj})
\bigg ] - \sum_{j=1}^{p_k} r_{\lambda_n}(\check \beta_{kj}) 
\bigg) \nonumber \\
& = & 
n \sum_{k=1}^{K_0} \sum_{s \in \mathcal{I}_k} 
\sum_{j>p_k}^{p} r_{\lambda_n}(\beta_{sj})
  + 
n \sum_{k=1}^{K_0} \left(\sum_{s \in \mathcal{I}_k} \sum_{j=1}^{p_k} 
r_{\lambda_n}(\beta_{sj})
 - \sum_{j=1}^{p_k} r_{\lambda_n}(\check \beta_{kj}) \right)
 \nonumber
\hspace{1cm}
\label{pen-sparse}
\end{eqnarray}

Also, by Condition (P1), the minimum-signal condition 
\eqref{min-signal}, and $W_2$-localization of $G$ and 
$\check G(G)$ around $G_0$, for sufficiently large $n$, 
all the coefficients $\beta_{sj}$ and $\check{\beta}_{kj}$
corresponding to the true nonzero coefficients $\beta^0_{kj} \not= 0, j \le p_k$, 
lie in the flat region of the penalty 
and hence 
$r_{\lambda_n}(\beta_{sj})- r_{\lambda_n}(\check \beta_{kj})
= c_n - c_n = 0$. Since $|{\cal I}_k| \ge 1$, for all $k \in [K_0]$, for sufficiently large $n$, 
\begin{equation*}
n \sum_{k=1}^{K_0} \left(\sum_{s \in \mathcal{I}_k} \sum_{j=1}^{p_k} 
r_{\lambda_n}(\beta_{sj})
 - \sum_{j=1}^{p_k} r_{\lambda_n}(\check \beta_{kj}) \right) 
 \ge 0.
\end{equation*}
Hence, for every $G \in \mathcal{G}_{K, \mathcal{S}^{c}}(b_\epsilon,\gamma_n)$, 
for sufficiently large $n$, with a high probability,
\begin{equation}
\label{pen-lower-bound}
  \mathcal{P}(G) - \mathcal{P}(\check{G}(G)) 
  \ge 
 n \sum_{k=1}^{K_0} \sum_{s \in \mathcal{I}_k} 
\sum_{j>p_k}^{p} r_{\lambda_n}(\beta_{sj}). 
\end{equation}

\noindent 
\textbf{Step 2: Bounding above the first term in equation \eqref{dif-plike}}:
Recall that the arbitrary mixing measure $G \in \mathcal{G}_{K, \mathcal{S}^{c}}(b_\epsilon,\gamma_n)$ 
is defined as $G = \sum_{j=1}^K \pi_j \delta_{\bPhi_j}$, 
where $\bPhi_j = (\bbeta_j, \phi_j)$. Let $H_k$ be a discrete measure whose atoms are the 
elements of the Voronoi cells ${\cal V}_k$, given by
\begin{equation}
\label{Hk_defn}
\displaystyle
H_k = \frac 1 {\varrho_k} \sum_{j \in {\cal I}_k} \pi_j \delta_{\bPhi_j}, \qquad k \in [K_0].
\end{equation}
which is a mixing measure in its own right. Then, we may rewrite the mixing measure $G$ as 
\begin{equation}
\label{GDef}
G\ = \sum_{k=1}^{K_0} \varrho_k  H_k.
\end{equation}
Using the inequality $\log (1+x) \le x - x^2/2 + x^3/3$ 
and the log-likelihood in equation~\eqref{joint-like}, 
\begin{equation}
\ell_n(G) - \ell_n(\check{G}) = \sum_{i=1}^{n} \log \left\{1 +  D_i\left(G,\check{G}\right)\right\} \leqslant \sum_{i=1}^{n} D_i\left(G,\check{G}\right) - \frac{1}{2} \sum_{i=1}^{n} D_i^2\left(G,\check{G}\right) + \frac{1}{3} \sum_{i=1}^{n} D_i^3\left(G,\check{G}\right),
\label{like-ineq1}
\end{equation}
where we denote
\begin{eqnarray*}
D_i\left(G,\check{G}\right) & = & 
\frac{p_{G}(\VV_i) - p_{\check{G}}(\VV_i)}{p_{\check{G}}(\VV_i)} 
 =  \frac{q_{G}(Y_i|\XX_i) - q_{\check{G}}(Y_i|\XX_i)}{q_{\check{G}}(Y_i|\XX_i)} \\ \\
& = &
\sum_{k=1}^{K_0} \varrho_k \int \frac{f(Y_i; \theta(\XX_i,\bbeta),\phi) 
- 
f(Y_i; \theta(\XX_i, (\check \bbeta^{(1)}_k, \bzero)),  \check \phi_k)}{q_{\check{G}}(Y_i|\XX_i)} \, d H_k(\bPhi),
\end{eqnarray*}
where  
$\VV_i = (\XX_i, Y_i)$, and by \eqref{joint-fmr} we have 
$p_G(\vv) \equiv p_G(\xx, y) = q_G(y|\xx) f_{\XX}(\xx)$.\\

Next, we assess the three terms on the RHS of equation \eqref{like-ineq1}
for large $n$, in three {\bf Steps 2A, 2B, 2C} that follow.  \\


\noindent
\textbf{Step 2A: Bounding the term $\sum_{i=1}^{n} D_i(G, \check G)$:}

Consider the parameter vector 
$\bPhi = (\beta_1, \ldots, \beta_p, \phi)^\top = (\eta_1, \ldots, \eta_{p+1})^\top$. 
Using a third-order Taylor's expansion, we have  
\begin{align}
\sum_{i=1}^{n} D_i(G, \check G)
& = 
\sum_{k=1}^{K_0} \varrho_k \sum_{i=1}^n \int \frac{f(Y_i; \theta(\XX_i,\bbeta),\phi) 
- 
f(Y_i; \theta(\XX_i, (\check \bbeta^{(1)}_k, \bzero)),  \check \phi_k)}{q_{\check{G}}(Y_i|\XX_i)}~ d H_k(\bPhi)
\nonumber\\
& = 
\sum_{k=1}^{K_0} \varrho_k \bigg\{\sum_{r=1}^{p+1} \int (\eta_r - \check{\eta}_{kr}) ~dH_k(\bPhi)\sum_{i=1}^{n} U_{i,r}( \check{\bPhi}_k; \check G) \nonumber\\ &
+ \frac{1}{2}\sum_{r=1}^{p+1}\sum_{l=1}^{p+1} 
\int (\eta_r - \check \eta_{kr})(\eta_l - \check{\eta}_{kl})~ dH_k(\bPhi)
\sum_{i=1}^{n} U_{i,rl}(\check{\bPhi}_{k}; \check{G}) \nonumber\\ & 
+ \frac{1}{6} \sum_{r=1}^{p+1}\sum_{l=1}^{p+1}\sum_{h=1}^{p+1} \int (\eta_r - \check{\eta}_{kr})(\eta_l - \check{\eta}_{kl})(\eta_h - \check{\eta}_{kh}) \sum_{i=1}^{n} U_{i,rlh}(\bxi_k; \check{G})~ dH_k(\bPhi)\bigg\}
\nonumber\\
& = 
\sum_{k=1}^{K_0} \varrho_k \sum_{r=1}^{p+1} m_{k,r} \sum_{i=1}^{n} 
 U_{i,r}( \check{\bPhi}_k; \check G) + 
 \frac{1}{2}\sum_{k=1}^{K_0} \rho_k \sum_{r=1}^{p+1}\sum_{l=1}^{p+1} 
 m_{k,rl}
\sum_{i=1}^{n} U_{i,rl}(\check{\bPhi}_{k}; \check{G}) \nonumber\\
& + 
\frac{1}{6} \sum_{k=1}^{K_0} \varrho_k 
\sum_{r=1}^{p+1}\sum_{l=1}^{p+1}\sum_{h=1}^{p+1} \int (\eta_r - \check{\eta}_{kr})(\eta_l - \check{\eta}_{kl})(\eta_h - \check{\eta}_{kh}) \sum_{i=1}^{n} U_{i,rlh}(\bxi_k; \check{G})~ dH_k(\bPhi)
\nonumber
\\
& =  (I) + (II) + (III),
\label{three-terms1}
\end{align}
where $\bxi_k$ is on the segment between $\bPhi$ 
and $\check{\bPhi}_{k}$, the moments 
\begin{equation}
\label{moments}
m_{k,r} = \int (\eta_r - \check{\eta}_{kr})~ dH_k(\bPhi)
~~,~~
m_{k,rl} = \int (\eta_r - \check{\eta}_{kr})(\eta_l - \check{\eta}_{kl})~ dH_k(\bPhi), 
\end{equation}
and $U_{i, \cdot}(\bPhi,G)\equiv U_{\cdot}(Y_i;\theta(\XX_i,\bbeta),\phi,G)$ are defined in equations \eqref{U0} and \eqref{U1}.

By construction, we know that $\check{G}$ is a stationary point of $L_n(G)$, so for $n$ sufficiently large, with high probability, we have the following equations, 
for $k \in [K_0]$,
\begin{eqnarray}
\label{Eq1thm3}
\varrho_k \sum_{i=1}^{n} U_{i,j}(\check{\bPhi}_{k}; \check{G}) 
& = &  
n \frac{\partial r_{\lambda_n}(\check{\beta}_{kj})}{\partial \beta} = 0~~,~~
j=1,\ldots,p_k 
\\
\label{Eq2thm3}
\left|\varrho_k \sum_{i=1}^{n} U_{i,j} (\check{\bPhi}_k; \check{G}) \right| 
& < & 
n r_{\lambda_n}^{'}(0^+)~~,~~
j = p_k + 1,\ldots,p\\
\varrho_k \sum_{i=1}^{n} U_{i,j}(\check{\bPhi}_{k}; \check{G}) 
& = &  0~~,~~j= p+1,
\label{Eq3thm3}
\end{eqnarray}
where equations \eqref{Eq1thm3} and \eqref{Eq2thm3} correspond to the regression 
parameters and equation \eqref{Eq3thm3} corresponds to the dispersion parameter. 
Condition (P1) on $r_{\lambda_n}$ results in equation \eqref{Eq1thm3}, for large $n$. 
The above equations are referred to as the 
Karush-Kuhn-Tucker (KKT) conditions. Thus, 
\begin{eqnarray*}
(I) & = &  
\sum_{k=1}^{K_0} \varrho_k \sum_{r=1}^{p+1} m_{k,r} \sum_{i=1}^{n} 
 U_{i,r}( \check{\bPhi}_k; \check G)\\
 & = & 
 \sum_{k=1}^{K_0} \varrho_k \sum_{r=1}^{p_k} m_{k,r} \sum_{i=1}^{n} 
 U_{i,r}( \check{\bPhi}_k; \check G) 
  +  
 \sum_{k=1}^{K_0} \varrho_k \sum_{r=p_k+1}^{p} m_{k,r} \sum_{i=1}^{n} 
 U_{i,r}( \check{\bPhi}_k; \check G) + 0
 \end{eqnarray*}
 and hence, by Condition (P1), for large $n$, 
 \begin{eqnarray}
(I) & = & 
 n \sum_{k=1}^{K_0} \sum_{r=1}^{p_k} m_{k,r} 
 \frac{\partial r_{\lambda_n}(\check{\beta}_{kr})}{\partial \beta} + 
 \sum_{k=1}^{K_0} \varrho_k \sum_{r=p_k+1}^{p} m_{k,r} \sum_{i=1}^{n} 
 U_{i,r}( \check{\bPhi}_k; \check G) \nonumber \\ 
 & = & 0+ 
 \sum_{k=1}^{K_0} \varrho_k \sum_{r=p_k+1}^{p} m_{k,r} \sum_{i=1}^{n} 
 U_{i,r}( \check{\bPhi}_k; \check G)  
 =  
 \sum_{k=1}^{K_0} \varrho_k \sum_{r=p_k+1}^{p} m_{k,r} \sum_{i=1}^{n} 
 U_{i,r}( \check{\bPhi}_k; \check G).
 \label{linear-term1}
\end{eqnarray}

Next, to assess (II) in equation \eqref{three-terms1}, we first assess 
the behaviour of $\sum_{i=1}^{n} U_{i,rl}(\check{\bPhi}_{k}; \check{G})$
for large $n$. Under Condition (R3), we have 
$\mathbb E \{U_{i,rl}(\bPhi; G_0) \} = 0$;  
thus by the Dominated Convergence Theorem, uniformly over
all $\bPhi$ in a neighborhood of a true atom of $G_0$, and 
all $r, l=1, 2, \dots, p+1$, we have 
\begin{equation}
\label{secondTerm1}
\sum_{i=1}^n U_{i,rl}(\bPhi; G_0) = O_p(n^{\frac 1 2}).
\end{equation}
 
Now, for all $k\in[K_0]$, we write,
\begin{equation}
\label{secondTerm2}
\sum_{i=1}^n U_{i, rl}(\check \bPhi_k; \check G) = 
\sum_{i=1}^n U_{i, rl}(\check \bPhi_k; G_0) + 
\sum_{i=1}^n  \left\{ U_{i, rl}(\check \bPhi_k; \check G) 
- U_{i, rl}(\check \bPhi_k; G_0) 
\right\}
\end{equation}

The first term in the above identity can be bounded as follows. Using equation \eqref{secondTerm1},
for some vectors $\tilde \bPhi^0_{ik}$ on the segment between $\bPhi^0_{k}$ and $\check \bPhi_k$,
\begin{align}
\sum_{i=1}^n U_{i, rl}(\check \bPhi_k; G_0) 
 &= \sum_{i=1}^n U_{i, rl}(\bPhi^0_{k}; G_0) 
+\sum_{i=1}^n \Big[U_{i, rl}(\check \bPhi_k; G_0)  - U_{i, rl}(\bPhi^0_{k}; G_0) \Big] \nonumber\\
 &= O_p(n^{1/2}) 
+\sum_{i=1}^n \sum_{s=1}^{p+1} U_{i, rls}(\tilde \bPhi^0_{ik}; G_0)
(\check \eta_{ks} - \eta^0_{ks}) 
=O_p(n\gamma_n^{\frac 1 2}),
\label{secondTerm22}
\end{align}
where to obtain the given order in \eqref{secondTerm22},
we invoked the uniform bound in condition (R3) on the third order derivative $U_{i, rls}(; G_0)$, 
$\|\check{\bPhi}_k - \bPhi^0_k\| = O_p(\gamma^{1/2}_n)$, and that the dimension 
$p$ is fixed. The above order assessment hold uniformly in $G \in \mathcal{G}_{K, \mathcal{S}^{c}}(b_\epsilon,\gamma_n)$.

By Cauchy-Schwarz inequality, 
we may bound the second term in equation \eqref{secondTerm2} as follows,
\begin{align*}
&\sum_{i=1}^n \left| U_{i, rl}
(\check \bPhi_k; \check G) - U_{i, rl}(\check \bPhi_k; G_0) \right| 
 = \sum_{i=1}^n|U_{i, rl}(\check \bPhi_k; G_0)| 
 \frac{1}{q_{\check G}(Y_i|\XX_i)} \left| q_{G_0}(Y_i|\XX_i) - q_{\check G}(Y_i|\XX_i) \right| \\
 &\leq \sum_{i=1}^n |U_{i, rl}(\check \bPhi_k; G_0)| 
   \bigg\{\frac {1}{q_{\check G}(Y_i|\XX_i)} \sum_{k=1}^{K_0} \pi^0_k \big| 
   f(Y_i; \theta(\XX_i,\check \bbeta_k),\check \phi_k) - f(Y_i;\theta(\XX_i, \bbeta^0_k),\phi^0_k)\big| \\
   &+ 
    \frac{1}{q_{\check G}(Y_i|\XX_i)} 
    \sum_{k=1}^{K_0} |\varrho_k-\pi^0_k| f(Y_i; \theta(\XX_i,\check \bbeta_k),\check \phi_k) 
    \bigg\}\\
     &\leq \sum_{i=1}^n |U_{i, rl}(\check \bPhi_k; G_0)| 
   \bigg\{\frac {1}{q_{\check G}(Y_i|\XX_i)} \sum_{k=1}^{K_0} \pi^0_k 
   \sum_{h=1}^{p+1} \bigg | \frac{\partial f(Y_i;\theta(\XX_i,\bzeta_{1k}),\bzeta_{2k})}{\partial \eta_h}\bigg |
   \big| \check \eta_{kh} - \eta^0_{kh} \big|\\
    &+ 
    \frac{1}{q_{\check G}(Y_i|\XX_i)} 
    \sum_{k=1}^{K_0} |\varrho_k-\pi^0_k| f(Y_i; \theta(\XX_i,\check\bbeta_k),\check\phi_k) 
    \bigg\}\\ 
    &\asymp
    \sum_{i=1}^n |U_{i, rl}(\check \bPhi_k; G_0)|
    \bigg\{ \sum_{k=1}^{K_0} \sum_{h=1}^{p+1} 
    \big |U_{i,h}(\bzeta_k; \check G ) \big | \big| \check \eta_{kh} - \eta^0_{kh} \big| + 
    \sum_{k=1}^{K_0} |\varrho_k-\pi^0_k| U_i(\check \bPhi_k; \check G)
    \bigg \}\\ 
    & \le 
\bigg\{ \sum_{i=1}^n |U_{i, rl}(\check \bPhi_k; G_0)|^2 \bigg \}^{1/2} 
\bigg\{\sum_{i=1}^n \bigg [\sum_{k=1}^{K_0} \sum_{h=1}^{p+1} 
    \big |U_{i,h}(\bzeta_k; \check G ) \big | \big| \check \eta_{kh} - \eta^0_{kh} \big| + 
    \sum_{k=1}^{K_0} |\varrho_k-\pi^0_k| U_i(\check \bPhi_k; \check G) \bigg ]^2 \bigg \}^{1/2}\\
    &\lesssim
    \bigg\{ \sum_{i=1}^n |U_{i, rl}(\check \bPhi_k; G_0)|^2 \bigg \}^{1/2}
    \bigg\{\sum_{k=1}^{K_0} \sum_{h=1}^{p+1} \bigg [
    \big| \check \eta_{kh} - \eta^0_{kh} \big|^2 \sum_{i=1}^n 
    \big |U_{i,h}(\bzeta_k; \check G ) \big |^2 + 
     |\varrho_k-\pi^0_k|^2 \sum_{i=1}^{n} U^2_i(\check \bPhi_k; \check G) \bigg ] \bigg \}^{1/2}
\end{align*}
where we denote $\bzeta_k:=(\bzeta_{1k},\bzeta_{2k})$, while $U_i(\cdot) \equiv U(\VV_i;\cdot)$ are given in equation \eqref{U0}, 
and we have used the Cauchy-Schwarz inequality in the second-to-last line. 
In view of condition (R3),  there exists $h_2 \in L^3(P_0)$ such that 
$$\frac 1 n \sum_{i=1}^n |U_{i,rl}(\check \bPhi_k; G_0)|^2 \leq \frac 1 n \sum_{i=1}^n h_2^2(\VV_i) \overset{a.s.}{\longrightarrow} E\left\{h_2^2(\VV)\right\},
~ \text{so, } \sum_{i=1}^n |U_{i,rl}(\check \bPhi_k; G_0)|^2 = O_p(n),$$
by Kolmogorov's Strong Law of Large Numbers. Similarly, 
by condition (R4),
\[
\sum_{i=1}^n\left|U_{i,h}(\bzeta_k; \check G)\right|^2 = O_p(n), 
\quad \sum_{i=1}^n U_i^2(\check \bPhi_k; \check G) = O_p(n).
\]
Since $\check G$ is in $\gamma_n^{1/2}$-neighborhood of the true $G_0$, we arrive at 
\begin{equation}
\sum_{i=1}^n  \left| U_{i, rl}
(\check \bPhi_k; \check G) - U_{i, rl}(\check \bPhi_k; G_0) \right| 
= 
O_p(n) \bigg\{\sum_{k=1}^{K_0} \sum_{h=1}^{p+1} \left [
    \big| \check \eta_{kh} - \eta_{0kh} \big|^2  + 
     |\varrho_k-\pi^0_k|^2  \right ] \bigg \}^{1/2} = O_p(n \gamma^{1/2}_n)
\end{equation}
Thus, together with the order assessments in equations \eqref{secondTerm2} and \eqref{secondTerm22}, 
\begin{equation}
\label{secondTerm4}
\sum_{i=1}^n U_{i, rl}(\check \bPhi_k; \check G) = O_p(n \gamma^{1/2}_n).
\end{equation}
Therefore, it follows that 
\begin{equation}
\label{secondTerm5}
  (II) = \frac{1}{2}\sum_{k=1}^{K_0} \varrho_k \sum_{r=1}^{p+1}\sum_{l=1}^{p+1} 
 m_{k,rl}
\sum_{i=1}^{n} U_{i,rl}(\check{\bPhi}_{k}; \check{G})
= \frac{O_p(n \gamma^{1/2}_n)}{2} 
\sum_{k=1}^{K_0} \varrho_k \sum_{r=1}^{p+1}\sum_{l=1}^{p+1} |m_{k,rl}|.
\end{equation}

In regard to (III) in equation \eqref{three-terms1}, under condition (R3), for all 
$r, l, h=1, \ldots, p+1$, 
\[
\sum_{i=1}^{n} U_{i,rlh}(\bxi_k; \check{G}) = O_p(n). 
\]
Now, since all the atoms of the mixing measures $H_k$ in equation \eqref{Hk_defn}
are in a $\gamma_n^{\frac 1 2}$-neighborhood of the true atoms of $G_0$,
\begin{eqnarray}
&& \frac{1}{6}  
\sum_{r=1}^{p+1}\sum_{l=1}^{p+1}\sum_{h=1}^{p+1} \int (\eta_r - \check{\eta}_{kr})(\eta_l - \check{\eta}_{kl})(\eta_h - \check{\eta}_{kh}) \sum_{i=1}^{n} U_{i,rlh}(\bxi_k; \check{G})~ dH_k(\bPhi) \nonumber\\ 
& = & \frac{1}{6} O_p(n) 
\sum_{r=1}^{p+1}\sum_{l=1}^{p+1}\sum_{h=1}^{p+1} \int |(\eta_r - \check{\eta}_{kr})(\eta_l - \check{\eta}_{kl})(\eta_h - \check{\eta}_{kh})| dH_k(\bPhi)\nonumber\\ 
& = & \frac{1}{6} O_P(n \gamma^{1/2}_n) 
\sum_{r=1}^{p+1}\sum_{l=1}^{p+1} \int |\eta_r - \check{\eta}_{kr}||\eta_l - \check{\eta}_{kl}| dH_k(\bPhi).
\label{ThirdTerm1}
\end{eqnarray}

Hence, by equations 
\eqref{linear-term1}, \eqref{secondTerm5} and \eqref{ThirdTerm1}, 
for some constant $C_0>0$, we arrive at 
\begin{eqnarray}
\label{linear-term2}
& & \sum_{i=1}^n D_i(G, \check G)  =  (I) + (II) + (III) = 
 \sum_{k=1}^{K_0} \varrho_k \sum_{r=p_k+1}^{p} m_{k,r} \sum_{i=1}^{n} 
 U_{i,r}( \check{\bPhi}_k; \check G) \nonumber \\ 
& + & O_p(n\gamma_n^{\frac 1 2} )
 \sum_{k=1}^{K_0} \varrho_k \left\{\frac 1 2 \sum_{r=1}^{p+1} \sum_{l=1}^{p+1} |m_{k, rl}| 
 + 
 \frac 1 6 \sum_{r=1}^{p+1}\sum_{l=1}^{p+1} 
 \int |\eta_r - \check{\eta}_{kr}||\eta_l - \check{\eta}_{kl}| dH_k(\bPhi)
 \right\} \nonumber \\ \nonumber\\
 &  \le &
 \sum_{k=1}^{K_0} \varrho_k \sum_{r=p_k+1}^{p} m_{k,r} \sum_{i=1}^{n} 
 U_{i,r}( \check{\bPhi}_k; \check G) + 
C_0 n\gamma_n^{\frac 1 2} \sum_{k=1}^{K_0} 
 \sum_{r=1}^{p+1}\sum_{l=1}^{p+1} 
 \int |\eta_r - \check{\eta}_{kr}||\eta_l - \check{\eta}_{kl}| ~dH_k(\bPhi). \quad \quad~~~
 \label{linear-term}
\end{eqnarray}
 

\noindent
\textbf{Step 2B: Bounding the quadratic term $\sum_{i=1}^{n} D^2_i(G, \check G)$}:

We have that, 
\begin{eqnarray*}
 &&   \sum_{i=1}^{n} D^2_i(G, \check G) 
 = 
 \sum_{i=1}^{n}
\bigg \{
\sum_{k=1}^{K_0} \varrho_k \sum_{r=1}^{p+1} m_{k,r} 
 U_{i,r}( \check{\bPhi}_k; \check G) + 
 \frac{1}{2}\sum_{k=1}^{K_0} \varrho_k \sum_{r=1}^{p+1}\sum_{l=1}^{p+1} 
 m_{k,rl}
 U_{i,rl}(\check{\bPhi}_{k}; \check{G}) \nonumber\\
& + & 
\frac{1}{6} \sum_{k=1}^{K_0} \varrho_k 
\sum_{r=1}^{p+1}\sum_{l=1}^{p+1}\sum_{h=1}^{p+1} \int (\eta_r - \check{\eta}_{kr})(\eta_l - \check{\eta}_{kl})(\eta_h - \check{\eta}_{kh})  U_{i,rlh}(\bxi_k; \check{G})~ dH_k(\bPhi) \bigg\}^2 \nonumber \\
& = & 
\sum_{i=1}^{n}
\bigg \{
\sum_{k=1}^{K_0} \varrho_k \sum_{r=1}^{p+1} m_{k,r} 
 U_{i,r}( \check{\bPhi}_k; \check G) + 
 \frac{1}{2}\sum_{k=1}^{K_0} \varrho_k \sum_{r=1}^{p+1}\sum_{l=1}^{p+1}  m_{k,rl}
 U_{i,rl}(\check{\bPhi}_{k}; \check{G}) \bigg \}^2 \nonumber \\
& + & 
\sum_{i=1}^{n}
\bigg \{
\frac{1}{6} \sum_{k=1}^{K_0} \varrho_k 
\sum_{r=1}^{p+1}\sum_{l=1}^{p+1}\sum_{h=1}^{p+1} \int (\eta_r - \check{\eta}_{kr})(\eta_l - \check{\eta}_{kl})(\eta_h - \check{\eta}_{kh})  U_{i,rlh}(\bxi_k; \check{G})~ dH_k(\bPhi) \bigg\}^2 \nonumber\\
& + & 
2 \sum_{i=1}^{n} \bigg \{
\sum_{k=1}^{K_0} \varrho_k \sum_{r=1}^{p+1} m_{k,r} 
 U_{i,r}( \check{\bPhi}_k; \check G) + 
 \frac{1}{2}\sum_{k=1}^{K_0} \varrho_k \sum_{r=1}^{p+1}\sum_{l=1}^{p+1} 
 m_{k,rl}
 U_{i,rl}(\check{\bPhi}_{k}; \check{G}) \bigg \} \times \nonumber \\
& & 
\bigg \{
\frac{1}{6} \sum_{k=1}^{K_0} \varrho_k 
\sum_{r=1}^{p+1}\sum_{l=1}^{p+1}\sum_{h=1}^{p+1} \int (\eta_r - \check{\eta}_{kr})(\eta_l - \check{\eta}_{kl})(\eta_h - \check{\eta}_{kh})  U_{i,rlh}(\bxi_k; \check{G})~ dH_k(\bPhi) \bigg\} \nonumber \\ \\
& = & (A) + (B) + (C).
\end{eqnarray*}
In what follows we conduct an order assessment of $(A)$-$(C)$ for large $n$. 
First, denote the following vectors, 
\begin{eqnarray*}
\MM_{k1} & = & (m_{k,1}, \ldots, m_{k,p+1})^\top~~,~~ 
\MM_{k2}  =  (m_{k,11}, \ldots, m_{k, p+1,p+1})^\top \\ 
\UU_{i,1}(\check \bPhi_k; \check G) & = & \bigg (U_{i,1}(\check \bPhi_k; \check G), \ldots, U_{i,p+1}(\check \bPhi_k; \check G)\bigg )^{\top}   \\
\UU_{i,2}(\check \bPhi_k; \check G) & = & \bigg (
U_{i,11}(\check{\bPhi}_{k}; \check{G}), \ldots, U_{i,p+1,p+1}(\check{\bPhi}_{k}; \check{G})
\bigg )^\top. 
\end{eqnarray*}
Also, for $l=1, 2$, let $\MM_l = (\MM_{1l}, \dots, \MM_{K_0l})^\top$, and 
$\MM = (\MM_1, \MM_2)^\top$, 
$$
{\cal U}_{il}(\check \bPhi; \check G) 
= 
\Big(\UU_{i,l}(\check \bPhi_1; \check G), \dots, \UU_{i,l}(\check \bPhi_{K_0}; \check G)\Big)^\top
~,~l=1, 2
$$
and
\(
{\cal U}_i(\check \bPhi; \check G) = \Big({\cal U}_{i1}(\check \bPhi; \check G), ~{\cal U}_{i2}(\check \bPhi; \check G)\Big)^\top
\),
where $\check \bPhi = (\bPhi_1, \dots, \bPhi_{K_0})$.

Since the $\varrho_k$ are bounded away from zero in probability, we have,
\begin{eqnarray*}
(A) & = & 
\sum_{i=1}^{n}
\bigg \{
\sum_{k=1}^{K_0} \varrho_k \big [\MM^\top_{k1} \UU_{i,1}(\check \bPhi_k; \check G) 
+ 
\frac{1}{2} \MM^\top_{k2} \UU_{i,2}(\check \bPhi_k; \check G)
\big ] \bigg \}^2 \nonumber \\ \\
& \asymp & 
\sum_{i=1}^{n}
\bigg \{
 \MM^\top_{1} {\cal U}_{i,1}(\check \bPhi; \check G) 
+ 
\MM^\top_{2} {\cal U}_{i,2}(\check \bPhi; \check G) \bigg \}^2 \\
& = & 
\sum_{i=1}^{n} 
\bigg \{
\MM^\top {\cal U}_i(\check \bPhi; \check G) ~{\cal U}^\top_i(\check \bPhi; \check G)
\MM \bigg \} = 
\MM^\top \bigg \{
\sum_{i=1}^n {\cal U}_i(\check \bPhi; \check G) ~{\cal U}^\top_i(\check \bPhi; \check G) 
\bigg \}
\MM.
\end{eqnarray*}
in probability, for large $n$. By \cite{serfling2002approximation} 
(Lemma A, p. 253), as $n \to \infty$, 
\[
\frac{1}{n}
\sum_{i=1}^n {\cal U}_i(\check \bPhi; \check G)~ {\cal U}^\top_i(\check \bPhi; \check G) 
\overset{p}{\longrightarrow} \bSigma = 
\mathbb{E} \bigg \{ {\cal U}_1(\bPhi^0; G_0)~ {\cal U}^\top_1(\bPhi^0; G_0) \bigg \}.
\]
It follows that for large $n$, the following holds in probability
\[
\varrho_{\text{min}}(\bSigma) \norm{\MM}^2 \lesssim (A)/n \lesssim 
\varrho_{\text{max}}(\bSigma) \norm{\MM}^2. 
\]
 where $\norm{\MM}$ denotes the Frobenius norm of $\MM$. On the other hand, 
 by the (SI) condition, ${\cal U}_1(\bPhi^0; G_0)$ is non-singular and thus 
 $\bSigma$ is positive definite with $\varrho_{\text{min}}(\bSigma) >0$.
 Hence, for large $n$, 
 \begin{equation}
 \label{A-term}
     (A) \asymp n \norm{\MM}^2
 \end{equation}
 in probability.     
Using a similar argument, and since $\big\lVert\bPhi - \check \bPhi_k\big\rVert = o_p(1)$, 
for all $\bPhi \in \bTheta$ in a $\gamma_n^{\frac 1 2}$-neighborhood of an atom of $G_0$, we have,
\begin{equation*}
(B) = 
o_p(n)\norm{\MM_2}^2 =
o_p(n) \norm{\MM}^2.
\end{equation*}
By the Cauchy-Schwarz inequality, we also have,
\begin{equation*}
|(C)| \leq \sqrt{(I)(II)} = o_p(n) \norm{\MM}^2.
\end{equation*}

By combining the above inequalities, for large $n$, we can choose a constant 
$C^*>0$ such that 
\begin{equation}
    \sum_{i=1}^{n} D^2_i(G, \check G) \ge C^* n 
    \sum_{k=1}^{K_0} \left\{\sum_{r=1}^{p+1} m_{k,r}^2 
    + 
    \sum_{r=1}^{p+1} \sum_{l=1}^{p+1} m_{k, rl}^2 \right\}.
    \label{quad-lower-bound}
\end{equation}
with probability tending to one. 


\noindent
\textbf{Step 2C: Bounding the cubic term $\sum_{i=1}^{n} D^3_i(G, \check G)$}:

By a Taylor expansion and also 
H\"older's inequality, there exist vectors $\bxi_{ik}$   
on the segment joining $\bPhi$ and $\check \bPhi_k$ such that
\begin{eqnarray}
  \sum_{i=1}^{n} D^3_i(G, \check G) & = & 
 \sum_{i=1}^{n} \bigg \{
\sum_{k=1}^{K_0} \varrho_k \sum_{r=1}^{p+1} m_{kr} U_{i,r}( \check{\bPhi}_k; \check G) 
\nonumber\\ 
& &
+ \frac{1}{2}\sum_{r=1}^{p+1}\sum_{l=1}^{p+1} 
\int (\eta_r - \check \eta_{kr})(\eta_l - \check{\eta}_{kl})~ dH_k(\bPhi)
 U_{i,rl}(\check{\bxi}_{ik}; \check{G}) 
 \bigg \}^3 \nonumber\\ 
 & = & O_p(1) 
 \sum_{k=1}^{K_0} \bigg \{
\sum_{r=1}^{p+1} |m_{kr}|^3 \sum_{i=1}^n |U_{i,r}( \check{\bPhi}_k; \check G)|^3 
\nonumber\\ 
& &
+ 
\sum_{r=1}^{p+1}\sum_{l=1}^{p+1} 
\int |(\eta_r - \check \eta_{kr})|^3|(\eta_l - \check{\eta}_{kl})|^3~ dH_k(\bPhi)
 \sum_{i=1}^n |U_{i,rl}(\check{\bxi}_{ik}; \check{G})|^3
\bigg \} \nonumber \\ \nonumber\\ 
& = & 
O_p(n)
\sum_{k=1}^{K_0} \bigg \{
\sum_{r=1}^{p+1} |m_{kr}|^3 + 
\sum_{r=1}^{p+1}\sum_{l=1}^{p+1} 
\int |(\eta_r - \check \eta_{kr})|^3|(\eta_l - \check{\eta}_{kl})|^3~ dH_k(\bPhi)
\bigg \} \nonumber\\  \nonumber\\ 
& = & o_p(n) \norm{\MM}^2. 
\label{cub-upper-bound}
 \end{eqnarray}

This ends the assessment of the three terms on the RHS of 
the log-likelihood difference in
equation \eqref{like-ineq1}, through the three {\bf Steps 2A, 2B, 2C}. 

Now, we continue the proof. Thus, using equations \eqref{quad-lower-bound} and \eqref{cub-upper-bound}, we can rewrite equation \eqref{like-ineq1}, for large $n$, as  
\begin{equation}
    \ell_n(G) - \ell_n(\check{G}) \le \sum_{i=1}^{n} D_i(G, \check G)
     - \bigg \{ \sum_{i=1}^{n} D^2_i(G, \check G) \bigg \} \bigg (1+ o_p(1) \bigg). 
     \label{like-ineq2}
\end{equation}

Now, by equations \eqref{linear-term} and \eqref{quad-lower-bound}, for large $n$, in probability, we have
\begin{align}
 &\hspace{15pt}\sum_{i=1}^{n} D_i(G, \check G)
     - \sum_{i=1}^{n} D^2_i(G, \check G) 
     \le  
 \sum_{k=1}^{K_0} \varrho_k \sum_{r=p_k+1}^{p} m_{k,r} \sum_{i=1}^{n} 
 U_{i,r}( \check{\bPhi}_k; \check G) \nonumber \\ 
  & 
  +  C_0 n\gamma_n^{\frac 1 2} \sum_{k=1}^{K_0} 
 \sum_{r=1}^{p+1}\sum_{l=1}^{p+1} 
 \int |\eta_r - \check{\eta}_{kr}||\eta_l - \check{\eta}_{kl}| ~dH_k(\bPhi)
 -  C^* n 
    \sum_{k=1}^{K_0} \left\{\sum_{r=1}^{p+1} m_{k,r}^2 
    + 
    \sum_{r=1}^{p+1} \sum_{l=1}^{p+1} m_{k, rl}^2 \right\}. 
\hspace{.1cm}
 \label{diff-lin-quad}
\end{align} 

Consider the last difference in equation~\eqref{diff-lin-quad},
\[
C_0 n\gamma_n^{\frac 1 2} \sum_{k=1}^{K_0} 
 \sum_{r=1}^{p+1}\sum_{l=1}^{p+1} 
 \int |\eta_r - \check{\eta}_{kr}||\eta_l - \check{\eta}_{kl}| ~dH_k(\bPhi)
 -  C^* n 
    \sum_{k=1}^{K_0} \left\{\sum_{r=1}^{p+1} m_{k,r}^2 
    + 
    \sum_{r=1}^{p+1} \sum_{l=1}^{p+1} m_{k, rl}^2 \right\}
\]
\[
= A_n - B_n,
\]
where both are non-negative. 
Recall the definition of the moments $m_{k,r}$ and 
$m_{k,rl}$ in equation~\eqref{moments}. 
Since 
$|\eta_l - \check{\eta}_{kl}|, |\eta_r - \check{\eta}_{kr}|=
O_p(\gamma^{1/2}_n)$, 
for all $l,r=1, \ldots, p+1$, and $k=1, \ldots, K_0$, we have 
that $A_n = o_p(B_n)$. Thus, for large $n$, 
\(
A_n - B_n = -B_n(1-o_p(1)) \le 0, 
\)
implying
\begin{align}
 &\hspace{15pt}\sum_{i=1}^{n} D_i(G, \check G)
     - \sum_{i=1}^{n} D^2_i(G, \check G)   \nonumber 
     \le   
 \sum_{k=1}^{K_0} \varrho_k \sum_{r=p_k+1}^{p} m_{k,r} \sum_{i=1}^{n} 
 U_{i,r}( \check{\bPhi}_k; \check G) \nonumber \\ 
   &\hspace{30pt}+  C_0 n\gamma_n^{\frac 1 2} \sum_{k=1}^{K_0} 
 \sum_{r=1}^{p+1}\sum_{l=1}^{p+1} 
 \int |\eta_r - \check{\eta}_{kr}||\eta_l - \check{\eta}_{kl}| ~dH_k(\bPhi)
 -  C^* n 
    \sum_{k=1}^{K_0} \left\{\sum_{r=1}^{p+1} m_{k,r}^2 
    + 
    \sum_{r=1}^{p+1} \sum_{l=1}^{p+1} m_{k, rl}^2 \right\} \nonumber\\
     &= 
    \sum_{k=1}^{K_0} \varrho_k \sum_{r=p_k+1}^{p} m_{k,r} \sum_{i=1}^{n} 
 U_{i,r}( \check{\bPhi}_k; \check G) +A_n - B_n 
 \le 
 \sum_{k=1}^{K_0} \varrho_k \sum_{r=p_k+1}^{p} m_{k,r} \sum_{i=1}^{n} 
 U_{i,r}( \check{\bPhi}_k; \check G)
 \hspace{1cm}
 \label{diff-lin-quad2}
\end{align}
in probability. 
Thus, by combining equations \eqref{dif-plike}, 
\eqref{pen-lower-bound} and \eqref{diff-lin-quad2}, for large $n$, we arrive at
\begin{eqnarray}
L_n(G) - L_n(\check{G})  & \le &  
\left\{\ell_n(G)-\ell_n(\check{G})\right\}   
    - \left\{\mathcal{P}(G) - \mathcal{P}(\check{G})\right\} 
    \nonumber \\
  &  \le  &
 \sum_{k=1}^{K_0} \varrho_k \sum_{r>p_k}^{p} |m_{k,r}| 
 \bigg | \sum_{i=1}^{n} 
 U_{i,r}( \check{\bPhi}_k; \check G) \bigg | 
 -n \sum_{k=1}^{K_0} \sum_{s \in \mathcal{I}_k} 
\sum_{j>p_k}^{p} r_{\lambda_n}(\beta_{sj}).
 \label{pen-like-diff2}
\end{eqnarray}

By using the definition of the moments 
$m_{k,r}$ from \eqref{moments} to equation~\eqref{pen-like-diff2}, for large $n$, 
\begin{eqnarray*}
L_n(G) - L_n(\check{G}(G))  & \le &  
\sum_{k=1}^{K_0}  \sum_{s \in {\cal I}_k} \sum_{j>p_k}^{p} 
|\beta_{sj}| ~ \bigg | \varrho_k \sum_{i=1}^{n} 
 U_{i,j}( \check{\bPhi}_k; \check G) \bigg |
 -n \sum_{k=1}^{K_0} \sum_{s \in \mathcal{I}_k} 
\sum_{j>p_k}^{p} r_{\lambda_n}(\beta_{sj}).
\hspace{1cm}
 \label{pen-like-diff3}
\end{eqnarray*}

By a similar derivation led to equation~\eqref{secondTerm4}, we have 
\begin{equation}
\label{score}
  \sum_{i=1}^{n} 
 U_{i,j}( \check{\bPhi}_k(G); \check G(G)) = O_p(n \gamma^{1/2}_n).  
\end{equation}
Thus, for any mixing measure $G \in \mathcal{G}_{K, \mathcal{S}^{c}}(b_\epsilon,\gamma_n)$, 
there exists some $\tilde M(G)>0$ such that, for sufficiently large $n$, 
\begin{eqnarray}
L_n(G) - L_n(\check{G}(G))  & \le &  
\sum_{k=1}^{K_0}  \sum_{s \in {\cal I}_k} \sum_{j>p_k}^{p} 
\tilde{M}(G) n \gamma^{1/2}_n  |\beta_{sj}| 
 -n \sum_{k=1}^{K_0} \sum_{s \in \mathcal{I}_k} 
\sum_{j>p_k}^{p} r_{\lambda_n}(\beta_{sj})  \nonumber
\\
&  = & 
n \sum_{k=1}^{K_0}  \sum_{s \in {\cal I}_k} \sum_{j>p_k}^{p} 
\bigg \{\tilde{M}(G)  \gamma^{1/2}_n |\beta_{sj}| - r_{\lambda_n}(\beta_{sj}) \bigg \}. \nonumber
\end{eqnarray}
By conditions (P1), $r_{\lambda_n}(0)= 0$, and (P2), with ${\cal N}_n =\big \{\beta: 0 < \beta \leq \gamma_n^{\frac 1 2} \log n \big \}$, we have 
\begin{align*}
    \lim_{n\to\infty}\inf_{\beta\in\mathcal{N}_n}\frac{r_{\lambda_n}(\beta)}{\gamma_n^{1/2}\beta}=\infty.
\end{align*}
Thus, for sufficiently large $n$, with a high probability, 
\(
L_n(G) - L_n(\check{G}(G)) < 0
\), for all $G \in \mathcal{G}_{K, \mathcal{S}^{c}}(b_\epsilon,\gamma_n)$, which implies that 
$\sup_{G \in \mathcal{G}_{K, \mathcal{S}^{c}}(b_\epsilon,\gamma_n)} 
\big \{ L_n(G) - L_n(\check{G}_{K_0,S}(G)) \big\} \le 0$.\\ This completes the proof.
$\hfill \blacksquare$

\section{Additional Results}
\label{sec:additional_results}
In this appendix, we exhibit two additional results, along with their proofs, 
that are used in the proof of our main results. 
First, we establish inequalities relating the Wasserstein distance between two mixing measures to the 
Hellinger distance between their corresponding mixture densities in Appendix~\ref{appendix:Hellinger_lower_bounds}. Next, the consistency of the density estimator under the Hellinger distance will be presented in Appendix~\ref{appendix:density_consistency}.  

\subsection{Relation between the Hellinger and Wasserstein Distances}
\label{appendix:Hellinger_lower_bounds}
\begin{customprop}{C.1}
\label{thm:ho_long_results}
Suppose that ${\cal F}$ satisfies conditions (SI) and (R2).  
Then, there exist constants $\delta_0, c_0 > 0$ depending only on $G_0, \Theta$ and ${\cal F}$ such that
the following 
holds.
\begin{enumerate}
\item[(i)] For all mixing measures $G$ with exactly $K_0$ atoms 
satisfying $W_1(G, G_0) < \delta_0$, we have $h(p_G, p_0) \geq c_0 W_1(G, G_0)$.
\item[(ii)] For all mixing measures $G \in {\cal G}_{K}$ 
satisfying $W_2(G, G_0) < \delta_0$, 
we have $h(p_G, p_0) \geq c_0 W_2^2(G, G_0)$.
\end{enumerate}
\end{customprop}
\textbf{Proof of Proposition~\ref{thm:ho_long_results}.} 
\textbf{Part (i).} Note that the Hellinger distance is bounded below by the Total Variation distance. Thus, it suffices to demonstrate that 
\begin{align}
    \label{eq:TV_lower_bound_1}
    \lim_{\epsilon\to0}\inf_{G\in\mathcal{G}_{K_0}:W_1(G,G_0)\leq\epsilon}V(p_G,p_0)/W_1(G,G_0)>0.
\end{align}
Assume by contrary that the above inequality does not hold true. Then, there exists a sequence of mixing measures $G_n\in\mathcal{G}_{K_0}$ such that 
\begin{align*}
    V(p_{G_n},p_0)/W_1(G_n,G_0)&\to0,\\
    W_1(G_n,G_0)&\to0,
\end{align*}
as $n\to\infty$. Since $G_n$ has exactly $K_0$ atoms, we may assume without loss of generality that $\mathcal{I}_k=\{k\}$ for any $k\in[K_0]$. Thus, we can represent $G_n$ as $G_n=\sum_{k=1}^{K_0}\pi^n_k\delta_{(\bbeta^n_k,\phi^n_k)}$, where $(\bbeta^n_k,\phi^n_k)\to(\bbeta^0_k,\phi^0_k)$ and $\pi^n_k\to\pi^0_k$ for any $k\in[K_0]$.

According to Lemma~\ref{lem:ho}, we have that
\begin{align}
    \label{eq:D1_formulation}
    W_1(G_n, G_0) \asymp \sum_{k=1}^{K_0}\pi^n_k \Big[\norm{\Delta\bbeta^n_k}+|\Delta\phi^n_k|\Big] + \sum_{k=1}^{K_0}| \pi^0_k -  \pi^n_k|:=D_1(G_n,G_0),
\end{align}
where $\Delta\bbeta^n_k=\bbeta^n_k-\bbeta^0_k$ and $\Delta\phi^n_k=\phi^n_k-\phi^0_k$ for any $k\in[K_0]$. As $n\to\infty$, since $V(p_{G_n},p_0)/W_1(G_n,G_0)\to0$, we also get that $V(p_{G_n},p_0)/D_1(G_n,G_0)\to0$. Now, we divide the following arguments into three stages.

\noindent
\textbf{Stage 1.} In this stage, we decompose the difference $p_{G_n}(\xx,y)-p_{G_0}(\xx,y)$ into a combination of linearly independent elements. By invoking a first-order Taylor expansion, we have
\begin{align*}
    &f(y;\theta(\xx,\bbeta^n_k),\phi^n_k)-f(y;\theta(\xx,\bbeta^0_k),\phi^0_k)\\
    &=\sum_{|\balpha|+\zeta=1}(\Delta\bbeta^n_k)^{\balpha}(\Delta\phi^n_k)^{\zeta}\cdot\frac{\partial^{|\balpha|+\zeta}f(y;\theta(\xx,\bbeta^0_k),\phi^0_k)}{\partial\boldsymbol{\bbeta}^{\balpha}\partial\phi^{\zeta}}+R_{k}(\xx,y),
\end{align*}
where $\balpha=(\alpha_1,\ldots,\alpha_p)\in\mathbb{N}^p$, $\zeta\in\mathbb{N}$ and $|\balpha|=\alpha_1+\ldots+\alpha_p$, while $R_{1k}(\xx,y)$ is the Taylor remainder such that
\begin{align*}
    R_{1k}(\xx,y)f_{\XX}(\xx)=\mathcal{O}(\norm{\Delta\bbeta^n_k}^{1+\gamma}+|\Delta\phi^n_k|^{1+\gamma}),
\end{align*}
for some universal constant $\gamma>0$ for all $k\in[K_0]$. Therefore, we get that
\begin{align}
    &p_{G_n}(\xx,y)-p_{G_0}(\xx,y)\nonumber\\
    &=\sum_{k=1}^{K_0}\pi^n_k\Big[f(y;\theta(\xx,\bbeta^n_k),\phi^n_k)-f(y;\theta(\xx,\bbeta^0_k),\phi^0_k)\Big]f_{\XX}(\xx)\nonumber\\
    &\qquad +\sum_{k=1}^{K_0}(\pi^n_k-\pi^0_k)f(y;\theta(\xx,\bbeta^0_k),\phi^0_k)f_{\XX}(\xx)\nonumber\\
    &=\sum_{k=1}^{K_0}\sum_{|\balpha|+\zeta=1}\pi^n_k(\Delta\bbeta^n_k)^{\balpha}(\Delta\phi^n_k)^{\zeta}\cdot\frac{\partial^{|\balpha|+\zeta}f(y;\theta(\xx,\bbeta^0_k),\phi^0_k)}{\partial\boldsymbol{\bbeta}^{\balpha}\partial\phi^{\zeta}}f_{\XX}(\xx)\nonumber\\
    &\qquad +\sum_{k=1}^{K_0}(\pi^n_k-\pi^0_k)f(y;\theta(\xx,\bbeta^0_k),\phi^0_k)f_{\XX}(\xx)+R_1(\xx,y)\nonumber\\
    \label{eq:density_decomposition}
    &=A_{1n}+B_{1n} + R_1(\xx,y),
\end{align}
where $R_1(\xx,y)=\sum_{k=1}^{K_0}R_{1k}(\xx,y)f_{\XX}(\xx)=O\Big(\sum_{k=1}^{K_0}\pi^n_k\Big[\norm{\Delta\bbeta^n_k}^{1+\gamma}+|\Delta\phi^n_k|^{1+\gamma}\Big]\Big)$ under the condition (R2). Moreover, from the formulation of $D_1(G_n,G_0)$ in equation~\eqref{eq:D1_formulation}, we have 
\begin{align*}
    R_1(\xx,y)/D_1(G_n,G_0)\lesssim\sum_{k=1}^{K_0}\Big[\norm{\Delta\bbeta^n_k}^{\gamma}+|\Delta\phi^n_k|^{\gamma}\Big]\to0,~~\text{as}~~n\to\infty.
\end{align*}

\noindent
\textbf{Stage 2.} In this stage, we will prove by contradiction that at least one among the coefficients in the representation of $A_{1n}/D_1(G_n,G_0)$ and $B_{1n}/D_1(G_n,G_0)$ does not vanish as $n$ goes to infinity. Assume that all of them converge to zero, then by taking the summation of the absolute values of the coefficients of $B_{1n}/D_1(G_n,G_0)$, we have
\begin{align}
    \label{eq:weight_limit}
    \sum_{k=1}^{K_0}|\pi^n_k-\pi^0_k|/D_1(G_n,G_0)\to0.
\end{align}
Next, let us denote $C_{n,k}({\balpha,\zeta})=\pi^n_k(\Delta\bbeta^n_k)^{\balpha}(\Delta\phi^n_k)^{\zeta}$ for any $\balpha\in\mathbb{N}^p$, $n,\zeta\in\mathbb{N}$ and $k\in[K_0]$. Then, we get
\begin{align*}
    A_{1n}/D_1(G_n,G_0)=\sum_{k=1}^{K_0}\sum_{|\balpha|+\zeta=1}\frac{C_{n,k}(\balpha,\zeta)}{D_1(G_n,G_0)}\cdot\frac{\partial^{|\balpha|+\zeta}f(y;\theta(\xx,\bbeta^0_k),\phi^0_k)}{\partial\boldsymbol{\bbeta}^{\balpha}\partial\phi^{\zeta}}f_{\XX}(\xx).
\end{align*}
From the hypothesis, we deduce that $C_{n,k}(\balpha,\zeta)/D_1(G_n,G_0)\to0$ as $n\to\infty$ for any $k\in[K_0]$, $\balpha\in\mathbb{N}^p$ and $\zeta\in\mathbb{N}$ such that $|\balpha|+\zeta=1$. 
Then, the following limits hold:
\begin{align}
    \sum_{k=1}^{K_0}\pi^n_k\|\Delta\bbeta^n_k\|_1/D_1(G_n,G_0)=\sum_{k=1}^{K_0}\sum_{|\balpha|=1}\left|\frac{C_{n,k}(\balpha,0)}{D_1(G_n,G_0)}\right|&\to0,\nonumber\\
    \label{eq:scale_limit}
    \sum_{k=1}^{K_0}\pi^n_k|\Delta\phi^n_k|/D_1(G_n,G_0)=\sum_{k=1}^{K_0}\left|\frac{C_{n,k}(\bzero_p,1)}{D_1(G_n,G_0)}\right|&\to0.
\end{align}
Due to the topological equivalence between norm-1 and norm-2, we also get that
\begin{align}
    \label{eq:location_limit}
    \sum_{k=1}^{K_0}\pi^n_k\|\Delta\bbeta^n_k\|/D_1(G_n,G_0)=\sum_{k=1}^{K_0}\sum_{|\balpha|=1}\left|\frac{C_{n,k}(\balpha,0)}{D_1(G_n,G_0)}\right|&\to0
\end{align}
Taking the summation of limits in equations~\eqref{eq:weight_limit}, \eqref{eq:scale_limit} and \eqref{eq:location_limit}, we obtain that $$D_1(G_n,G_0)/D_1(G_n,G_0)\to0$$ as $n\to\infty$, which is a contradiction. Therefore, not all the coefficients in the representation of $A_{1n}/D_1(G_n,G_0)$ and $B_{1n}/D_1(G_n,G_0)$ approach zero as $n$ goes to infinity.

\noindent
\textbf{Stage 3.} In this stage, we use the Fatou's lemma to point out a contradiction to the results in Stage 2. In particular, let us denote by $m_n$ the maximum of the absolute values of the coefficients in the representation of $A_{1n}/D_1(G_n,G_0)$ and $B_{1n}/D_1(G_n,G_0)$. Recall from Stage 2 that not all the coefficients of $A_{1n}/D_1(G_n,G_0)$ and $B_{1n}/D_1(G_n,G_0)$ vanish, we deduce that $1/m_n\not\to\infty$ as $n\to\infty$. Additionally, we also denote 
\begin{align*}
    (\pi^n_k-\pi^0_k)/m_nD_1(G_n,G_0)&\to\rho_k,\\
    \pi^n_k(\Delta\bbeta^n_k)/m_nD_1(G_n,G_0)&\to\boeta_k,\\
    \pi^n_k(\Delta\phi^n_k)/m_nD_1(G_n,G_0)&\to\tau_k,
\end{align*}
as $n\to\infty$ for all $k\in[K_0]$. Above, at least one among $\rho_k,\boeta_k$ and $\tau_k$ is different from $0$ or $\bzero_p$. By applying the Fatou's lemma, it follows that
\begin{align*}
    0=\lim_{n\to\infty}\frac{V(p_{G_n},p_{G_0})}{m_nD_1(G_n,G_0)}\geq\int\liminf_{n\to\infty}\frac{|p_{G_n}(\xx,y)-p_{G_0}(\xx,y)|}{2m_nD_1(G_n,G_0)}d\nu(\xx,y)\geq 0.
\end{align*}
The above result implies that $[p_{G_n}(\xx,y)-p_{G_0}(\xx,y)]/m_nD_1(G_n,G_0)\to 0$ as $n\to\infty$ for $\nu$-almost every $(\xx,y)$. On the other hand, we have
\begin{align*}
    &\frac{p_{G_n}(\xx,y)-p_{G_0}(\xx,y)}{m_nD_1(G_n,G_0)}\\
    &\to\sum_{k=1}^{K_0}\Big[\rho_kf(y;\theta(\xx,\bbeta^0_k),\phi^0_k)+\tau_k\frac{\partial f(y;\theta(\xx,\bbeta^0_k),\phi^0_k)}{\partial\phi}+\boeta_k^{\top}\frac{\partial f(y;\theta(\xx,\bbeta^0_k),\phi^0_k)}{\partial\bbeta}\Big]f_{\XX}(\xx),
\end{align*}
which implies that 
\begin{align*}
    \sum_{k=1}^{K_0}\Big[\rho_kf(y;\theta(\xx,\bbeta^0_k),\phi^0_k)+\tau_k\frac{\partial f(y;\theta(\xx,\bbeta^0_k),\phi^0_k)}{\partial\phi}+\boeta_k^{\top}\frac{\partial f(y;\theta(\xx,\bbeta^0_k),\phi^0_k)}{\partial\bbeta}\Big]=0,
\end{align*}
for $\nu$-almost every $(\xx,y)$. Since the family $\mathcal{F}$ is strongly identifiable (see Definition~\ref{def:SI}), we deduce that $\rho_k=\tau_k=0$ and $\boeta_k=\bzero_p$, which contradicts the fact that at least one among $\rho_k,\boeta_k$ and $\tau_k$ is nonzero (vector). Hence, we reach the claim in equation~\eqref{eq:TV_lower_bound_1}.

\noindent
\textbf{Part (ii).} Similar to the proof of Part (i), we need to show that
\begin{align}
    \label{eq:TV_lower_bound_2}
    \lim_{\epsilon\to0}\inf_{G\in\mathcal{G}_{K}:W_2(G,G_0)\leq\epsilon}V(p_G,p_0)/W_2^2(G,G_0)>0.
\end{align}
Assume that the above claim is not true. Then, we can find a sequence of mixing measure $G_n\in\mathcal{G}_{K}$ such that 
\begin{align*}
    V(p_{G_n},p_0)/W_2^2(G_n,G_0)&\to0,\\
    W_2(G_n,G_0)&\to0,
\end{align*}
as $n\to\infty$. 
Then, we can represent $G_n$ as $G_n=\sum_{k=1}^{K_0}\sum_{j\in\mathcal{I}_k}\pi^n_{kj}\delta_{(\bbeta^n_{kj},\phi^n_{kj})}$, where $(\bbeta^n_{kj},\phi^n_{kj})\to(\bbeta^0_k,\phi^0_k)$ and $\sum_{j\in\mathcal{I}_k}\pi^n_{kj}\to\pi^0_k$ for any $k\in[K_0]$.

Let $\Delta\bbeta^n_{kj}=\bbeta^n_{kj}-\bbeta^0_k$ and $\Delta\phi^n_{kj}=\phi^n_{kj}-\phi^0_k$ for any $k\in[K_0]$ and $j\in\mathcal{I}_k$. According to Lemma~\ref{lem:ho}, we have 
\begin{align}
    \label{eq:D2_formulation}
    W_2^2(G_n, G_0) \asymp \sum_{k=1}^{K_0}\sum_{j\in\mathcal{I}_k}\pi^n_{kj} \Big[\norm{\Delta\bbeta^n_{kj}}^2+|\Delta\phi^n_{kj}|^2\Big] + \sum_{k=1}^{K_0}| \pi^0_k -  \sum_{j\in\mathcal{I}_k}\pi^n_{kj}|:=D_2(G_n,G_0).
\end{align}
Since $V(p_{G_n},p_0)/W_2^2(G_n,G_0)\to0$ as $n\to\infty$, we also get that $V(p_{G_n},p_0)/D_2(G_n,G_0)\to0$. In the sequel, our arguments are separated into three stages.

\noindent
\textbf{Stage 1.} In this stage, we decompose the difference $p_{G_n}(\xx,y)-p_{G_0}(\xx,y)$ into a combination of linearly independent elements. By applying a second-order Taylor expansion, it follows that
\begin{align*}
    &f(y;\theta(\xx,\bbeta^n_{kj}),\phi^n_{kj})-f(y;\theta(\xx,\bbeta^0_k),\phi^0_k)\\
    &=\sum_{1\leq |\balpha|+\zeta\leq 2}\frac{1}{\balpha!\zeta!}(\Delta\bbeta^n_{kj})^{\balpha}(\Delta\phi^n_{kj})^{\zeta}\cdot\frac{\partial^{|\balpha|+\zeta}f(y;\theta(\xx,\bbeta^0_k),\phi^0_k)}{\partial\boldsymbol{\bbeta}^{\balpha}\partial\phi^{\zeta}}+R_{2kj}(\xx,y),
\end{align*}
where $\balpha!=\alpha_1!\ldots\alpha_p!$. Under the condition (R2), $R_{2kj}(\xx,y)$ is the Taylor remainder such that
\begin{align*}
    R_{2kj}(\xx,y)f_{\XX}(\xx)=\mathcal{O}(\norm{\Delta\bbeta^n_{kj}}^{2+\gamma}+|\Delta\phi^n_{kj}|^{2+\gamma}),
\end{align*}
for some universal constant $\gamma>0$ for all $k\in[K_0]$ and $j\in\mathcal{I}_k$. By arguing similarly to equation~\eqref{eq:density_decomposition}, we have
\begin{align*}
    &p_{G_n}(\xx,y)-p_{G_0}(\xx,y)\\
    &=\sum_{k=1}^{K_0}\sum_{j\in\mathcal{I}_k}\sum_{1\leq|\balpha|+\zeta\leq2}\frac{1}{\balpha!\zeta!}\cdot\pi^n_{kj}(\Delta\bbeta^n_{kj})^{\balpha}(\Delta\phi^n_{kj})^{\zeta}\cdot\frac{\partial^{|\balpha|+\zeta}f(y;\theta(\xx,\bbeta^0_k),\phi^0_k)}{\partial\boldsymbol{\bbeta}^{\balpha}\partial\phi^{\zeta}}f_{\XX}(\xx)\\
    &\qquad +\sum_{k=1}^{K_0}\left(\sum_{j\in\mathcal{I}_k}\pi^n_{kj}-\pi^0_k\right)f(y;\theta(\xx,\bbeta^0_k),\phi^0_k)f_{\XX}(\xx)+R_2(\xx,y)\\
    &=A_{2n}+B_{2n} + R_2(\xx,y),
\end{align*}
where $R_2(\xx,y)=\sum_{k=1}^{K_0}\sum_{j\in\mathcal{I}_k}R_{2kj}(\xx,y)f_{\XX}(\xx)=O\Big(\sum_{k=1}^{K_0}\sum_{j\in\mathcal{I}_k}\pi^n_{kj}\Big[\norm{\Delta\bbeta^n_{kj}}^{2+\gamma}+|\Delta\phi^n_{kj}|^{2+\gamma}\Big]\Big)$. From the condition (R2) and the formulation of $D_2(G_n,G_0)$ in equation~\eqref{eq:D2_formulation}, it follows that 
\begin{align*}
    R_2(\xx,y)/D_2(G_n,G_0)\lesssim\sum_{k=1}^{K_0}\sum_{j\in\mathcal{I}_k}\Big[\norm{\Delta\bbeta^n_{kj}}^{\gamma}+|\Delta\phi^n_{kj}|^{\gamma}\Big]\to0,
\end{align*}
as $n\to\infty$.

\noindent
\textbf{Stage 2.} In this stage, we will prove by contradiction that at least one among the coefficients in the representation of $A_{2n}/D_2(G_n,G_0)$ and $B_{2n}/D_2(G_n,G_0)$ does not vanish as $n$ goes to infinity. Assume that all of them converge to zero, then by taking the summation of the absolute values of the coefficients of $B_{2n}/D_2(G_n,G_0)$, we have
\begin{align}
    \label{eq:weight_limit_2}
    \sum_{k=1}^{K_0}\left|\sum_{j\in\mathcal{I}_k}\pi^n_{kj}-\pi^0_k\right|/D_2(G_n,G_0)\to0.
\end{align}
Next, let us denote $C_{n,k}({\balpha,\zeta})=\sum_{j\in\mathcal{I}_k}\frac{1}{\balpha!\zeta!}\cdot\pi^n_{kj}(\Delta\bbeta^n_{kj})^{\balpha}(\Delta\phi^n_{kj})^{\zeta}$ for any $\balpha\in\mathbb{N}^p$, $n,\zeta\in\mathbb{N}$ and $k\in[K_0]$. Then, we get
\begin{align*}
    A_{2n}/D_2(G_n,G_0)=\sum_{k=1}^{K_0}\sum_{1\leq|\balpha|+\zeta\leq2}\frac{C_{n,k}(\balpha,\zeta)}{D_2(G_n,G_0)}\cdot\frac{\partial^{|\balpha|+\zeta}f(y;\theta(\xx,\bbeta^0_k),\phi^0_k)}{\partial\boldsymbol{\bbeta}^{\balpha}\partial\phi^{\zeta}}f_{\XX}(\xx).
\end{align*}
From the hypothesis, we deduce that $C_{n,k}(\balpha,\zeta)/D_2(G_n,G_0)\to0$ as $n\to\infty$ for any $k\in[K_0]$, $\balpha\in\mathbb{N}^p$ and $\zeta\in\mathbb{N}$ such that $1\leq|\balpha|+\zeta\leq 2$. 
For any $j=1,2,\ldots,p$, let $\boldsymbol{e}_j\in\mathbb{R}^p$ be a one-hot vector whose $j$-th entry is equal to one, whereas other entries are zero. 
Then, the following limits hold:
\begin{align}
    \label{eq:location_limit_2} \sum_{k=1}^{K_0}\sum_{j\in\mathcal{I}_k}\pi^n_{kj}\|\Delta\bbeta^n_{kj}\|^2/D_2(G_n,G_0)=2\sum_{k=1}^{K_0}\sum_{\balpha\in\{2\boldsymbol{e}_1,\ldots,2\boldsymbol{e}_p\}}\left|\frac{C_{n,k}(\balpha,0)}{D_2(G_n,G_0)}\right|&\to0,\\
    \label{eq:scale_limit_2}
    \sum_{k=1}^{K_0}\sum_{j\in\mathcal{I}_k}\pi^n_{kj}|\Delta\phi^n_{kj}|^2/D_2(G_n,G_0)=2\sum_{k=1}^{K_0}\left|\frac{C_{n,k}(\bzero_p,2)}{D_2(G_n,G_0)}\right|&\to0.
\end{align}
Taking the summation of limits in equations~\eqref{eq:weight_limit_2}, \eqref{eq:location_limit_2} and \eqref{eq:scale_limit_2}, we obtain that $$D_2(G_n,G_0)/D_2(G_n,G_0)\to0$$ as $n\to\infty$, which is a contradiction. Therefore, not all the coefficients in the representation of $A_{2n}/D_2(G_n,G_0)$ and $B_{2n}/D_2(G_n,G_0)$ approach zero as $n$ goes to infinity.

\noindent
\textbf{Stage 3.} In this stage, we use the Fatou's lemma to point out a contradiction to the results in Stage 2. In particular, let us denote by $m_n$ the maximum of the absolute values of the coefficients in the representation of $A_{2n}/D_2(G_n,G_0)$ and $B_{2n}/D_2(G_n,G_0)$. Recall from Stage 2 that not all the coefficients of $A_{2n}/D_2(G_n,G_0)$ and $B_{2n}/D_2(G_n,G_0)$ vanish, we deduce that $1/m_n\not\to\infty$ as $n\to\infty$. Additionally, we also denote 
\begin{align*}
    \left(\sum_{j\in\mathcal{I}_k}\pi^n_{kj}-\pi^0_k\right)/m_nD_2(G_n,G_0)&\to\rho_k, \quad \left(\sum_{j\in\mathcal{I}_k}\pi^n_{kj}(\Delta\bbeta^n_{kj})\right)/m_nD_2(G_n,G_0)\to\boeta_k,\\
    \left(\sum_{j\in\mathcal{I}_k}\pi^n_{kj}(\Delta\phi^n_{kj})\right)/m_nD_2(G_n,G_0)&\to\tau_{1k}, \quad \left(\frac{1}{2}\sum_{j\in\mathcal{I}_k}\pi^n_{kj}(\Delta\phi^n_{kj})^2\right)/m_nD_2(G_n,G_0)\to\tau_{2k},\\
    \left(\sum_{j\in\mathcal{I}_k}\pi^n_{kj}(\Delta\bbeta^n_{kj})^{\balpha}(\Delta\phi^n_{kj})\right)/m_nD_2(G_n,G_0)&\to \xi_{\balpha,k}, \ \text{for all } \balpha\in\mathbb{N}^p:|\balpha|=1\\
     \left(\sum_{j\in\mathcal{I}_k}\frac{1}{\balpha!}\cdot\pi^n_{kj}(\Delta\bbeta^n_{kj})^{\balpha}\right)/m_nD_2(G_n,G_0)&\to \gamma_{\balpha,k}, \ \text{for all } \balpha\in\mathbb{N}^p:|\balpha|=2
\end{align*}
as $n\to\infty$ for all $k\in[K_0]$. Above, at least one among $\rho_k,\boeta_k,\tau_{1k},\tau_{2k},\xi_{\balpha,k}$ and $\gamma_{\balpha,k}$ is different from $0$ or $\bzero_p$. 
By applying the Fatou's lemma, it follows that
\begin{align*}
    0=\lim_{n\to\infty}\frac{V(p_{G_n},p_{G_0})}{m_nD_2(G_n,G_0)}\geq\int\liminf_{n\to\infty}\frac{|p_{G_n}(\xx,y)-p_{G_0}(\xx,y)|}{2m_nD_2(G_n,G_0)}d\nu(\xx,y)\geq 0.
\end{align*}
The above result implies that $[p_{G_n}(\xx,y)-p_{G_0}(\xx,y)]/m_nD_2(G_n,G_0)\to0$ as $n\to\infty$ for $\nu$-almost every $(\xx,y)$. On the other hand, we have
\begin{align*}
    &\frac{p_{G_n}(\xx,y)-p_{G_0}(\xx,y)}{m_nD_2(G_n,G_0)}\\
    &\to\sum_{k=1}^{K_0}\Big[\rho_kf(y;\theta(\xx,\bbeta^0_k),\phi^0_k)+\tau_{1k}\frac{\partial f(y;\theta(\xx,\bbeta^0_k),\phi^0_k)}{\partial\phi}+\boeta_k^{\top}\frac{\partial f(y;\theta(\xx,\bbeta^0_k),\phi^0_k)}{\partial\bbeta}\\
    &\hspace{3cm} +\tau_{2k}\frac{\partial^2 f(y;\theta(\xx,\bbeta^0_k),\phi^0_k)}{\partial\phi^2}+\sum_{|\balpha|=1}\xi_{\balpha,k}\frac{\partial^2f(y;\theta(\xx,\bbeta^0_k),\phi^0_k)}{\partial\bbeta^{\balpha}\partial\phi}\\
    &\hspace{7cm}+\sum_{|\balpha|=2}\gamma_{\balpha,k}\frac{\partial^2f(y;\theta(\xx,\bbeta^0_k),\phi^0_k)}{\partial\bbeta^{\balpha}}\Big]f_{\XX}(\xx).
\end{align*}
Consequently, the above right hand side equals zero for $\nu$-almost every $(\xx,y)$.
Recall that the family $\mathcal{F}$ is strongly identifiable (see Definition~\ref{def:SI}), we obtain that $\rho_k=\tau_{1k}=\tau_{2k}=\xi_{\balpha,k}=\gamma_{\balpha,k}=0$ and $\boeta_k=\bzero_p$, which contradicts the fact that at least one among them is nonzero (vector). Hence, we reach the claim in equation~\eqref{eq:TV_lower_bound_2}.
$\hfill \blacksquare$

\subsection{Consistency of Density Estimator}
\label{appendix:density_consistency}
Before presenting the main result of this appendix, let us introduce the concept of bracket entropy integral. Recall that the set of all joint mixture densities with respect to mixing measures $G$ in ${\cal G}_K$ is defined as ${\cal P}_K = \left\{p_G(\xx, y): G \in{\cal G}_K\right\}$.
Additionally, we also define two additional sets of densities 
related to the function $\bar p_G=\frac{p_G + p_0}{2}$ as follows: 
\begin{align*}
    \bar {\cal P}_K^{\frac 1 2} &= 
\left\{\bar p_G^{\frac 1 2} : p_G \in {\cal P}_K\right\},\\
\bar{\cal P}_K^{\frac 1 2}(\delta) 
&= 
\left\{\bar p_G^{\frac 1 2} \in 
\bar{\cal P}_K^{\frac 1 2}: h(\bar p_G, p_0) \leq \delta\right\},
\end{align*}
for any $\delta > 0$. Then, the bracket entropy integral is given by
\begin{align*}
    \mathcal{J}_B(\gamma,\overline{\mathcal{P}}^{\frac{1}{2}}(\gamma),\nu)=\int_{0}^{\gamma}H_B^{1/2}(u,\overline{\mathcal{P}}^{\frac{1}{2}}_K(u),\nu)du,
\end{align*}
for all $\gamma>0$, where $H_B(u,\overline{\mathcal{P}}^{\frac{1}{2}}_K(u),\nu)$ stands for the $u$-bracket with respect to the $L^2(\nu)$ metric of the family of densities $\overline{\mathcal{P}}^{\frac{1}{2}}_K(u)$.

Now, we are ready to exhibit the result on the density convergence under the Hellinger distance in the following proposition.
\begin{customprop}{C.2}
\label{densityConsistency}
  Suppose that the penalty function $r_{\lambda_n}$ meets conditions (P1) and (P2). In addition, for a universal constant $J>0$, assume there exists a sequence of real numbers $\gamma_n\gtrsim(\log(n)/n)^{1/2}$ such that for all $\gamma\geq\gamma_n$, $\mathcal{J}_B\left(\gamma,\overline{\mathcal{P}}^{\frac{1}{2}}(\gamma),\nu\right)\leq J\sqrt{n}\gamma^2$.
 Then, under Conditions (SI), (P0), and (R1)-(R2), we obtain that
 \begin{align*}
     h(p_{\widehat{G}_n},p_0)=\mathcal{O}_p(\gamma_n).
 \end{align*}
 \end{customprop}
It should be noted that given the bounded parameter space $\Theta$, if we choose $\gamma_n \asymp (\log n/n)^{\frac 1 2}$, then the bracketing entropy condition stated in Proposition~\ref{densityConsistency} holds for many strongly identifiable {\sc fmr} models. For example, one can validate that {\sc fmr} models based on Gaussian distributions with mean function $\theta(\xx, \bbeta) = \mu(\xx) = \xx^\top \bbeta$ and known variance, 
or mean function with zero intercept and unknown variance $\phi$ satisfy this bracketing entropy condition \citep{HOJMLR2022}. 

\textbf{Proof of Proposition \ref{densityConsistency}:} Firstly, it is important to control the penalty differences $\Delta_{1n}(G)$ and $\Delta_{2n}(G)$ for all $G$ in a suitable neighborhood of $G_0$, which will be established in the following lemma.
\begin{customlem}{C.3}
    \label{prop1}
    Let $\kappa_n\geq\gamma_n\gtrsim(\log(n)/n)^{1/2}$ and $0<b_0<\min_{1\leq k\leq K_0}\pi^0_k$. Under the conditions (P0) and (P1) on the penalties, there exist constants $c,M>0$ depending on $G_0$ such that, if $\kappa_n\leq M$, then the following hold
    \begin{align}
        \label{eq:prop1_claim_2}
        \sup\left\{\Delta_{1n}(G):G\in\mathcal{G}_K(b_0;\kappa_n)\right\}&\leq\frac{c\kappa_n}{n},\\
        \label{eq:prop1_claim_1}
        \sup\left\{\Delta_{2n}(G):G\in\mathcal{G}_K(b_0;\kappa_n)\right\}&\leq 0.
    \end{align}
    \end{customlem}
{\bf Proof of Lemma~\ref{prop1}}. In this proof, we consider mixing measures $G$ belonging to the set $\mathcal{G}_K(b_0;\kappa_n)$. Before going into further details, let us lay out the setup for our proof. Firstly, for ease of presentation, we omit the dependence of $G$, on the sample size $n$. It suffices to demonstrate that for sufficiently large $n$,
\begin{align*}
    \Delta_{2n}(G)\leq 0,
\end{align*}
for any $G\in\mathcal{G}_K(b_0;\kappa_n)$.

Recall that $G=\sum_{j=1}^{K}\pi_j\delta_{\bPhi_j}$ and the Voronoi diagram is given by
\begin{align*}
    \mathcal{V}_k=\{\bPhi_j:\|\bPhi_j-\bPhi^0_{k}\|\leq\|\bPhi_j-\bPhi^0_{\ell}\|, \ \forall\ell\neq k\}, \quad k\in[K_0],
\end{align*}
associated with index sets $\mathcal{I}_{k}=\{1\leq j\leq K:\bPhi_j\in\mathcal{V}_{k}\}$. Note that Lemma~\ref{lem:ho} indicates that we can find sufficiently small positive constants $c_1,M_1$, which depend on $G_0$ rather than on $G$, such that if $\kappa_n\leq M_1$, then
\begin{align}
    \label{eq:Wasserstein_lower_bound}
    W_2^2(G,G_0)>c_1\left\{\sum_{k=1}^{K_0}\sum_{j\in\mathcal{I}_{k}}\pi_j\|\bPhi_j-\bPhi^0_{k}\|^2+\sum_{k=1}^{K_0}\left|\pi^0_k-\sum_{j\in\mathcal{I}_{k}}\pi_j\right|\right\}.
\end{align}
Since $\pi_j\geq b_0$ for any $j$, it follows that 
\begin{align}
    \label{eq:parameter_bound}
    \|\bPhi_j-\bPhi^0_{k}\|<\frac{W_2(G,G_0)}{\sqrt{c_1b_0}}<\frac{\kappa_n^{\frac{1}{2}}}{c_0\sqrt{c_1b_0}}, \quad \forall j\in\mathcal{I}_{k}, \ k\in[K_0].
\end{align}
Above, $c_0$ is the constant mentioned in Proposition~\ref{thm:ho_long_results}. Next, we denote $c_2=\frac{1}{c_0\sqrt{c_1b_0}}$, and
\begin{align*}
    \epsilon_0=\inf\{\|\bPhi^0_{j}-\bPhi^0_{k}\|:1\leq j<k\leq K_0\}.
\end{align*}
Then, by setting $M_2=\Big(\frac{\epsilon_0\wedge\delta_0}{4c_2}\Big)^2$, we fix $M=M_1\wedge M_2$ throughout this proof. Moreover, we also assume that $\kappa_n\leq M$ in the sequel.

From the above definitions, we get that $\|\bPhi_j-\bPhi^0_{k}\|<\epsilon_0/4$, for all $j\in\mathcal{I}_{k}$ and $k\in[K_0]$. This result implies that
\begin{align*}
    \|\bPhi_j-\bPhi_{\ell}\|\leq\|\bPhi_j-\bPhi^0_{k}\|+\|\bPhi^0_{k}-\bPhi_{\ell}\|<\frac{\epsilon_0}{2}.
\end{align*}
Additionally, for all $k,k'\in[K_0]$ and $k\neq k'$, if $j\in\mathcal{I}_{k}$ and $i\in\mathcal{I}_{k'}$,
\begin{align*}
    \epsilon_0\leq\|\bPhi^0_{k}-\bPhi^0_{k'}\|\leq\|\bPhi^0_{k}-\bPhi_j\|+\|\bPhi_j-\bPhi_i\|+\|\bPhi_{i}-\bPhi^0_{k'}\|<\frac{\epsilon_0}{2}+\|\bPhi_j-\bPhi_i\|.
\end{align*}
As a consequence, we obtain that
\begin{align*}
    \max_{j,\ell\in\mathcal{I}_{k}}\|\bPhi_j-\bPhi_{\ell}\|<\min_{j\in\mathcal{I}_{k},i\in\mathcal{I}_{k'}}\|\bPhi_j-\bPhi_i\|,
\end{align*}
for all $k\neq k'$. This suggests that $\{\mathcal{V}_1,\ldots,\mathcal{V}_{K_0}\}$ is a cluster partition. 


Subsequently, we divide our arguments into three following stages.



\noindent
\textbf{Stage 1.} In this stage, we provide an upper bound of the difference between parameters $\bbeta$ of $G$ and $G_0$. 
Since $\{\mathcal{V}_{1},\mathcal{V}_{2},\ldots,\mathcal{V}_{K_0}\}$ is a cluster partition of $\bPhi$, we can find a permutation $\tau\in S_{K_0}$ that satisfies
\begin{align*}
    (\bPhi_{1},\ldots,\bPhi_{K})=(\mathcal{V}_{\tau(1)},\ldots,\mathcal{V}_{\tau(K_0)}),
\end{align*}
where we use block matrix notation to display the above right-hand side. 
This result together with equation~\eqref{eq:parameter_bound} implies that $\|\bPhi_{j}-\bPhi^0_{k}\|\leq c_2\kappa_n^{\frac{1}{2}}$ for any $k\in[K_0]$. Consequently, for all $k\in[K_0]$ and $j\in\mathcal{I}_{k}$, 
we obtain 
\begin{align}
    \label{eq:beta_bound}
    \|\bbeta_{j}-\bbeta^0_{k}\|\leq\|\bPhi_{j}-\bPhi^0_{k}\|\leq c_2\kappa_n^{\frac{1}{2}}.
\end{align}

\noindent
\textbf{Stage 2.} In this stage, we provide an upper bound for $\Delta_{2n}(G)$. Note that we have $r_{\lambda_n}\geq 0$, then it follows from the condition (P1) that, for sufficiently large $n$, 
\begin{align*}
    \Delta_{2n}(G)&=\sum_{k=1}^{K_0}\sum_{\ell=1}^{p}r_{\lambda_n}(\beta^0_{k\ell})-\sum_{k=1}^{K}\sum_{\ell=1}^{p}r_{\lambda_n}(\beta_{k\ell})\\
    \text{($j_k$ is an element of the index set $\mathcal{I}_{k}$)}&\leq\sum_{k=1}^{K_0}\sum_{\ell=1}^{p}r_{\lambda_n}(\beta^0_{k\ell})-\sum_{k=1}^{K_0}\sum_{\ell=1}^{p}r_{\lambda_n}(\beta_{j_k\ell})\\
    &=\sum_{k=1}^{K_0}\sum_{\ell=1}^{p}\Big[r_{\lambda_n}(\beta^0_{k\ell})-r_{\lambda_n}(\beta_{j_k\ell})\Big]\\
  (\text{corresponding to the nonzero}~\beta^0_{kl})
  &\leq\sum_{k=1}^{K_0}\sum_{\ell=1}^{p_k}\Big[r_{\lambda_n}(\beta^0_{k\ell})-r_{\lambda_n}(\beta_{j_k\ell})\Big]\\
  \text{(according to condition (P1) and equation~\eqref{eq:beta_bound}) }&\leq\sum_{k=1}^{K_0}\sum_{\ell=1}^{p_k}\Big[c_n-c_n\Big] =0.
\end{align*}
Thus, we achieve the claim in equation~\eqref{eq:prop1_claim_1}. 

Regarding the claim in equation~\eqref{eq:prop1_claim_2}, let $\varrho_k=\sum_{j\in\mathcal{I}_{k}}\pi_j$ for all $k\in[K_0]$. 
It is worth noting that for all $G\in\mathcal{G}_K(b_0;\kappa_n)$, the result in equation~\eqref{eq:Wasserstein_lower_bound} indicates that 
$|\pi^0_k-\varrho_k|\lesssim\kappa_n$ for all $k\in[K_0]$. Furthermore, it follows from condition (P0) that
\begin{align*}
    \Delta_{1n}(G)=\frac{1}{n}[\rho(\bpi^0)-\rho(\bpi)]\leq\frac{1}{n}[\rho(\bpi_0)-\rho(\varrho_1,\ldots,\varrho_{K_0})]\leq \frac{\ell_{\rho}}{n}\sum_{k=1}^{K_0}|\pi^0_k-\varrho_k|\lesssim\frac{\kappa_n}{n}.
\end{align*}
Therefore, we obtain the claim~\eqref{eq:prop1_claim_2}, and the proof of Lemma~\ref{prop1} is completed.
$\hfill \blacksquare$

Now, we are going back to the proof of Proposition \ref{densityConsistency}. From Theorem~\ref{thm:paramConsistency}(ii), for any $\epsilon>0$ 
we can find $b_\epsilon>0$ and an integer $N_{1, \epsilon}>0$ such that for every $n\geq N_{1, \epsilon}$,
\begin{align*}
    \bbP(\widehat{G}_n\in\mathcal{G}_K(b_\epsilon))&>1-\epsilon.
\end{align*}
Let $M>0$ be the positive constant mentioned in Lemma~\ref{prop1}, and let $N_2>0$ be a sufficiently large integer that satisfies $\gamma_n\leq M$ for all $n\geq N_2$. In the sequel, we assume that 
$n\geq N_{1,\epsilon}\vee N_2$. Based on the results in Theorem~\ref{thm:paramConsistency}, it is sufficient to demonstrate that
\begin{align*}
    \bbP(\gamma_n<h(\hat{p}_n,p_0)<M)\to0, \quad \text{as } n\to\infty.
\end{align*}
Note that the above left hand side can be written as
\begin{align}
    &\bbP(\gamma_n<h(\hat{p}_n,p_0)<M)=\bbP\Big(\gamma_n<h(\hat{p}_n,p_0)<M,\{\widehat{G}_n\not\in\mathcal{G}_K(b_0)\}\Big)\nonumber\\
    &\hspace{5cm}+\bbP\Big(\gamma_n<h(\hat{p}_n,p_0)<M,\widehat{G}_n\in\mathcal{G}_K(b_0)\Big)\nonumber\\
    &\leq\bbP\Big(\widehat{G}_n\not\in\mathcal{G}_K(b_0)\Big)+\bbP\Big(\gamma_n<h(\hat{p}_n,p_0)<M,\widehat{G}_n\in\mathcal{G}_K(b_0)\Big)\nonumber\\
    &\leq \epsilon+\bbP\Big(\gamma_n<h(\hat{p}_n,p_0)<M,\widehat{G}_n\in\mathcal{G}_K(b_0)\Big)\nonumber\\
    &\leq\epsilon+\bbP\Big(\gamma_n/4<h\Big(\frac{\hat{p}_n+p_0}{2},p_0\Big)<M/\sqrt{2},\widehat{G}_n\in\mathcal{G}_K(b_0)\Big)\nonumber\\
    \label{eq:rhs}
    &\leq \epsilon+\bbP\left(\sup_{\substack{G\in\mathcal{G}_K(b_0),\\ \gamma_n/4<h(\overline{p}_G,p_0)<M/\sqrt{2}}}n^{-\frac{1}{2}}\nu_n(G)+\frac{1}{4}[\Delta_{1n}(G)+\Delta_{2n}(G)]-h^2(\overline{p}_{G},p_0)\geq 0\right),
\end{align}
where the last two inequalities follow from the inequalities 
\begin{align*}
    h^2(\overline{p}_G,p_0)\leq\frac{1}{2}h^2(p_G,p_0), \quad h^2(p_G,p_0)\leq 16h^2(\overline{p}_G,p_0),
\end{align*}
and the inequality in equation \eqref{lemma_bound}, respectively. Thus, we need to show that the term in equation~\eqref{eq:rhs} goes to zero. For that purpose, let us denote $\mathcal{S}_n=\min\{s:2^{s+1}\gamma_n>M/\sqrt{2}\}$. Then, we have
\begin{align}
    \bbP&\left(\sup_{\substack{G\in\mathcal{G}_K(b_0),\\ \gamma_n/4<h(\overline{p}_G,p_0)<M\sqrt{2}}}n^{-\frac{1}{2}}\nu_n(G)+\frac{1}{4}[\Delta_{1n}(G)+\Delta_{2n}(G)]-h^2(\overline{p}_{G},p_0)\geq 0\right)\nonumber\\
    &\leq\sum_{s=0}^{\mathcal{S}_n}\bbP\left(\sup_{\substack{G\in\mathcal{G}_K(b_0;2^{s+1}\gamma_n/4)}}\nu_n(G)+\frac{\sqrt{n}}{4}[\Delta_{1n}(G)+\Delta_{2n}(G)]\geq\sqrt{n}2^{2s}\Big(\frac{\gamma_n}{4}\Big)^2\right)\nonumber\\
    \label{eq:probability_bound}
    &\leq\sum_{s=0}^{\mathcal{S}_n}\bbP\left(\sup_{G\in\mathcal{G}_K(b_0;2^{s+1}\gamma_n/4)}\nu_n(G)\geq\sqrt{n}2^{2s}\Big(\frac{\gamma_n}{4}\Big)^2-\frac{c\gamma_n2^{s}}{4\sqrt{n}}\right)
\end{align}
where the last inequality is obtained by applying Lemma~\ref{prop1}. 

Subsequently, we aim to utilize Lemma~\ref{thm:incrementGeer}. In particular, by setting 
\begin{align}
    \label{eq:a_formula}
    &a_n=\sqrt{n}2^{2s}\Big(\frac{\gamma_n}{4}\Big)^2-\frac{c\gamma_n2^{s}}{4\sqrt{n}},\nonumber\\
    R&=2^{s+1}\gamma_n, \qquad C_1=15,
\end{align}
we can verify that the condition~\eqref{thm511_2} is met. Then, it is necessary to validate the condition~\eqref{thm511_5}. Note that we have
\begin{align}
    4C&\int_{0}^{2^{s+1}\gamma_n}\sqrt{H_B\left(\frac{u}{\sqrt{2}},\overline{\mathcal{P}}_K^{\frac{1}{2}}\Big(2^{s+1}\frac{\gamma_n}{4}\Big),\nu\right)}du\vee 2^{s+1}\gamma_n\nonumber\\
    &\leq 4C\sqrt{2}\int_{0}^{2^{s+\frac{1}{2}}\gamma_n}\sqrt{H_B\left(u,\overline{\mathcal{P}}_K^{\frac{1}{2}}\Big(2^{s+\frac{1}{2}}\gamma_n\Big),\nu\right)}du\vee 2^{s+1}\gamma_n\nonumber\\
    &\leq 4C\Big(J\sqrt{2n}2^{2s+\frac{1}{2}}\gamma_n^2\vee 2^{s+1}\gamma_n\Big)\nonumber\\
    \label{eq:integral_bound}
    &=4C\Big(J\sqrt{n}2^{2s+1}\gamma_n^2\vee 2^{s+1}\gamma_n\Big).
\end{align}
Furthermore, it can be validated that $a_n\geq 2^{s+1}\gamma_n$ for large enough $n$, and
\begin{align*}
    4CJ\sqrt{n}2^{2s+1}\gamma_n^2&=a_n-\sqrt{n}2^{2s}\Big(\frac{\gamma_n}{4}\Big)^2+4CJ\sqrt{n}2^{2s+1}\gamma_n^2+\frac{c\gamma_n2^{s}}{4\sqrt{n}}\\
    &=a_n+\sqrt{n}2^{2s}\gamma_n^2\left(8CJ-\frac{1}{16}\right)+\frac{c\gamma_n2^{s}}{4\sqrt{n}}.
\end{align*}
It can be seen that if we select $J$ that satisfies $8CJ<\frac{1}{16}$, then for sufficiently large $n$, since $\gamma_n\gtrsim\sqrt{\log(n)/n}$, we have
\begin{align*}
    \left(8CJ-\frac{1}{16}\right)2^s+\frac{c\gamma_n2^{s}}{4\sqrt{n}}\leq 0.
\end{align*}
Consequently, the condition~\eqref{thm511_5} is satisfied. Then, Lemma~\ref{thm:incrementGeer} indicates that
\begin{align}
    \sum_{s=0}^{\mathcal{S}_n}\bbP&\left(\sup_{G\in\mathcal{G}_K(b_0;2^{s+1}\gamma_n/4)}\nu_n(G)\geq\sqrt{n}2^{2s}\Big(\frac{\gamma_n}{4}\Big)^2-\frac{c\gamma_n2^{s-1}}{4\sqrt{n}}\cdot\gamma_n^{\frac{1}{2}}2^{s}\right)
    \leq C\sum_{s=0}^{\mathcal{S}_n}\exp\left(-\frac{a_n^2}{16^2R^2}\right)
    \label{eq:sum_bound_1}
\end{align}
In the formulation of $a_n$ in equation~\eqref{eq:a_formula}, it can be checked that the first term dominates the second one. Therefore, we can find a constant $c_1>0$ that satisfies $a_n^2\geq c_1n2^{4s}\gamma_n^4$ for sufficiently large $n$. By denoting $c_n=\frac{c_1n\gamma_n^2}{64C^2}$, we have $c_n\to\infty$ as $\gamma_n\gtrsim\sqrt{\log(n)/n}$. Thus, we achieve that
\begin{align}
    C\sum_{s=0}^{\mathcal{S}_n}\exp\left(-\frac{a_n^2}{16^2R^2}\right)&\lesssim\sum_{s=0}^{\infty}\exp(-c_n2^{2s}) 
    \leq\exp(-c_n)-1+\sum_{s=0}^{\infty}\exp(-c_ns)\nonumber\\
    \label{eq:sum_bound_2}
    &=\exp(-c_n)-1+\frac{1}{1-\exp(-c_n)}\to0,
\end{align}
as $n\to\infty$. Putting the results in equations~\eqref{eq:probability_bound}, \eqref{eq:sum_bound_1} and \eqref{eq:sum_bound_2}, we obtain the conclusion of Proposition~\ref{densityConsistency}. Hence, the proof is completed.
$\hfill \blacksquare$


\section{Auxiliary Results}
To start with, we introduce a result on the likelihood ratio statistics from \cite{DACUNHA1999} with notations adapted to the setting of our manuscript.
\begin{customlem}{D.1}[Theorem 3.2, \cite{DACUNHA1999}] 
\label{thm:lrs}
Suppose that the conditions (SI) and (R3) hold true, then the log-likelihood ratio statistic over the class ${\cal G}_K$ satisfies 
\(
\sup_{G \in {\cal G}_K}\ell_n(G) - \ell_n(G_0) = O_p(1).
\)
\end{customlem}
Next, we utilize the following result from empirical process theory provided in \cite{GEER2000} for the proof of Proposition~\ref{densityConsistency}.
\begin{customlem}{D.2}[Theorem 5.11, \cite{GEER2000}]
\label{thm:incrementGeer}
Given universal constants $R,C>0$, let $a, C_1 > 0$ be chosen such that
\begin{equation}
\label{thm511_2}
a \leq C_1\sqrt n R^2 \wedge 8\sqrt n R,
\end{equation}
and
\begin{equation}
\label{thm511_5}
a \geq\sqrt{C^2(C_1+1)} \left(\int_0^R \sqrt{H_B 
\left(\frac u {\sqrt 2}, \left\{p_G:G \in N(R)\right\}, \nu \right)} du \vee  R\right), 
\end{equation}
where we define $N(R) = \left\{G \in {\cal G}_K: h(\bar p_G, p_0) \leq R\right\}$. Then, we obtain
\[
\bbP 
\left\{\sup_{G \in N(R)} |\nu_n(G)| \geq a\right\}
 \leq C \exp\left(-\frac{a^2}{C^2(C_1+1)R^2}\right),
 \]
where $\nu_n(G)$ is defined in equation \eqref{nuDef}.
\end{customlem}

Lastly, to connect the convergence of a mixing measure under the Wasserstein and the convergence of its atoms and mixing proportions, we invoke a result in previous work \citep{HO2016singularity}.
\begin{customlem}{D.3}[Lemma 3.1, \cite{HO2016singularity}]
\label{lem:ho}
For any mixing measure $G = \sum_{j=1}^K \pi_j\delta_{\bPhi_j} \in {\cal G}_K(b_0)$ with some $b_0 > 0$, we have
$$
W_r^r(G, G_0) \asymp \sum_{k=1}^{K_0} \sum_{j \in {\cal I}_k} \pi_j \norm{\bPhi_j - \bPhi^0_{k}}^r +
                \sum_{k=1}^{K_0} \Bigg| \pi^0_{k} - \sum_{j \in {\cal I}_k} \pi_j \Bigg|,
                $$
as $W_r(G, G_0) \downarrow 0$, for any $r \geq 1$.
\end{customlem}    

\newpage

\section{Additional Numerical Details}
\label{sec:add-numeric}

In this section, we provide formulas for updating the regression coefficients $\bbeta_k$ under 
the SCAD and MCP penalties in the M-step of the modified EM algorithm. We also describe our 
initialization procedure for the algorithm and the selection of the tuning parameter 
$\lambda$ for the variable selection penalty $r_{\lambda}$.

\subsection{Computational expressions}
\label{sec:additional_computation}

Since we use the quadratic approximation of the weighted log-likelihood in the M-step, 
together with the proximal coordinate descent method, there are analytical solutions 
for the updates when the regularization function is Lasso, SCAD, or MCP.
In Section \ref{sec:numeric} of the paper, we have explicitly presented the 
update of $\bbeta_k$'s when implementing the Lasso. 
Recall that for any finite constant $\eta>0$, the soft-threshold operator is defined as
$S(x; \eta) =  \text{sign}(x) ~(|x| - \eta)_+$. Moreover, in Section \ref{sec:numeric}, we mentioned that each coordinate $\beta_{kj}$ is updated by minimizing \eqref{CoordObj} in the paper, which is restated below,
\begin{equation*}
    -\nabla_j \mathcal{L}_k(\tilde{\bPhi}_k)(\beta_{kj} - \tilde{\beta}_{kj}) - \frac{1}{2} \nabla^2_{jj} \mathcal{L}_k(\tilde{\bPhi}_k)(\beta_{kj} - \tilde{\beta}_{kj})^2 +\frac{\xi}{2}(\beta_{kj} - \tilde{\beta}_{kj})^2 +  r_{\lambda_n}(\beta_{kj}).
\end{equation*}
Define $\tilde{r}_{kj} = (\xi - \nabla_{jj}^2 \mathcal{L}_k(\tilde{\bPhi}_k))\tilde{\beta}_{kj} + \nabla_j \mathcal{L}_k(\tilde{\bPhi}_k)$.
When $r_{\lambda_n}$ is the SCAD penalty, the update of $\beta_{kj}$, $k = 1,\ldots,K$ and $j = 1,\ldots,p$ is

\begin{equation*}
    \beta_{kj}^{(m, s+1)} = \begin{cases}
        \frac{S \left(\tilde{r}_{kj}; \lambda_n\right)}{\xi - \nabla_{jj}^2 \mathcal{L}_k(\tilde{\bPhi}_k)}, & \text{if } \left|\tilde{r}_{kj} \right| \leqslant (\xi - \nabla_{jj}^2 \mathcal{L}_k(\tilde{\bPhi}_k) +1)\lambda_n \\
        \frac{S (\tilde{r}_{kj}; \frac{a \lambda_n}{a-1})}{\xi - \nabla_{jj}^2 \mathcal{L}_k(\tilde{\bPhi}_k) - \frac{1}{a-1}}, & \text{if } (\xi - \nabla_{jj}^2 \mathcal{L}_k(\tilde{\bPhi}_k) +1 )\lambda_n  < \left|\tilde{r}_{kj}\right| 
        \leqslant a \lambda_n (\xi - \nabla_{jj}^2 \mathcal{L}_k(\tilde{\bPhi}_k)) \\
        \frac{\tilde{r}_{kj}}{\xi - \nabla_{jj}^2 \mathcal{L}_k(\tilde{\bPhi}_k)}, & \text{if } \left|\tilde{r}_{kj}\right| > a \lambda_n (\xi - \nabla_{jj}^2 \mathcal{L}_k(\tilde{\bPhi}_k)) 
    \end{cases}
    \label{SCADUp}
\end{equation*}
When $r_{\lambda_n}$ is the MCP penalty, the update of $\beta_{kj}$, $k=1,\ldots,K$ and $j = 1,\ldots,p$ is
\begin{equation*}
    \beta_{kj}^{(m, s+1)} = \begin{cases}
        \frac{S\left(\tilde{r}_{kj}; \lambda_n \right)}{\xi - \nabla_{jj}^2 \mathcal{L}_k(\tilde{\bPhi}_k) - \frac{1}{a}}, & \text{if } \left|\tilde{r}_{kj}\right| \leqslant a \lambda_n (\xi - \nabla_{jj}^2 \mathcal{L}_k(\tilde{\bPhi}_k))\\
        \frac{\tilde{r}_{kj}}{\xi - \nabla_{jj}^2 \mathcal{L}_k(\tilde{\bPhi}_k)}, & \text{if } \left|\tilde{r}_{kj}\right| > a \lambda_n (\xi - \nabla_{jj}^2 \mathcal{L}_k(\tilde{\bPhi}_k))
    \end{cases}
    \label{MCPUpBeta}
\end{equation*}

One remark is that if the components of the FMR belong to exponential family (Gaussian, Binomial, Poisson) 
and the link function is the canonical link, each $\mathcal{L}_k(\bPhi_k)$ is concave in the coefficient $\bbeta_k$. In this situation, each diagonal element of the Hessian matrix $\nabla_{jj}^2 \mathcal{L}_k(\bPhi_k)$ taken w.r.t.\ $\bbeta_k$ is non-positive, that is, we have $\nabla_{jj}^2 \mathcal{L}_k(\tilde{\bPhi}_k) \leq 0$ in the denominator of the update expressions. With a positive $\xi$, the denominator $\xi - \nabla_{jj}^2 \mathcal{L}_k(\tilde{\bPhi}_k)$ is strictly greater than $0$, guaranteeing the stability of the updates. On the other hand, we choose $\xi$ sufficiently large to ensure the denominator $\xi - \nabla_{jj}^2 \mathcal{L}_k(\tilde{\bPhi}_k) - \frac{1}{a-1} > 0$ in the SCAD updates and the denominator $\xi - \nabla_{jj}^2 \mathcal{L}_k(\tilde{\bPhi}_k) - \frac{1}{a} > 0$ in the MCP updates.\\

\subsection{Initialization of the modified EM algorithm}
\label{sec:inital}

We implement a data-adaptive method for the initialization of the modified EM algorithm. 

First, we cluster the paired data points $\{(\xx_i, y_i)\}_{i=1}^{n}$ into $K$ clusters by using the \texttt{Mclust} function in the \texttt{mclust} package in \texttt{R}, from which we obtain the 
initial cluster labels for all the observations. Suppose there are $n_k$ data points 
in the $k$-th cluster, the mixing proportions are initialized as 
\(
    \pi_k^{(0)} = \frac{n_k}{n}~~\text{or}~~\frac{1}{K}; 
    \quad k = 1,\ldots,K.
\)

As for the initialization of the $\bbeta_k$, for each cluster $k$, we fit a GLM to the data points in this cluster by using \texttt{glmnet} with a specific choice of the GLM family, e.g., Gaussian or Binomial. 
The estimates of $\bbeta_k$, $k = 1,\ldots, K$, are used as the initial values for the modified EM algorithm. For the mixture of Gaussian regressions, we use the sample variance of $y_i$'s in the $k$-th cluster as the initial value for each dispersion parameter $\phi_k$. Furthermore, from our simulations for mixture of Gaussian regressions, we find that initializing all the regression coefficients, except the intercept, at zero also works well.

In the case that some clusters among the $K$ initial clusters are empty, we initialize their class-specific parameters to be the estimates based on all the data points $\{(\xx_i, y_i)\}_{i=1}^{n}$ plus a class-specific perturbation to make sure each cluster has its own initial values that are different 
from other clusters.

\subsection{Tuning parameter selection for
\texorpdfstring{$r_{\lambda_n}$}{r(lambda n)}}
\label{sec:tuning}

Remark~\ref{rmk1} establishes that, for the SCAD and MCP penalties, 
a theoretically valid order for the tuning parameter is
\begin{equation*}
\lambda_n\asymp \gamma_n^{1/2}\log n,
\qquad
\gamma_n\asymp
\left(\frac{\log n}{n}\right)^{1/2}.
\end{equation*}
We therefore take
\begin{equation*}
a_n
:=
\left(\frac{\log n}{n}\right)^{1/4}\log n
=
\frac{(\log n)^{5/4}}{n^{1/4}},
\end{equation*}
and construct the candidate sequence
\begin{equation*}
\lambda_{n,\ell}=C_\ell a_n,
\qquad
C_1>C_2>\cdots>C_L>0,
\qquad
\ell=1,\ldots,L,
\end{equation*}
where the constants $C_1,\ldots,C_L$ do not depend on $n$. Thus, every candidate has the theoretically prescribed order, while the multiplicative constant is selected from the data.

The candidates are fitted sequentially from the largest to the smallest value of $\lambda_n$. We first fit the model using $\lambda_{n,1}$, which imposes the strongest regularization and generally produces the sparsest solution. For each subsequent candidate $\lambda_{n,\ell}$, $\ell\geq 2$, the estimate obtained under $\lambda_{n,\ell-1}$ is used as the initial value, that is, $\widehat G_{n,\ell}^{(0)} = \widehat G_{n,\ell-1}$. This warm-start strategy improves computational efficiency and stability along the regularization path \citep{friedman2010regularization, simon2011regularization}. As $\lambda_n$ decreases, the strength of regularization is gradually relaxed, allowing additional features to enter the fitted model. 

For each candidate, let $\widehat G_{n,\ell}$ be the resulting OMPLE and let $d_\ell$ denote its effective number of estimated parameters, with coefficients estimated as zero excluded from the count. We compute the ordinary BIC
\begin{equation*}
\operatorname{BIC}(\lambda_{n,\ell})
=
-2\ell_n(\widehat G_{n,\ell})
+
d_\ell\log n,
\end{equation*}
where $\ell_n(\widehat G_{n,\ell})$ is the unpenalized observed-data log-likelihood 
evaluated at the penalized estimator. The selected tuning parameter and final estimator are
\begin{equation*}
\widehat\ell
=
\underset{1\leq\ell\leq L}{\arg\min} \, 
\operatorname{BIC}(\lambda_{n,\ell}),
\qquad
\widehat\lambda_n
=
\lambda_{n,\widehat\ell},
\qquad
\widehat G_n
=
\widehat G_{n,\widehat\ell} \,.
\end{equation*}
We find this strategy effective in our simulations and real data example. 

\section{Additional Simulation Results}
\label{sec:additional-simul}

\begin{figure}[htbp]
\centering
\subfloat{%
\includegraphics[width=0.8\textwidth, height=6.5cm, keepaspectratio=false]{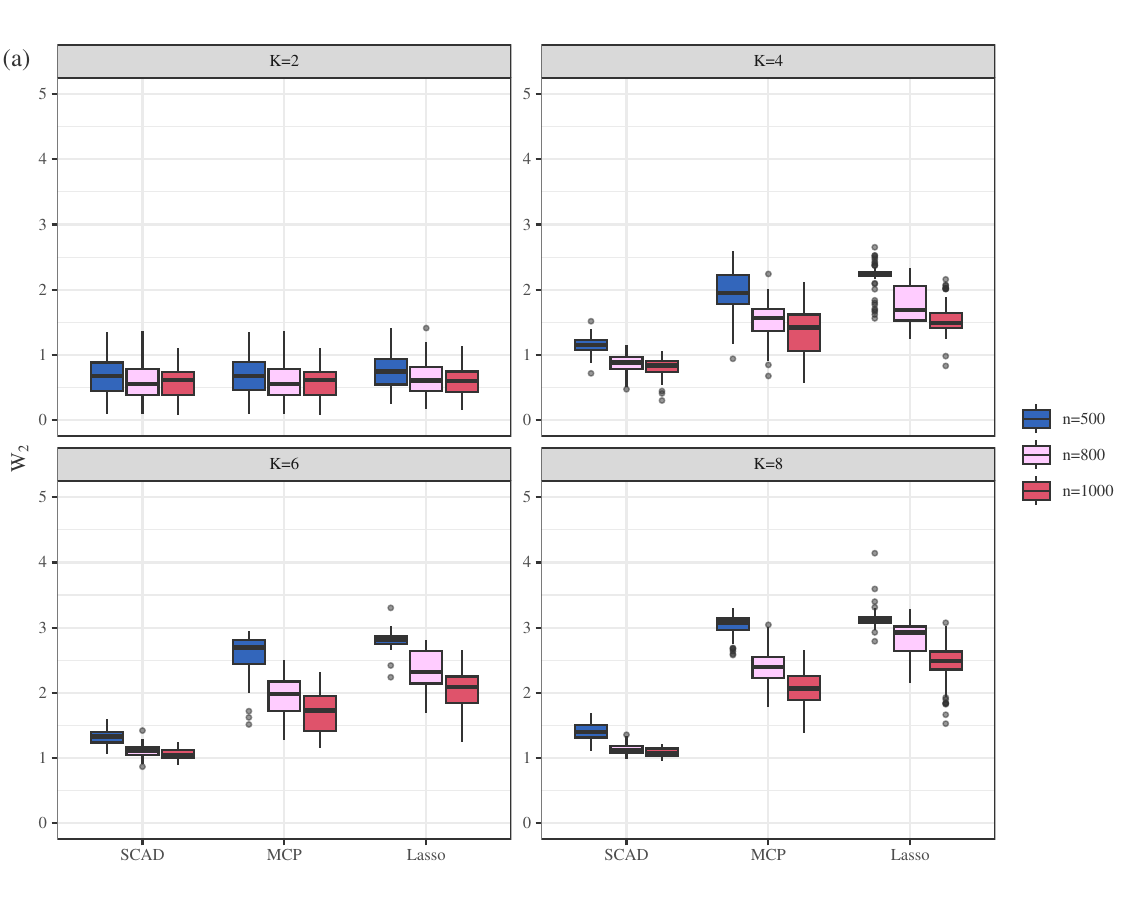}
}    
\vspace*{0.0001cm} 
\subfloat{%
\includegraphics[width=0.8\textwidth, height=6.5cm, keepaspectratio=false]{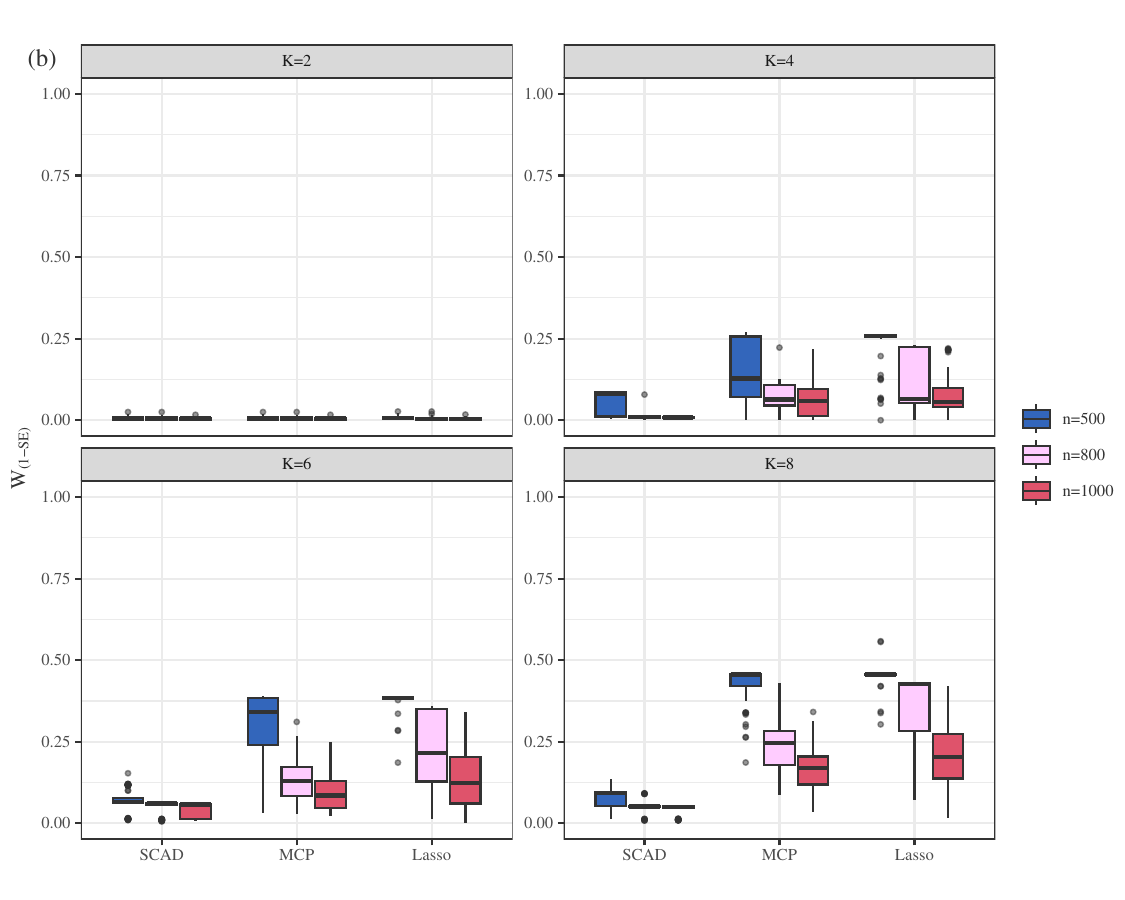}
}
\vspace*{0.0001cm} 
\subfloat{%
\includegraphics[width=0.8\textwidth, height=6.5cm, keepaspectratio=false]{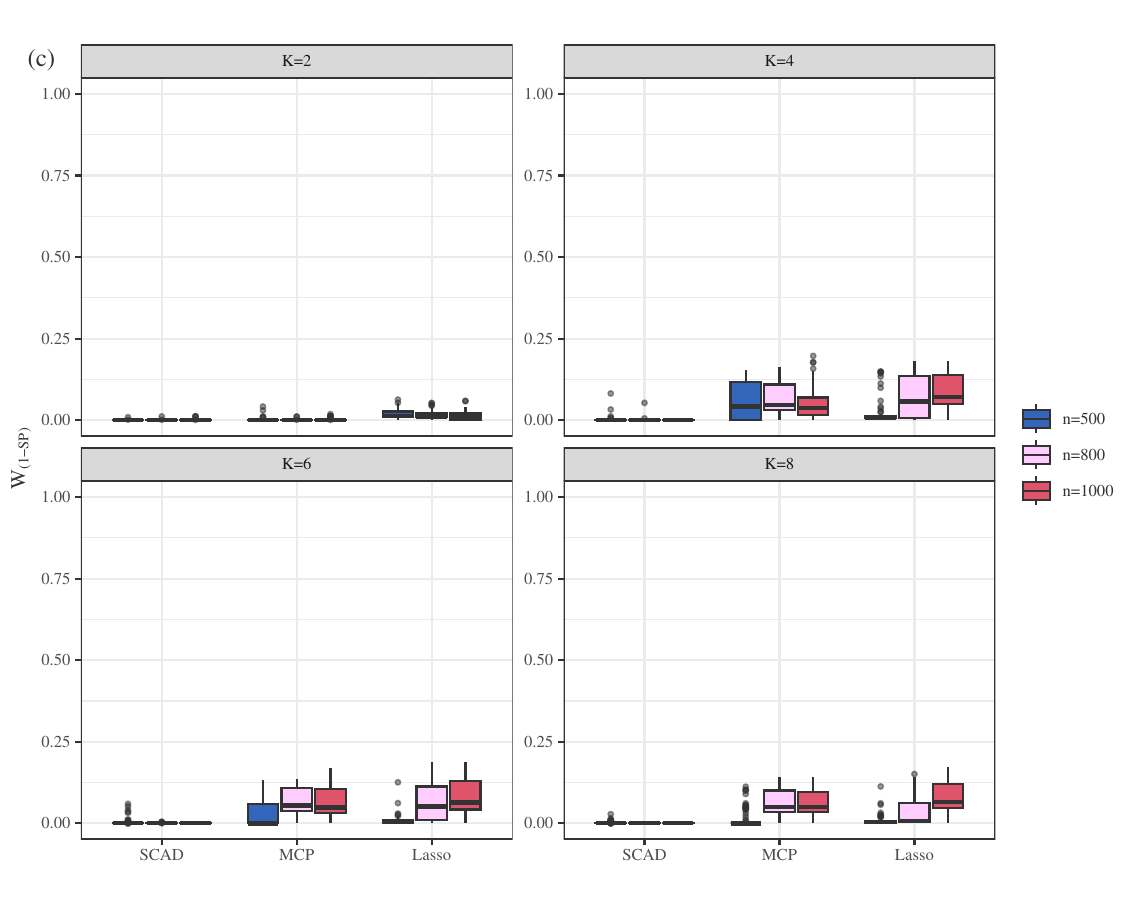}
}
 \caption{Simulation results for Gaussian FMR with $K_0 = 2$ and $p = 50$. Boxplots of:\\ (a) $W_2$; (b) $W_{\text{(1-SE)}}$; 
 (c) $W_{\text{(1-SP)}}$. The mixture order upper bounds are: 
 $K = 2, 4, 6, 8$.
}
 \label{Gaussian_overfit_plot_p50}
\end{figure}

\begin{figure}[htbp]
\centering
\subfloat{%
\includegraphics[width=0.8\textwidth, height=6.5cm, keepaspectratio=false]{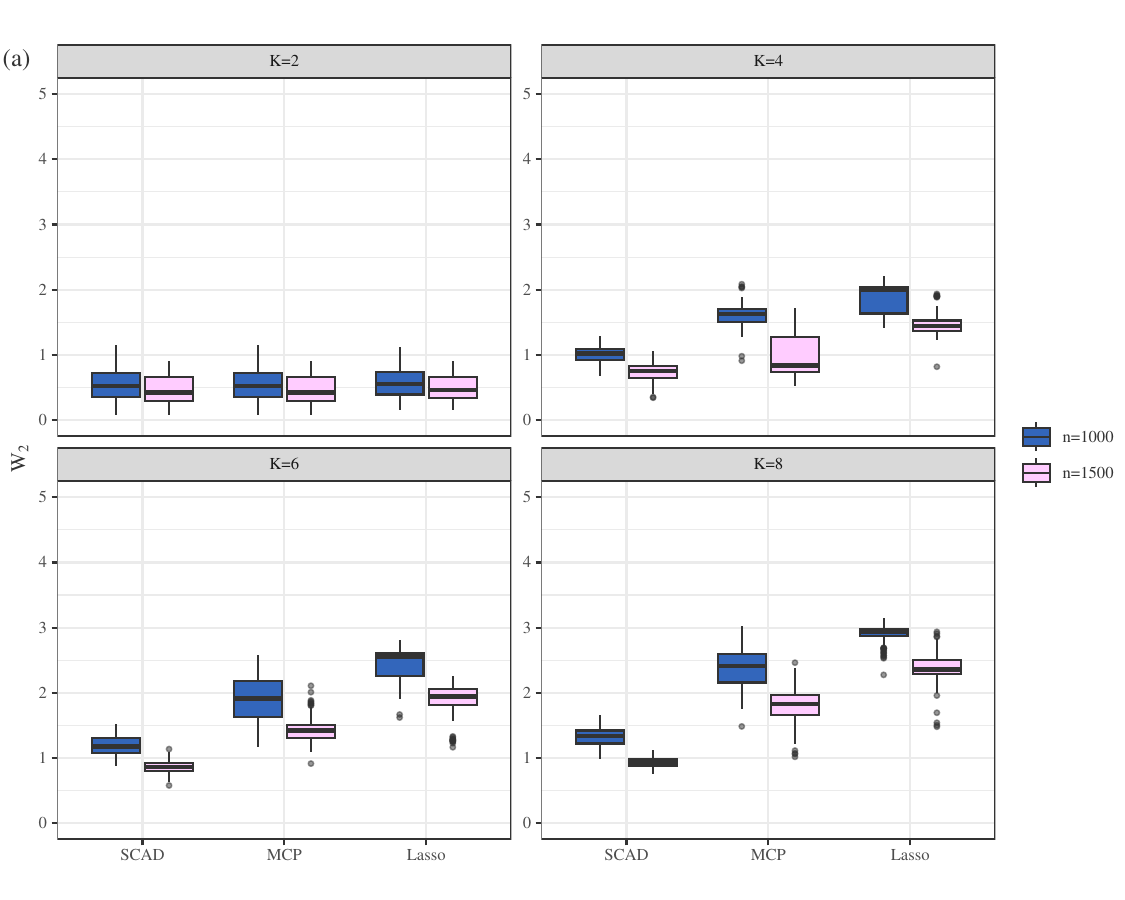}
}    
\vspace*{0.0001cm} 
\subfloat{%
\includegraphics[width=0.8\textwidth, height=6.5cm, keepaspectratio=false]{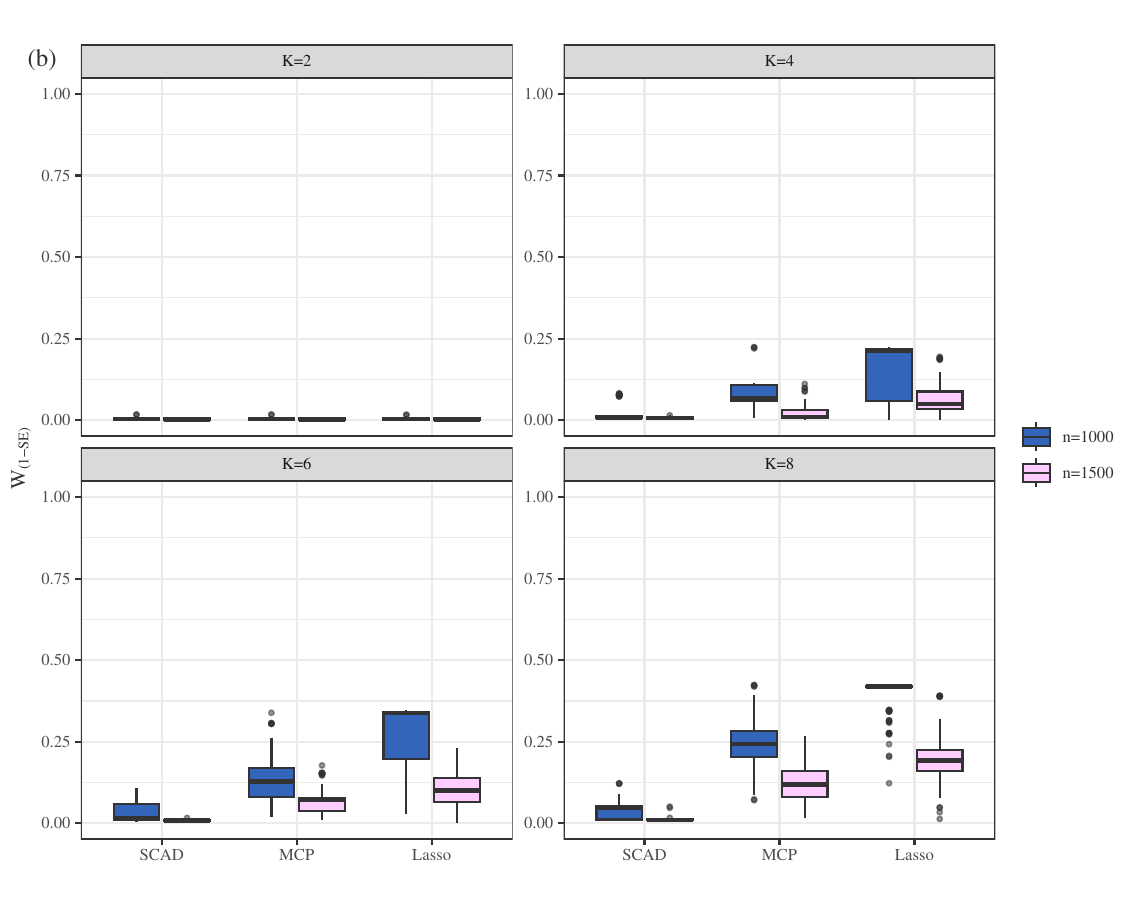}
}
\vspace*{0.0001cm} 
\subfloat{%
\includegraphics[width=0.8\textwidth, height=6.5cm, keepaspectratio=false]{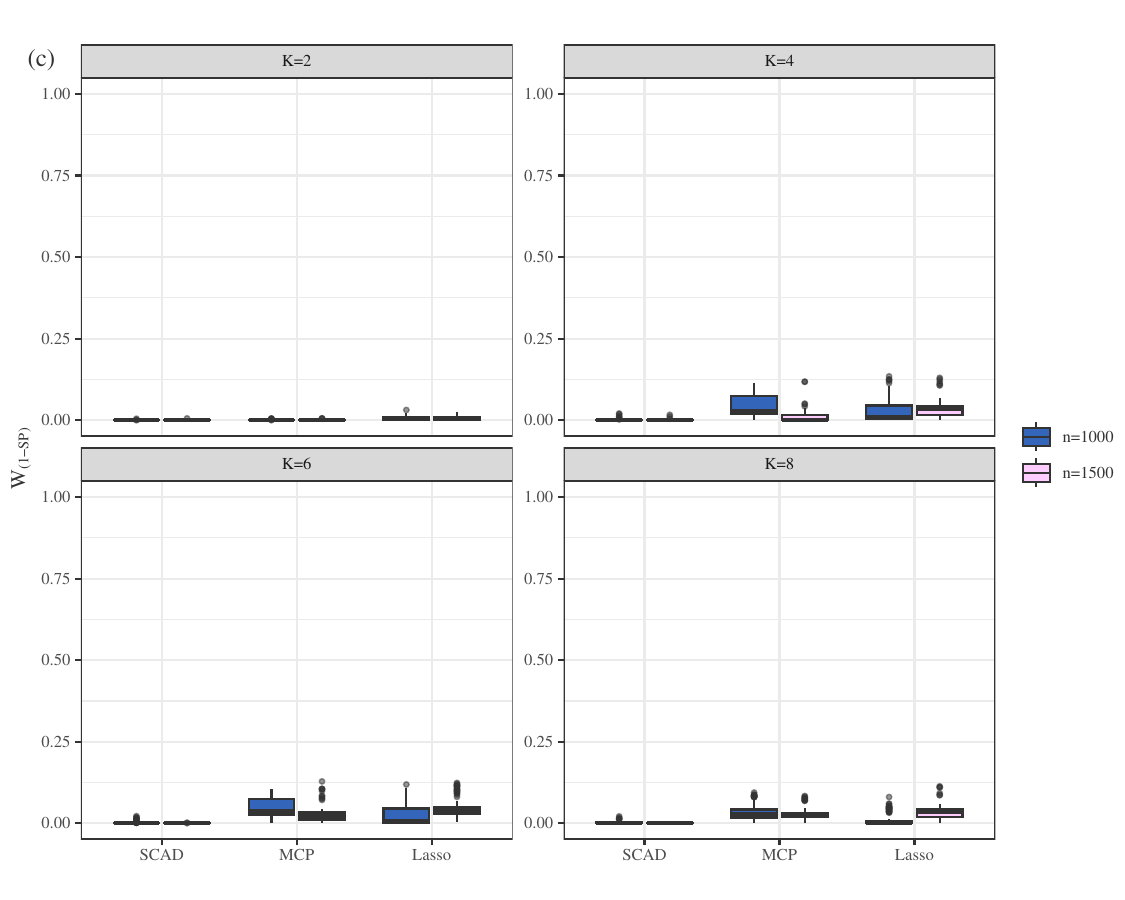}
}
 \caption{Simulation results for Gaussian FMR with $K_0 = 2$ 
 and $p = 100$. Boxplots of:\\ (a) $W_2$; (b) $W_{\text{(1-SE)}}$; (c) $W_{\text{(1-SP)}}$. The mixture order upper bounds are: $K = 2, 4, 6, 8$.}
 \label{Gaussian_overfit_plot_p100}
\end{figure}

\noindent
{\bf Example 1 (Gaussian FMR)}.
Figures~\ref{Gaussian_overfit_plot_p50} and \ref{Gaussian_overfit_plot_p100} present boxplots of the performance metrics \(W_2\), \(W_{(1-\text{SE})}\), and \(W_{(1-\text{SP})}\) for dimensions \(p = 50\) and \(p = 100\), respectively. The results are based on \(R = 100\) simulated samples from the Gaussian FMR model in Example~1 of Section~\ref{sec:simul}, and for three penalties.

\newpage

Tables~\ref{K02_p10}, \ref{K02_p50}, and \ref{K02_p100} report the empirical means and standard deviations of the performance metrics \(W_2\), \(W_{(1-\text{SE})}\), and \(W_{(1-\text{SP})}\) for all three dimensions $p=10, 50, 100$. 
The results are based on \(R = 100\) simulated samples from the Gaussian FMR  in Example~1 of Section~\ref{sec:simul}.

\begin{table}[H]
\small
\centering
\begin{tabular}{|c|ccc|ccc|ccc|}
\hline
$K$ & & $n=200$  & & &$n=500$ & && $n=800$ & \\
\hline
{\sc scad} & $W_2$ & $W_{(1-\text{SE})}$ & $W_{(1-\text{SP})}$&  
$W_2$ & $W_{(1-\text{SE})}$ & $W_{(1-\text{SP})}$&  
$W_2$ & $W_{(1-\text{SE})}$ & $W_{(1-\text{SP})}$\\
\hline
 $2$  & $.846_{(.335)}$ & $.011_{(.008)}$ & $.005_{(.007)}$
 &$.689_{(.270)}$ & $.007_{(.005)}$ & $.003_{(.002)}$&
 $.617_{(.231)}$ & $.006_{(.004)}$ & $.002_{(.002)}$\\
 $4$  & $1.30_{(.125)}$ & $.062_{(.041)}$ & $.035_{(.056)}$&
 $.957_{(.114)}$ & $.011_{(.003)}$ & $.005_{(.005)}$&
 $.862_{(.131)}$ & $.009_{(.003)}$ & $.003_{(.001)}$
 \\
 $6$  & $1.57_{(.145)}$ & $.100_{(.039)}$ & $.029_{(.047)}$
 &$1.05_{(.078)}$ & $.013_{(.002)}$ & $.006_{(.004)}$
 & $.981_{(.086)}$ & $.012_{(.002)}$ & $.004_{(.001)}$
 \\
$8$  & $1.65_{(.141)}$ & $.116_{(.037)}$ & $.023_{(.028)}$
&$1.11_{(.069)}$ & $.014_{(.001)}$ & $.008_{(.008)}$
& $1.05_{(.065)}$ & $.013_{(.002)}$ & $.005_{(.002)}$
\\
\hline
{\sc mcp} & & $n=200$ & &&$n=500$&&& $n=800$ &\\
\hline
$2$  & $.850_{(.337)}$ & $.010_{(.009)}$ & $.015_{(.045)}$&
$.690_{(.269)}$ & $.007_{(.005)}$ & $.010_{(.025)}$
& $.615_{(.232)}$ & $.006_{(.004)}$ & $.005_{(.013)}$
\\
$4$  & $2.18_{(.372)}$ & $.170_{(.099)}$ & $.192_{(.186)}$
&$1.54_{(.369)}$ & $.063_{(.049)}$ & $.143_{(.109)}$
& $.912_{(.190)}$ & $.011_{(.011)}$ & $.044_{(.077)}$
\\
$6$  & $2.89_{(.291)}$ & $.354_{(.095)}$ & $.098_{(.143)}$&
$1.79_{(.314)}$ & $.099_{(.073)}$ & $.218_{(.176)}$
& $1.42_{(.219)}$ & $.051_{(.035)}$ & $.140_{(.127)}$
\\
$8$  & $3.30_{(.221)}$ & $.457_{(.073)}$ & $.055_{(.086)}$&
$2.25_{(.274)}$ & $.179_{(.080)}$ & $.194_{(.129)}$
& $1.68_{(.269)}$ & $.080_{(.052)}$ & $.182_{(.121)}$
\\
\hline
Lasso & & $n=200$ & &&$n=500$&&& $n=800$ &\\
\hline
$2$  & $.935_{(.384)}$ & $.009_{(.009)}$ & $.154_{(.175)}$&
$.710_{(.268)}$ & $.007_{(.005)}$ & $.088_{(.082)}$ & $.611_{(.246)}$ & $.005_{(.005)}$ & $.090_{(.070)}$\\
$4$  & $2.40_{(.339)}$ & $.229_{(.099)}$ & $.144_{(.160)}$&
$1.70_{(.229)}$ & $.064_{(.050)}$ & $.266_{(.147)}$ & $1.46_{(.246)}$ & $.044_{(.035)}$ & $.268_{(.169)}$\\
$6$  & $3.03_{(.234)}$ & $.380_{(.076)}$ & $.105_{(.128)}$&
$1.97_{(.413)}$ & $.100_{(.083)}$ & $.320_{(.170)}$ & $1.53_{(.311)}$ & $.039_{(.050)}$ & $.368_{(.156)}$\\
$8$  & $3.45_{(.261)}$ & $.488_{(.067)}$ & $.058_{(.075)}$&
$2.47_{(.377)}$ & $.185_{(.118)}$ & $.273_{(.156)}$ & $1.75_{(.382)}$ & $.056_{(.074)}$ & $.381_{(.149)}$\\
\hline
\end{tabular}
\caption{Simulation results for 
Gaussian FMR with $K_0 = 2$ and $p = 10$.}
\label{K02_p10}
\end{table}

\begin{table}[H]
\small
\centering
\begin{tabular}{|c|ccc|ccc|ccc|}
\hline
$K$ & & $n=500$  & & &$n=800$ & && $n=1000$ & \\
\hline
{\sc scad} &  $W_2$ & $W_{(1-\text{SE})}$ & $W_{(1-\text{SP})}$&  
$W_2$ & $W_{(1-\text{SE})}$ & $W_{(1-\text{SP})}$&  
$W_2$ & $W_{(1-\text{SE})}$ & $W_{(1-\text{SP})}$\\
\hline
 $2$  & $.676_{(.288)}$ & $.007_{(.006)}$ & $.001_{(.001)}$
 &$.591_{(.265)}$ & $.005_{(.005)}$ & $.001_{(.001)}$&
 $.582_{(.232)}$ & $.005_{(.004)}$ & $.000_{(.000)}$\\
 $4$  & $1.15_{(.129)}$ & $.051_{(.037)}$ & $.002_{(.009)}$&
 $.871_{(.139)}$ & $.010_{(.008)}$ & $.001_{(.005)}$&
 $.829_{(.136)}$ & $.009_{(.003)}$ & $.001_{(.000)}$
 \\
 $6$  & $1.32_{(.118)}$ & $.068_{(.035)}$ & $.003_{(.010)}$
 &$1.10_{(.087)}$ & $.050_{(.021)}$ & $.001_{(.001)}$
 & $1.08_{(.081)}$ & $.058_{(.007)}$ & $.001_{(.000)}$
 \\
$8$  & $1.40_{(.132)}$ & $.078_{(.033)}$ & $.002_{(.004)}$
&$1.14_{(.084)}$ & $.051_{(.015)}$ & $.001_{(.001)}$
& $1.13_{(.067)}$ & $.050_{(.008)}$ & $.001_{(.000)}$
\\
\hline
{\sc mcp} & & $n=500$ & &&$n=800$&&& $n=1000$ &\\
\hline
$2$  & $.676_{(.288)}$ & $.007_{(.006)}$ & $.001_{(.005)}$&
$.591_{(.265)}$ & $.005_{(.005)}$ & $.001_{(.002)}$
& $.579_{(.234)}$ & $.005_{(.004)}$ & $.001_{(.003)}$
\\
$4$  & $1.97_{(.285)}$ & $.156_{(.085)}$ & $.056_{(.057)}$
&$1.51_{(.296)}$ & $.069_{(.039)}$ & $.064_{(.046)}$
& $1.37_{(.362)}$ & $.056_{(.043)}$ & $.054_{(.051)}$
\\
$6$  & $2.58_{(.298)}$ & $.308_{(.092)}$ & $.034_{(.044)}$&
$1.94_{(.321)}$ & $.136_{(.065)}$ & $.068_{(.038)}$
& $1.70_{(.329)}$ & $.098_{(.055)}$ & $.063_{(.044)}$
\\
$8$  & $3.04_{(.155)}$ & $.429_{(.052)}$ & $.012_{(.027)}$&
$2.41_{(.269)}$ & $.241_{(.078)}$ & $.064_{(.040)}$
& $2.08_{(.259)}$ & $.167_{(.064)}$ & $.064_{(.036)}$
\\
\hline
Lasso & & $n=500$ & &&$n=800$&&& $n=1000$ &\\
\hline
$2$  & $.733_{(.264)}$ & $.007_{(.006)}$ & $.018_{(.015)}$&
$.619_{(.256)}$ & $.005_{(.005)}$ & $.015_{(.012)}$ & 
$.599_{(.221)}$ & $.005_{(.004)}$ & $.012_{(.012)}$\\
$4$  & $2.22_{(.176)}$ & $.240_{(.055)}$ & $.018_{(.033)}$&
$1.75_{(.267)}$ & $.111_{(.082)}$ & $.069_{(.059)}$ &
$1.55_{(.214)}$ & $.071_{(.053)}$ & $.087_{(.052)}$\\
$6$  & $2.81_{(.116)}$ & $.380_{(.025)}$ & $.008_{(.015)}$&
$2.34_{(.283)}$ & $.221_{(.108)}$ & $.062_{(.055)}$ & 
$2.04_{(.329)}$ & $.135_{(.093)}$ & $.080_{(.051)}$\\
$8$  & $3.14_{(.141)}$ & $.454_{(.027)}$ & $.007_{(.015)}$&
$2.82_{(.246)}$ & $.351_{(.101)}$ & $.039_{(.050)}$ & 
$2.48_{(.301)}$ & $.219_{(.107)}$ & $.077_{(.047)}$\\
\hline
\end{tabular}
\caption{Simulation results for 
Gaussian FMR with $K_0 = 2$ and $p = 50$.}
\label{K02_p50}
\end{table}

\begin{table}[H]
\small
\centering
\begin{tabular}{|c|ccc|ccc|}
\hline
$K$ & & $n=1000$  & & &$n=1500$ & \\
\hline
{\sc scad} & $W_2$ & $W_{(1-\text{SE})}$ & $W_{(1-\text{SP})}$&  
$W_2$ & $W_{(1-\text{SE})}$ & $W_{(1-\text{SP})}$\\
\hline
 $2$  & $.539_{(.230)}$ & $.004_{(.004)}$ & $.000_{(.000)}$
 &$.475_{(.219)}$ & $.004_{(.003)}$ & $.000_{(.000)}$\\
 $4$  & $1.01_{(.119)}$ & $.019_{(.025)}$ & $.001_{(.003)}$&
 $.753_{(.154)}$ & $.009_{(.011)}$ & $.000_{(.000)}$
 \\
 $6$  & $1.19_{(.152)}$ & $.035_{(.026)}$ & $.002_{(.004)}$
 &$.883_{(.126)}$ & $.013_{(.015)}$ & $.000_{(.000)}$
 \\
$8$  & $1.33_{(.150)}$ & $.039_{(.030)}$ & $.003_{(.004)}$
&$.986_{(.110)}$ & $.014_{(.010)}$ & $.001_{(.000)}$
\\
\hline
{\sc mcp} & & $n=1000$  & & &$n=1500$ & \\
\hline
$2$  & $.535_{(.231)}$ & $.004_{(.004)}$ & $.000_{(.000)}$&
$.475_{(.219)}$ & $.004_{(.003)}$ & $.000_{(.000)}$
\\
$4$  & $1.60_{(.191)}$ & $.081_{(.039)}$ & $.042_{(.031)}$
&$.972_{(.312)}$ & $.021_{(.025)}$ & $.011_{(.020)}$
\\
$6$  & $1.90_{(.356)}$ & $.139_{(.077)}$ & $.047_{(.030)}$&
$1.45_{(.224)}$ & $.067_{(.039)}$ & $.027_{(.026)}$
\\
$8$  & $2.39_{(.304)}$ & $.253_{(.080)}$ & $.033_{(.025)}$&
$1.80_{(.277)}$ & $.122_{(.058)}$ & $.027_{(.018)}$
\\
\hline
Lasso & & $n=1000$  & & &$n=1500$ & \\
\hline
$2$  & $.565_{(.219)}$ & $.004_{(.004)}$ & $.007_{(.006)}$&
$.498_{(.210)}$ & $.003_{(.003)}$ & $.007_{(.006)}$\\
$4$  & $1.85_{(.237)}$ & $.147_{(.078)}$ & $.029_{(.035)}$&
$1.46_{(.169)}$ & $.059_{(.044)}$ & $.035_{(.027)}$\\
$6$  & $2.41_{(.248)}$ & $.265_{(.092)}$ & $.027_{(.031)}$&
$1.88_{(.267)}$ & $.102_{(.055)}$ & $.047_{(.030)}$\\
$8$  & $2.89_{(.148)}$ & $.391_{(.060)}$ & $.012_{(.020)}$&
$2.37_{(.246)}$ & $.202_{(.078)}$ & $.036_{(.023)}$\\
\hline
\end{tabular}
\caption{Simulation results for Gaussian FMR with $K_0 = 2$ and $p = 100$.}
\label{K02_p100}
\end{table}

We also conducted simulations for a Gaussian FMR model 
with $K_0 = 3$, and the parameter values: mixing probabilities 
$\pi_{1}^{0} = 0.30, \pi_{2}^{0} = 0.30, \pi_{3}^{0} = 0.40$, dispersion parameters 
$\phi_{1}^{0} = \phi_{2}^{0} = \phi_{3}^{0} = 1$, and sparse regression vectors  
\begin{align*}
    \bbeta_{1}^{0} &= (2, 2.5, 3, 0, \ldots, 0)^\top~,~ 
    \bbeta_{2}^{0} = (-2, 0, 2.5, 0, 1.5, 0, \ldots, 0)^\top \\
    \bbeta_{3}^{0} &= (1.5, 0, 0, 2, -2.5, 1.7, 0, \ldots, 0)^\top
\end{align*}
with dimensions $p=5, 10, 50, 100$. As the dimension
$p$ increases, the number of nonzero regression coefficients remains fixed while the number of zero coefficients grows.
The design points $\xx_i, i= 1,\ldots, n$ are generated 
from a Gaussian distribution as before. 

We generate the data $(\xx_1, Y_1), \ldots,(\xx_n, Y_n)$ 
from the above model with different covariate dimensions 
and sample sizes. For each random sample, 
we fit FMR models using the OMPLE $\hat G_n$ in equation \eqref{MPLE-Est},
considering the true order $K = K_0= 3$ and 
also overspecified orders $K=6, 9, 12$. 
We report results based only on the {\sc scad} penalty, as its
overall performance is 
generally better 
than the 
other two penalties in the scenarios considered here.
The results corresponding to dimensions
$p=5, 10$ are given in Table \ref{SCADK03_p5_10}, and those 
for dimensions $p=50, 100$ are given in Table 
\ref{SCADK03_p50_100}. The results indicate that as long as 
the sample size is sufficiently large relative to the total number of 
parameters in the overspecified FMR, OMPLE performs 
reasonably well in 
both parameter estimation 
accuracy and feature selection.

\begin{table}[H]
\centering
\begin{tabular}{|c|ccc|ccc|}
\hline
{\sc scad}&  & $p=5$ & & &$p=10$&\\
\hline
$(K, n)$& $W_2$ & $W_{(1-\text{SE})}$ & $W_{(1-\text{SP})}$
& $W_2$ & $W_{(1-\text{SE})}$ & $W_{(1-\text{SP})}$
\\
\hline
$(3,500)$ & $1.04_{(.289)}$ & $.017_{(.009)}$ & $.021_{(.012)}$&
$.987_{(.254)}$ & $.015_{(.009)}$ & $.007_{(.005)}$
\\
$(6,800)$ & $1.37_{(.085)}$ & $.030_{(.008)}$ & $.061_{(.043)}$&
$1.36_{(.090)}$ & $.030_{(.004)}$ & $.011_{(.004)}$\\
$(9, 1500)$ & $1.41_{(.067)}$ & $.023_{(.006)}$ & $.136_{(.062)}$&
$1.41_{(.068)}$ & $.027_{(.004)}$ & $.023_{(.013)}$\\
$(12,1500)$ & $1.53_{(.060)}$ & $.024_{(.008)}$ & $.138_{(.060)}$&
$1.49_{(.057)}$ & $.031_{(.005)}$ & $.027_{(.017)}$\\
\hline
\end{tabular}
\caption{Simulation results for Gaussian FMR with 
$K_0 = 3$.}
\label{SCADK03_p5_10}
\end{table}

\begin{table}[H]
\centering
\begin{tabular}{|c|ccc||c|ccc|}
\hline
{\sc scad}&  & $p=50$ & &&&$p=100$&\\
\hline
$(K, n)$& $W_2$ & $W_{(1-\text{SE})}$ & $W_{(1-\text{SP})}$&
$(K, n)$& $W_2$ & $W_{(1-\text{SE})}$ & $W_{(1-\text{SP})}$
\\
\hline
$(3,500)$ & $1.07_{(.345)}$ & $.017_{(.011)}$ & $.016_{(.017)}$&
$(3,800)$ & $.914_{(.265)}$ & $.014_{(.008)}$ & $.002_{(.005)}$
\\
$(6,1000)$ & $1.49_{(.078)}$ & $.078_{(.031)}$ & $.002_{(.001)}$&
$(6,1500)$ & $1.33_{(.098)}$ & $.058_{(.023)}$ & $.004_{(.004)}$
\\
$(9 ,2000)$ & $1.41_{(.069)}$ & $.031_{(.008)}$ & $.004_{(.001)}$&
$(9,2500)$ & $1.44_{(.078)}$ & $.053_{(.015)}$ & $.004_{(.006)}$
\\
$(12,2000)$ & $1.51_{(.068)}$ & $.044_{(.015)}$ & $.004_{(.001)}$&
$(12,3000)$ & $1.44_{(.081)}$ & $.050_{(.012)}$ & $.002_{(.002)}$
\\
\hline
\end{tabular}
\caption{Simulation results for Gaussian FMR with $K_0 = 3$.}
\label{SCADK03_p50_100}
\end{table}


\noindent
{\bf Example 2 (Binomial FMR)} 

Table~\ref{K02_p50_Binomial} reports the empirical means and standard deviations of the performance metrics \(W_2\), \(W_{(1-\text{SE})}\), and \(W_{(1-\text{SP})}\) for dimension \(p = 50\). The results are based on \(R = 100\) simulated samples from the Binomial FMR model in Example~2 of Section~\ref{sec:simul}, for three penalties.

\begin{table}[H]
\small
\centering
\begin{tabular}{|c|ccc|ccc|ccc|}
\hline
$K$ & & $n=500$  & & &$n=800$ & && $n=1000$ & \\
\hline
{\sc scad} &  $W_2$ & $W_{(1-\text{SE})}$ & $W_{(1-\text{SP})}$&  
$W_2$ & $W_{(1-\text{SE})}$ & $W_{(1-\text{SP})}$&  
$W_2$ & $W_{(1-\text{SE})}$ & $W_{(1-\text{SP})}$\\
\hline
 $2$  & $.636_{(.259)}$ & $.006_{(.005)}$ & $.000_{(.000)}$
 &$.600_{(.236)}$ & $.005_{(.004)}$ & $.000_{(.000)}$&
 $.571_{(.239)}$ & $.005_{(.004)}$ & $.000_{(.000)}$\\
 $4$  & $1.17_{(.091)}$ & $.071_{(.014)}$ & $.001_{(.001)}$&
 $.872_{(.132)}$ & $.009_{(.004)}$ & $.001_{(.001)}$&
 $.797_{(.135)}$ & $.007_{(.003)}$ & $.000_{(.000)}$
 \\
 $6$  & $1.33_{(.050)}$ & $.110_{(.002)}$ & $.001_{(.001)}$
 &$1.15_{(.078)}$ & $.057_{(.002)}$ & $.001_{(.001)}$
 & $1.13_{(.094)}$ & $.055_{(.003)}$ & $.002_{(.002)}$
 \\
$8$  & $1.35_{(.033)}$ & $.092_{(.001)}$ & $.001_{(.001)}$
&$1.26_{(.053)}$ & $.051_{(.002)}$ & $.003_{(.001)}$
& $1.28_{(.056)}$ & $.050_{(.002)}$ & $.005_{(.002)}$
\\
\hline
{\sc mcp} & & $n=500$ & &&$n=800$&&& $n=1000$ &\\
\hline
$2$  & $.632_{(.263)}$ & $.006_{(.005)}$ & $.000_{(.000)}$&
$.597_{(.245)}$ & $.005_{(.004)}$ & $.000_{(.000)}$
& $.571_{(.242)}$ & $.005_{(.004)}$ & $.000_{(.000)}$
\\
$4$  & $.977_{(.110)}$ & $.010_{(.003)}$ & $.004_{(.002)}$
&$.931_{(.171)}$ & $.009_{(.004)}$ & $.009_{(.008)}$
& $.788_{(.148)}$ & $.007_{(.003)}$ & $.001_{(.002)}$
\\
$6$  & $1.38_{(.105)}$ & $.061_{(.002)}$ & $.015_{(.006)}$&
$1.18_{(.082)}$ & $.058_{(.003)}$ & $.003_{(.002)}$
& $1.23_{(.104)}$ & $.056_{(.003)}$ & $.010_{(.005)}$
\\
$8$  & $1.33_{(.053)}$ & $.054_{(.002)}$ & $.005_{(.002)}$&
$1.38_{(.084)}$ & $.051_{(.002)}$ & $.009_{(.004)}$
& $1.29_{(.053)}$ & $.050_{(.002)}$ & $.006_{(.002)}$
\\
\hline
Lasso & & $n=500$ & &&$n=800$&&& $n=1000$ &\\
\hline
$2$  & $1.35_{(.199)}$ & $.010_{(.008)}$ & $.019_{(.029)}$&
$.976_{(.150)}$ & $.006_{(.005)}$ & $.000_{(.000)}$ & 
$.858_{(.162)}$ & $.005_{(.004)}$ & $.000_{(.000)}$\\
$4$  & $1.13_{(.058)}$ & $.009_{(.003)}$ & $.014_{(.008)}$&
$1.09_{(.071)}$ & $.007_{(.003)}$ & $.004_{(.004)}$ &
$.987_{(.071)}$ & $.006_{(.003)}$ & $.004_{(.005)}$\\
$6$  & $1.39_{(.070)}$ & $.035_{(.025)}$ & $.007_{(.005)}$&
$1.37_{(.076)}$ & $.034_{(.025)}$ & $.001_{(.002)}$ & 
$1.27_{(.056)}$ & $.013_{(.016)}$ & $.004_{(.003)}$\\
$8$  & $1.46_{(.102)}$ & $.048_{(.025)}$ & $.002_{(.002)}$&
$1.50_{(.088)}$ & $.072_{(.021)}$ & $.001_{(.001)}$ & 
$1.35_{(.047)}$ & $.020_{(.018)}$ & $.002_{(.002)}$\\
\hline
\end{tabular}
\caption{Simulation results for 
Binomial FMR with $K_0 = 2$ and $p = 50$.}
\label{K02_p50_Binomial}
\end{table}

\newpage

\section{Additional Plots and Experiments for the Real Data}
\label{appendix:plots_real_data}

In this section, we present additional details on the real data analysis discussed in Section~\ref{sec:real_data} of the main manuscript. 

\begin{figure}[H]       
    \centering                
    \includegraphics[width=0.78\linewidth, height=5.5cm]{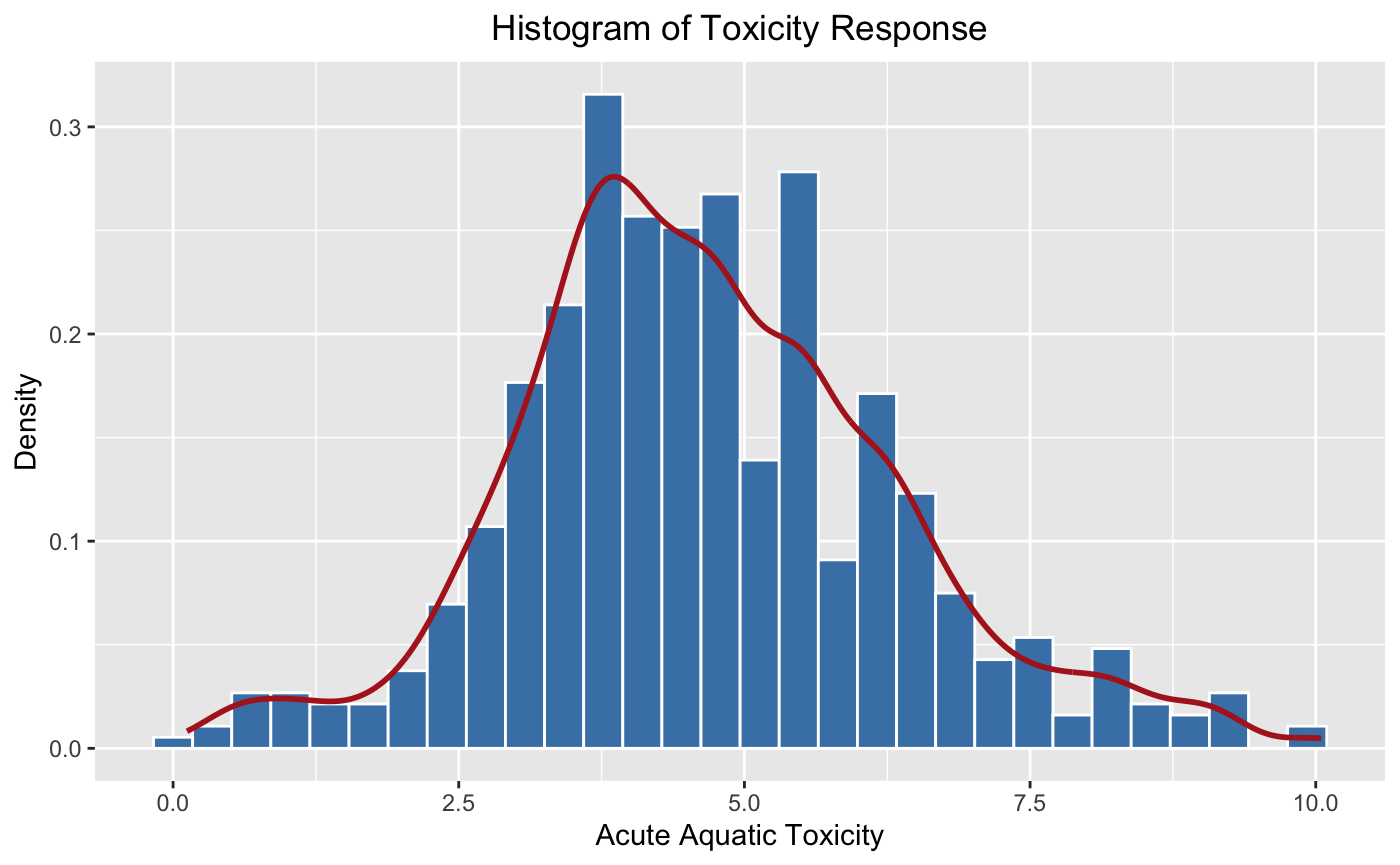} 
    \caption{Histogram of the response $y$ in the Aquatic Toxicity dataset.}
    \label{real_data_hist}       
    \end{figure}

\noindent
{\bf Histogram}:
Figure~\ref{real_data_hist} displays the histogram of the toxicity concentration, revealing a multimodal pattern in its marginal empirical distribution. Moreover, the distribution's shape suggests that a Gaussian FMR model could be a reasonable choice for this data. 


\begin{figure}[H]
    \begin{subfigure}{0.49\textwidth}
     \includegraphics[width=\textwidth]{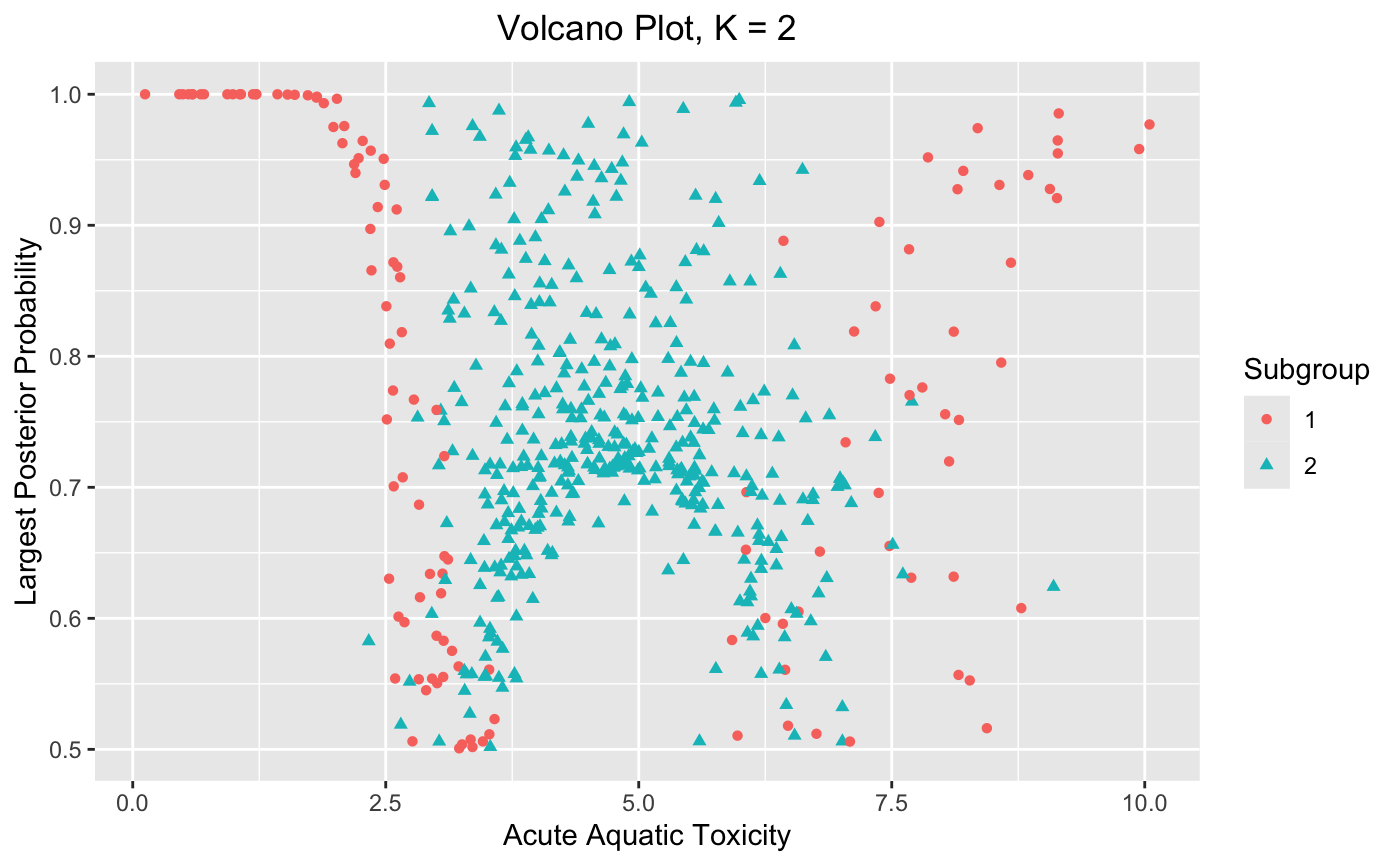}
     \caption{}
     \label{VolK2}
    \end{subfigure}
    \hfill
    \begin{subfigure}{0.49\textwidth}
     \includegraphics[width=\textwidth]{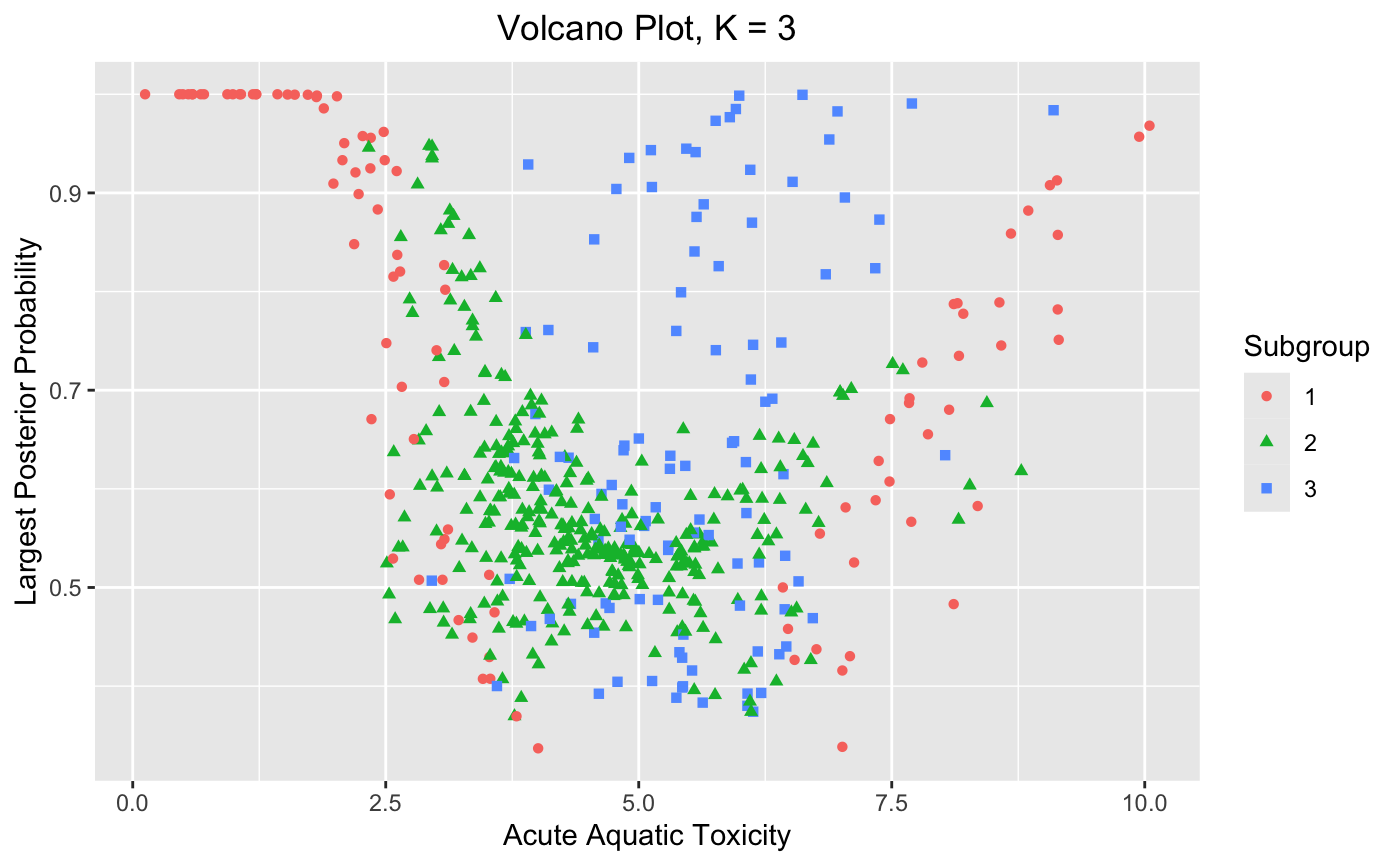}
     \caption{}
     \label{VolK3}
    \end{subfigure}
 
    \medskip
    \begin{subfigure}{0.49\textwidth}
     \includegraphics[width=\textwidth]{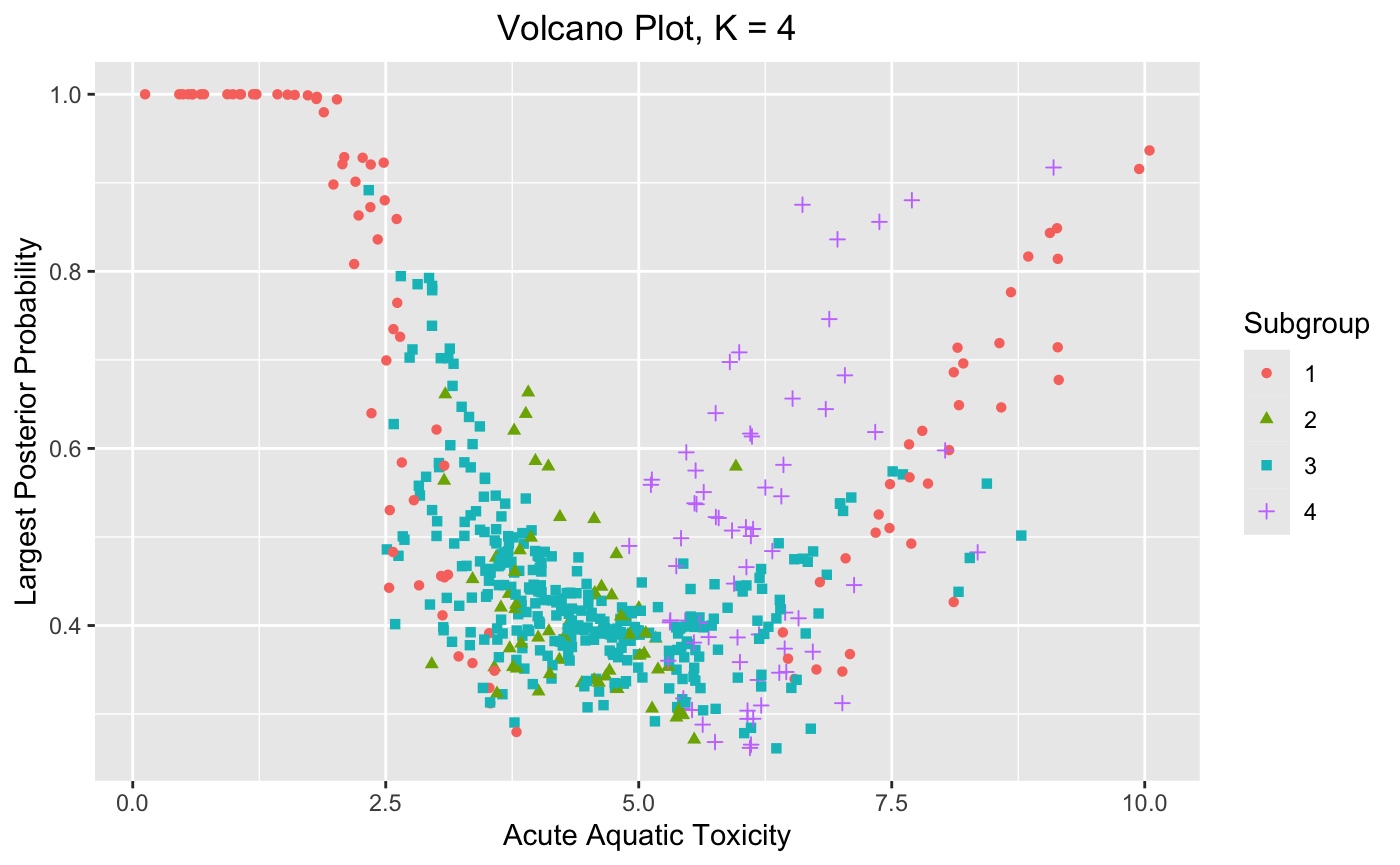}
     \caption{}
     \label{VolK4}
    \end{subfigure}
    \hfill
    \begin{subfigure}{0.49\textwidth}
     \includegraphics[width=\textwidth]{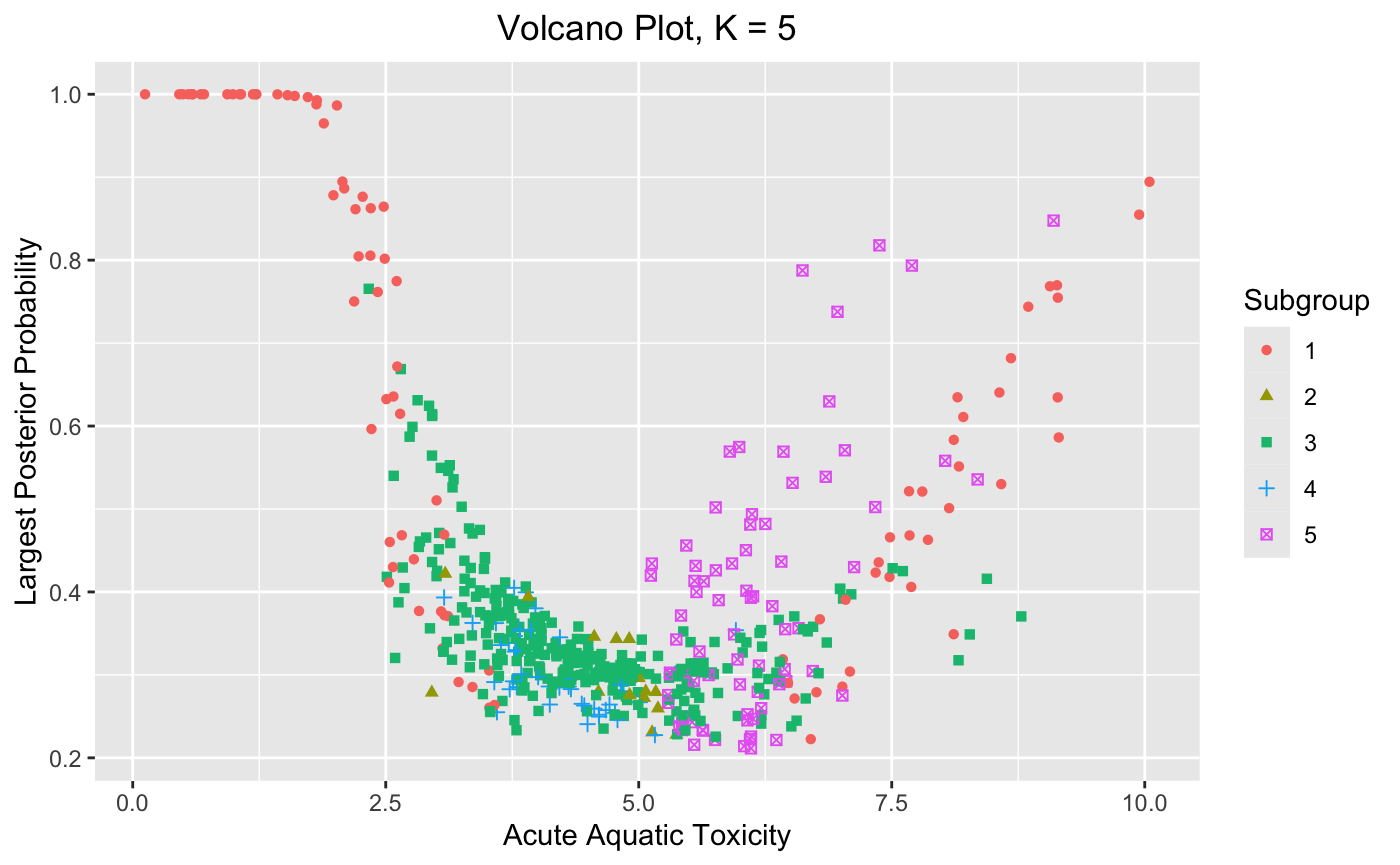}
     \caption{}
     \label{VolK5}
    \end{subfigure}
 \caption{Volcano plots for the Aquatic Toxicity data with $K = 2, 3, 4, 5$. (a) $K = 2$; (b) $K = 3$; (c) $K = 4$; (d) $K = 5$.}
 \label{Volcano2345}
\end{figure}

\noindent
{\bf Volcano plots (maximum-posterior-probability plots)}: 
Figure~\ref{Volcano2345} presents the volcano plots obtained based on the fitted Gaussian FMR models with orders \(K = 2, \ldots, 5\). For each fitted model, we first compute the posterior probabilities in equation~\eqref{weight} for every observation \((\xx_i, y_i)\), 
\(i = 1, \ldots, n\). For each observation, we then compute the largest posterior probability of belonging to any component of the fitted FMR model with order \(K\). Finally, we plot these maximum posterior probabilities (MPP) against the corresponding \(y_i\). From the four volcano plots in Figure \ref{Volcano2345}, 
we observe that for the fitted Gaussian FMR with $K = 4$ and $K = 5$, many of the observations have low MPP, 
which may be an indication of overfitting the number of components. In contrast, for the fitted FMR with $K = 2, 3$, many of the 
observations have MPP
higher than $50\%$. Note that as discussed 
in Section~\ref{sec:real_data}, 
BIC selected the FMR model with $K=3$. 

\noindent
{\bf Selected covariates in the fitted FMR model with $K=3$}:
For the fitted Gaussian FMR model with \(K = 3\), the covariates selected in the first component (subgroup) are TPSA(tot), MLOGP, 
and GATS1p; in the second component, only MLOGP is selected; 
and in the third component, no covariates are selected.
The covariates TPSA(tot) and MLOGP have both positive 
estimated regression coefficients implying that 
in component 1 (with toxicity concentration approximately 
less than $2.5$ or above $7.5$, Figure \ref{Volcano2345} (b)), greater polar surface area
and lipophilicity 
are associated with higher fitted toxicity concentration. 
On the other hand, covariate GATS1p in this component has 
a negative estimated coefficient suggesting that big contrasts in polarizability between adjacent atoms can lead to lower toxicity concentrations. As for subgroup 2 (intermediate toxicity concentration levels approximately in the range $[2.5,5]$), the only selected covariate is MLOGP with a positive estimated coefficient also indicating that for the compounds in this group, greater lipophilicity can increase the toxicity concentration that causes $50\%$ mortality. For subgroup 3 (with toxicity concentration approximately in the range $[5,7.5]$), no molecular descriptor is 
selected suggesting the $8$ descriptors provide limited explanation beyond the component intercept or variance, possibly because the effects are weak and absorbed by residual variance.

\begin{figure}[H]
    \begin{subfigure}{0.49\textwidth}
     \includegraphics[width=\textwidth]{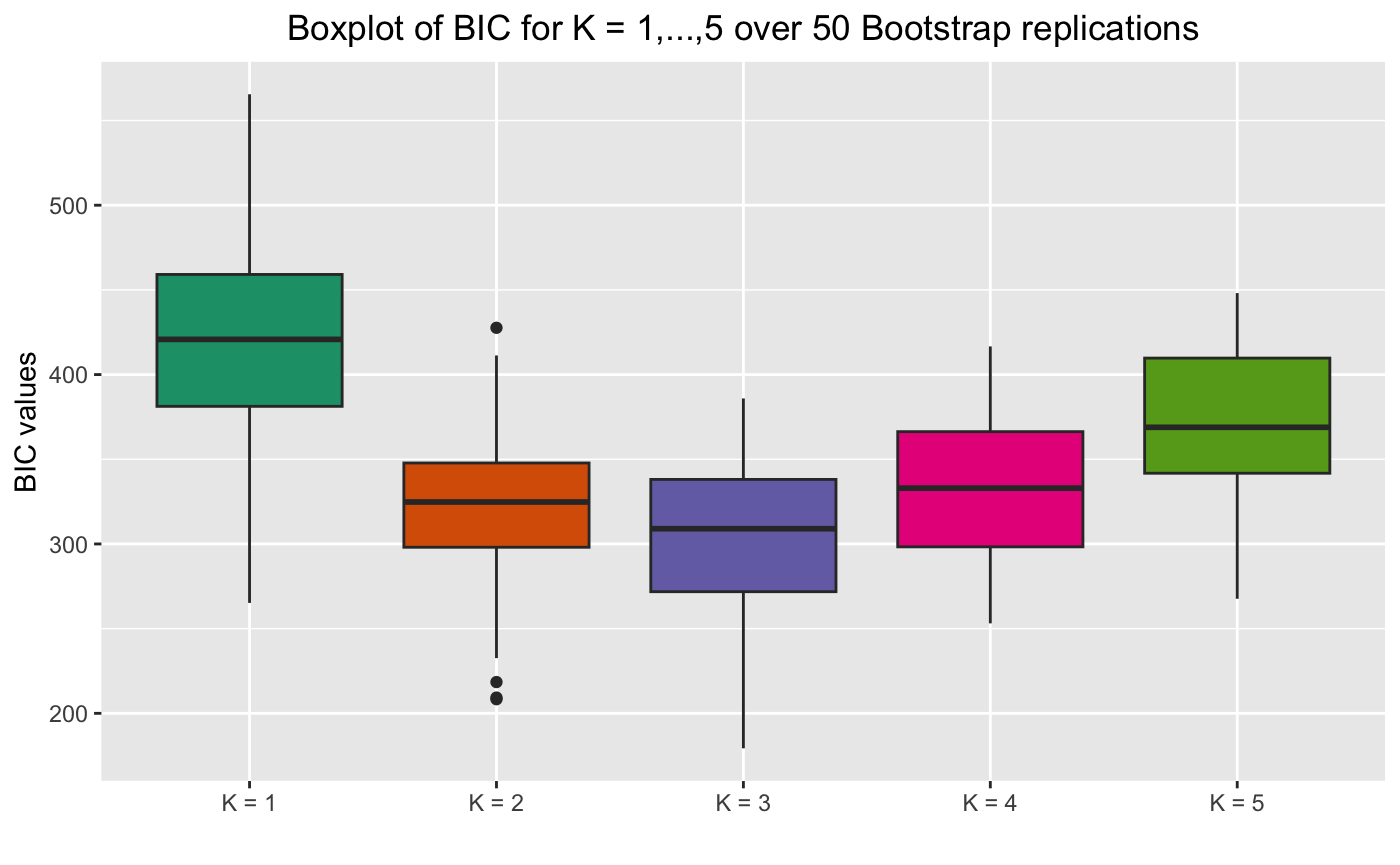}
     \caption{}
     \label{real_data_boot_BIC_box}
    \end{subfigure}
    \hfill
    \begin{subfigure}{0.49\textwidth}
     \includegraphics[width=\textwidth]{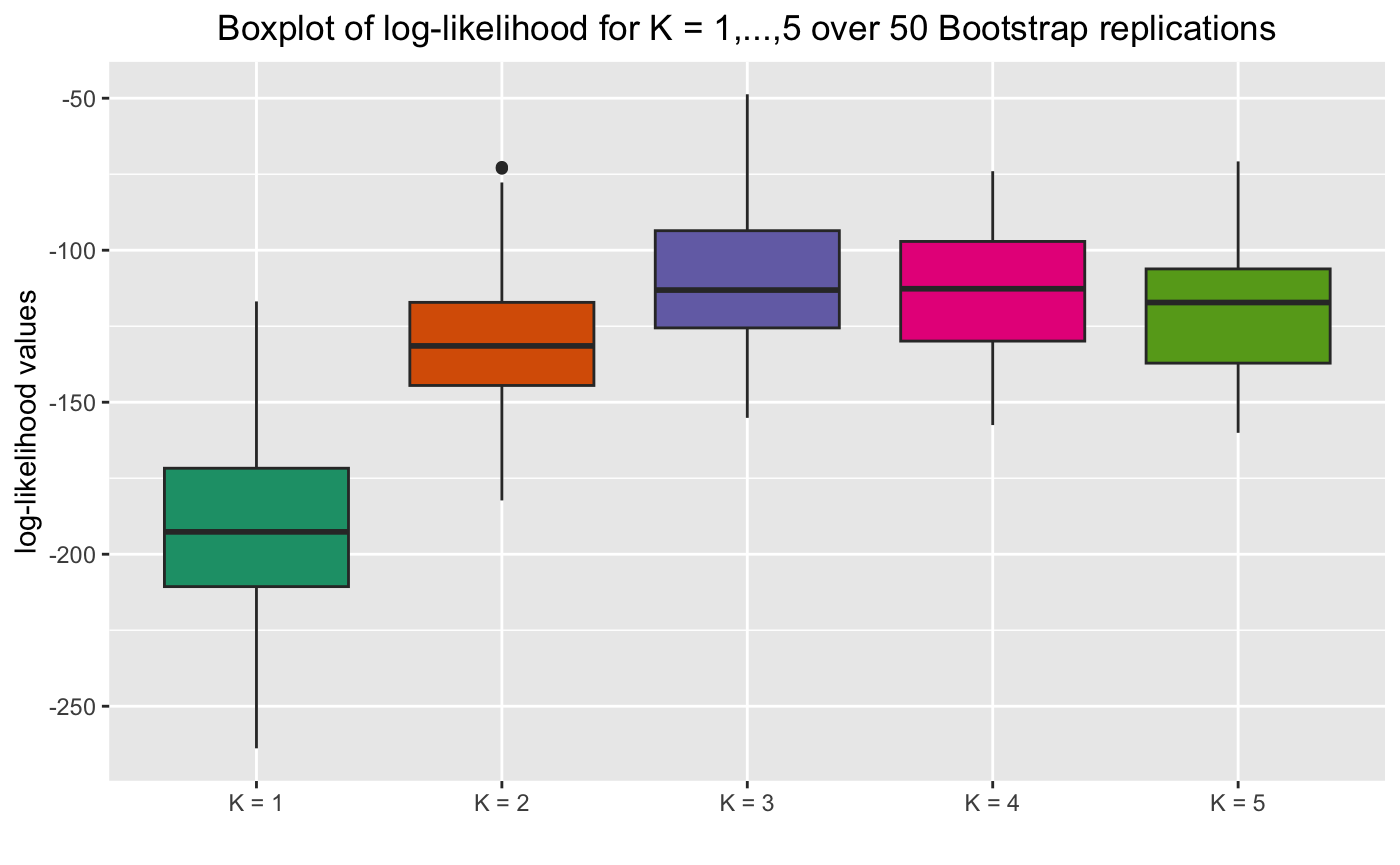}
     \caption{}
     \label{real_data_boot_loglike_box}
    \end{subfigure}
 \caption{Boxplots of BIC (a) and log-likelihood (b) values 
 of the $50$ fitted Gaussian FMR
 models, over $B = 50$ bootstrap samples, 
 for each $K = 1,\ldots,5$. 
 }
 \label{real_data_boot_BIC_loglike_boxes}
\end{figure}

\noindent
{\bf Bootstrap experiments}: 
To assess the stability of selecting \(K = 3\) based on the BIC, as well as the stability of the covariates chosen in the final model, we performed a bootstrap analysis as follows. We generated \(B = 50\) bootstrap samples (with replacement) from the original dataset. For each sample, we fitted Gaussian FMR models with \(K = 1, \ldots, 5\), following the same procedure as for the original data, and computed the corresponding BIC and log-likelihood values. Table~\ref{real_data_BIC_boot} reports the average BIC and log-likelihood over the 50 bootstrap samples for each \(K\). The averages indicate that the BIC attains its minimum at \(K = 3\) while the log-likelihood attains its maximum at \(K = 3\), consistent with the results from the original dataset (see Table~\ref{real_data_BIC} in Section~\ref{sec:real_data}). Furthermore, Figure~\ref{real_data_boot_BIC_loglike_boxes} presents boxplots of the BIC and log-likelihood values for the 50 fitted models across \(K = 1, \ldots, 5\). In the BIC plot, the box for \(K = 3\) lies below those for \(K = 1, 2, 4, 5\), whereas in the log-likelihood plot, the box for \(K = 3\) lies above the others. These patterns further confirm the stability of selecting \(K = 3\).

\begin{table}[h]
    \centering
    \begin{tabular}{|c|c|c|c|c|c|}
    \hline
     $K$ & 1 & 2 & 3 & 4 & 5  \\
    \hline
     Average BIC & 421.66 & 320.99 & {\bf 303.70} & 333.14 & 371.30\\
     \hline
     Average log-likelihood & -191.92 & -129.36 & {\bf -109.75} & -113.00 & -118.97\\
     \hline
    \end{tabular}
    \caption{Average BIC and log-likelihood values 
    corresponding to different fitted Gaussian FMRs 
    over $B = 50$ bootstrap samples.}
    \label{real_data_BIC_boot}
\end{table}

\begin{figure}[H]
 \begin{subfigure}{0.49\textwidth}
     \includegraphics[width=\textwidth]{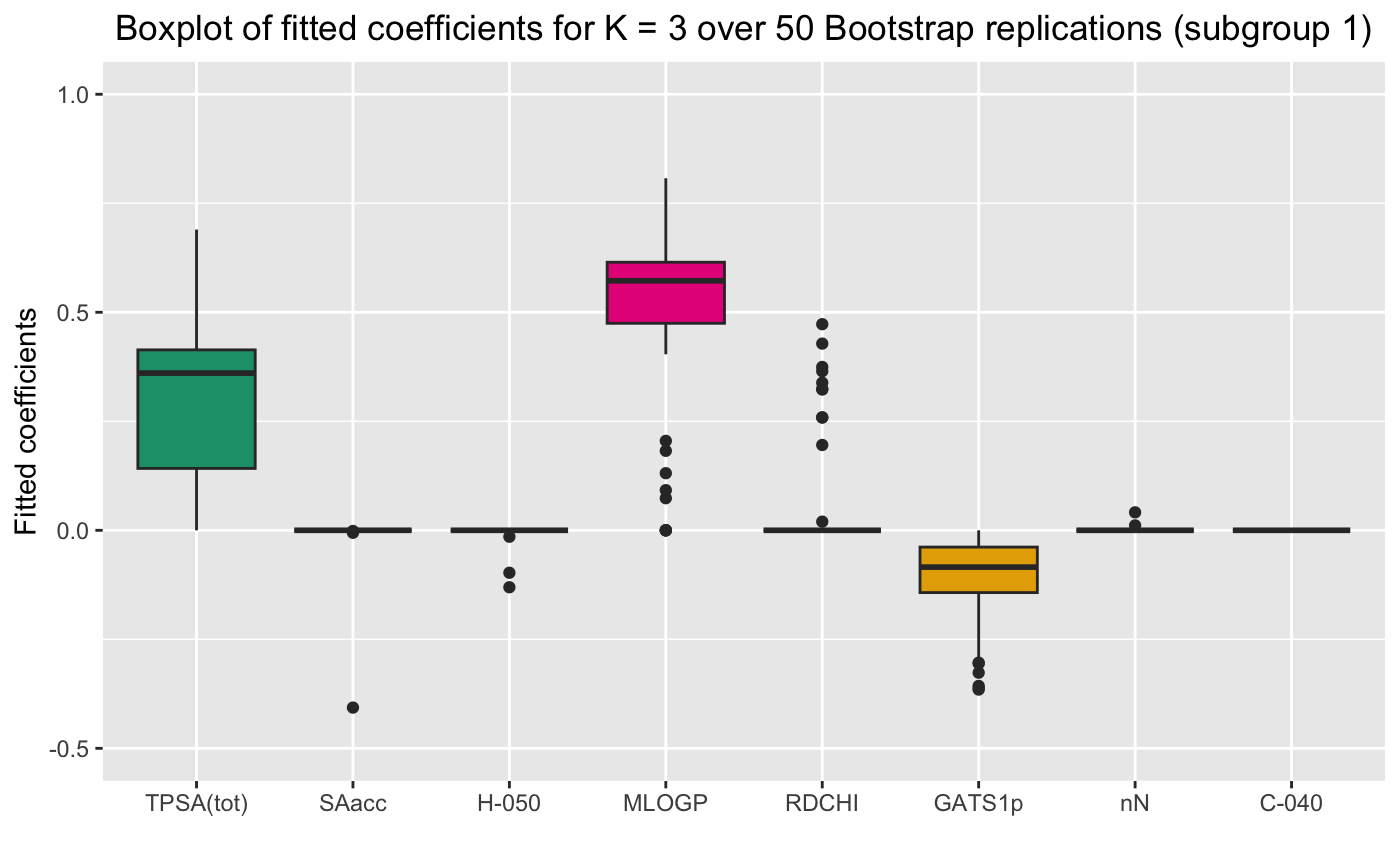}
     \caption{}
     \label{real_data_boot_subgroup1_plot}
 \end{subfigure}
 \hfill
 \begin{subfigure}{0.49\textwidth}
     \includegraphics[width=\textwidth]{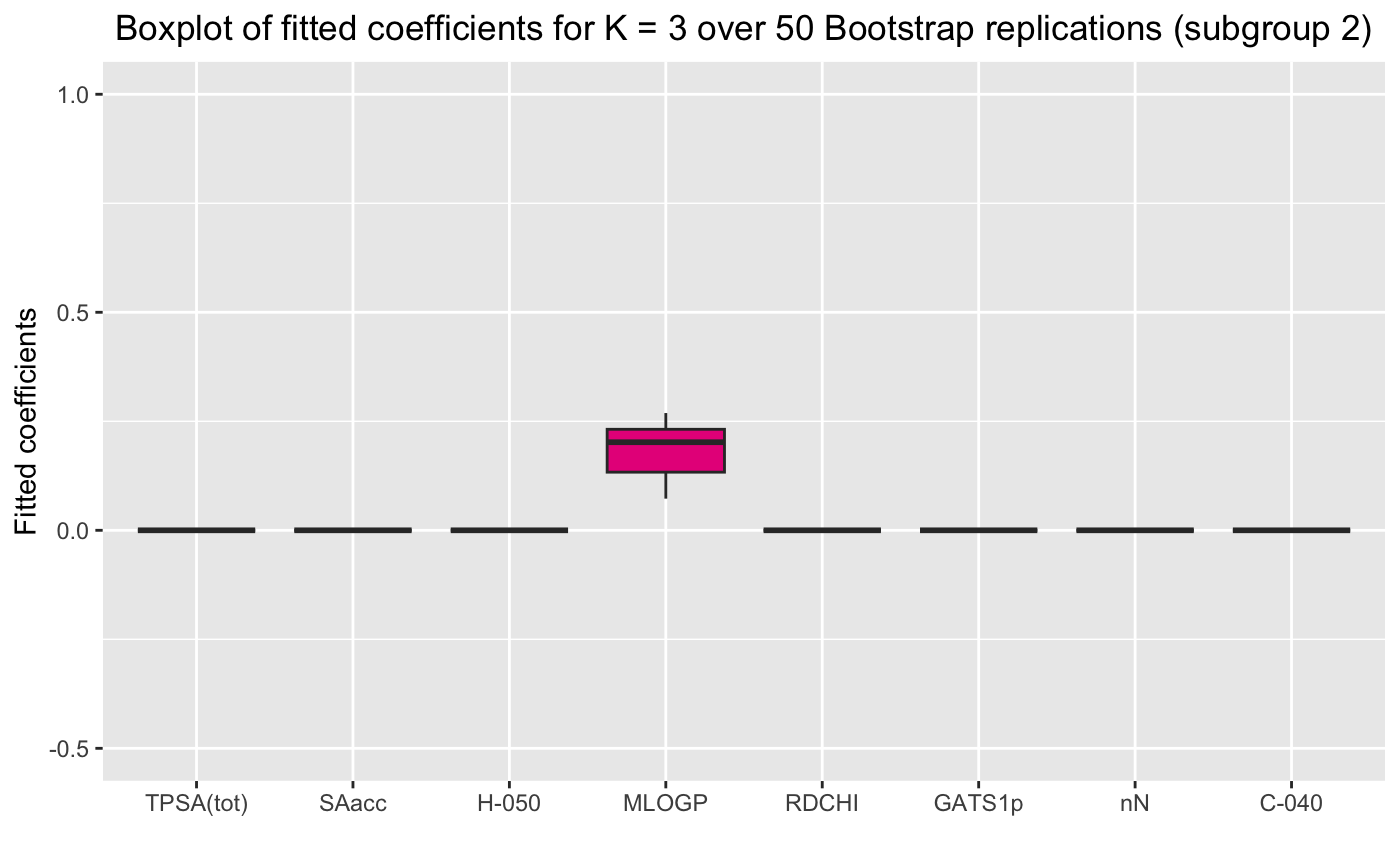}
     \caption{}
     \label{real_data_boot_subgroup2_plot}
 \end{subfigure}
 
 \medskip
 \centering
 \begin{subfigure}{0.55\textwidth}
     \includegraphics[width=\textwidth]{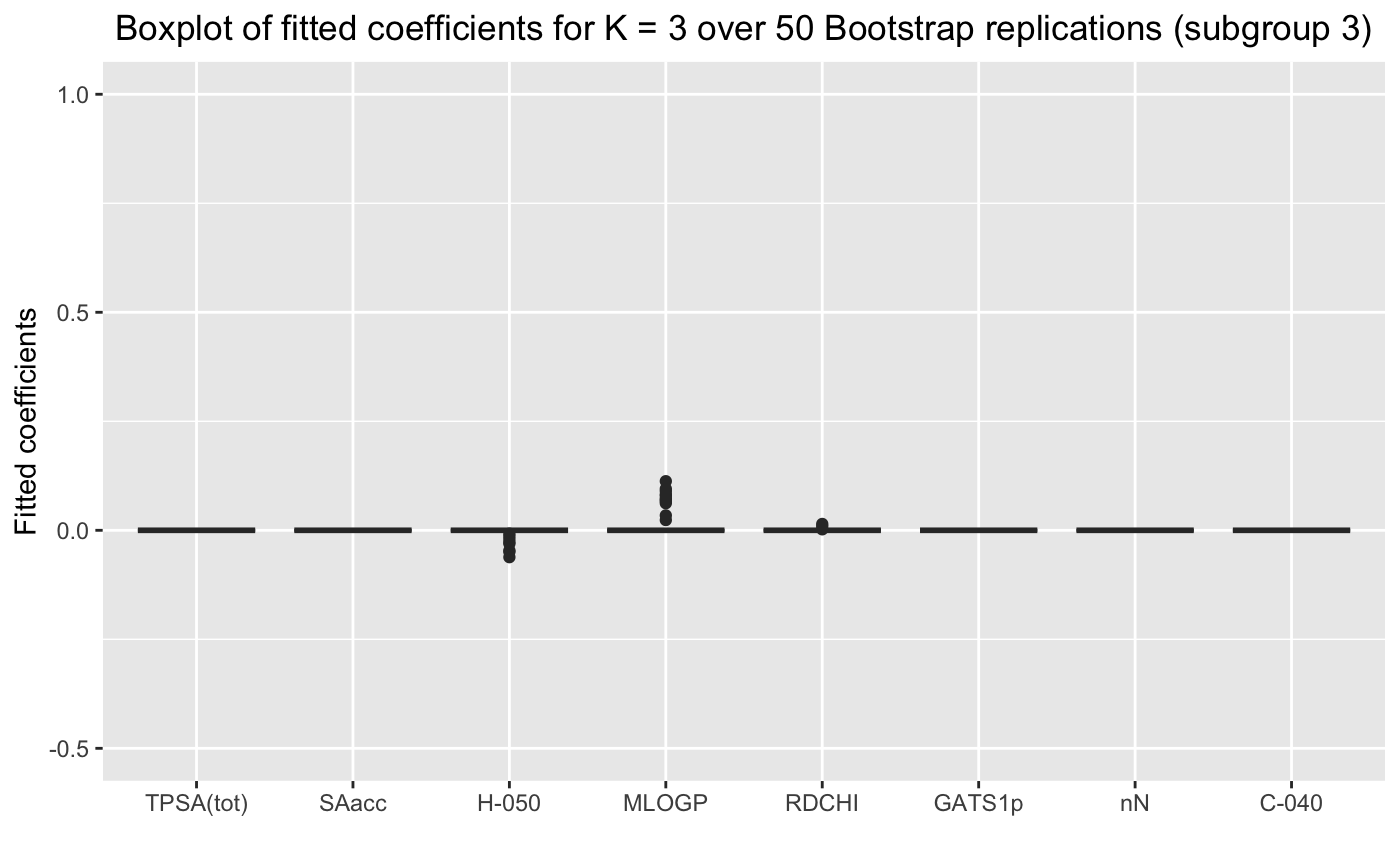}
     \caption{}
     \label{real_data_boot_subgroup3_plot}
 \end{subfigure}
 
 \caption{Boxplots of the fitted coefficients in each of the three components of the $K = 3$ models fitted to the $50$ bootstrap samples. (a) Boxplot of fitted coefficients for subgroup 1; (b) boxplot of fitted coefficients for subgroup 2; (c) boxplot of fitted coefficients for subgroup 3.}
 \label{real_data_boot_subgroups_plot}
\end{figure}

We also examined the stability of feature selection for 
the fitted FMR model with \(K = 3\). For each of the three mixture components in this model, we constructed boxplots of the estimated regression coefficients for the eight covariates, based on the \(B = 50\) bootstrap samples. The resulting boxplots are shown in Figure~\ref{real_data_boot_subgroups_plot}. 
As seen in panels (a)--(c), the patterns are largely 
consistent with the covariates selected in the fitted FMR 
model with \(K = 3\) based on the original data. 
Specifically, the covariates selected based on the 
original data have boxplots clearly separated from zero, 
whereas the remaining covariates have boxplots that are 
nearly flat and centered around zero.

\bibliographystyle{abbrvnat}
\bibliography{manuscript}

\end{document}